\documentclass[10pt,rmp,twocolumn,nofootinbib,amsmath,floatfix,aps,longbibliography]{revtex4-1}

\usepackage{graphicx}
\usepackage{epsfig}
\usepackage{amsmath}
\usepackage{amssymb}
\usepackage{enumerate}
\newcommand{\ket}[1]{|#1\rangle}
\begin{document}
\title{Delayed-choice gedanken experiments and their realizations}

\author{Xiao-song Ma}
\email{xiaosong.ma@nju.edu.cn}
\affiliation{Institute for Quantum Optics and Quantum Information
(IQOQI), Austrian Academy of Sciences, Boltzmanngasse 3, 1090
Vienna, Austria}
\affiliation{Department of Electrical
Engineering,~Yale University,~15 Prospect Street,~New Haven,~CT~06520,~USA}
\affiliation{National Laboratory of Solid State Microstructures, School of Physics, Collaborative Innovation Center of Advanced Microstructures, Nanjing University, Nanjing 210093, China}

\author{Johannes Kofler}
\email{johannes.kofler@mpq.mpg.de} \affiliation{Max Planck Institute
of Quantum Optics (MPQ), Hans-Kopfermann-Strasse\ 1, 85748 Garching,
Germany}

\author{Anton Zeilinger}
\email{anton.zeilinger@univie.ac.at} \affiliation{Vienna Center of
Quantum Science and Technology (VCQ), University of Vienna,
Boltzmanngasse 5, 1090 Vienna, Austria} \affiliation{Institute for
Quantum Optics and Quantum Information (IQOQI), Austrian Academy of
Sciences, Boltzmanngasse 3, 1090 Vienna, Austria}

\begin{abstract}
The wave-particle duality dates back to Einstein's explanation of
the photoelectric effect through quanta of light and de Broglie's
hypothesis of matter waves. Quantum mechanics uses an abstract
description for the behavior of physical systems such as photons,
electrons, or atoms. Whether quantum predictions for single systems
in an interferometric experiment allow an intuitive understanding in
terms of the particle or wave picture, depends on the specific
configuration which is being used. In principle, this leaves open
the possibility that quantum systems always behave either definitely
as a particle or definitely as a wave in every experimental run by a
priori adapting to the specific experimental situation. This is
precisely what is tried to be excluded by delayed-choice
experiments, in which the observer chooses to reveal the particle or
wave character of a quantum system -- or even a continuous
transformation between the two -- at a late stage of the experiment.
The history of delayed-choice gedanken experiments, which
can be traced back to the early days of quantum mechanics, is reviewed. Their experimental realizations, in particular Wheeler's
delayed choice in interferometric setups as well as delayed-choice
quantum erasure and entanglement swapping are discussed. The latter is
particularly interesting, because it elevates the wave-particle
duality of a single quantum system to an entanglement-separability
duality of multiple systems.
\end{abstract}

\maketitle

\tableofcontents

\section{Introduction}
\label{introduction} In the 17th century, two different theories of
light were developed. While Huygens explained optical phenomena
by a theory of waves, Newton put forward a corpuscular description
where light consists of a stream of fast particles. At first, the
large authority of Newton led to the general acceptance of the
corpuscular theory. However, at the beginning of the 19th century,
Young demonstrated the wave character of light, in particular by
showing interference fringes in the shadow of a ``slip of card,
about one-thirtieth of an inch in breadth," formed by the ``portions
of light passing on each side" \cite{Young1804}. Many other
subsequent experiments further established the wave nature of light,
in particular the discovery of electromagnetic waves with light
being a special case.

The picture changed again in 1905, when Einstein explained the
photoelectric effect with his hypothesis that light consists of
``energy quanta which move without splitting and can only be
absorbed or produced as a whole" \cite{Eins1905}. These massless
corpuscles of light, called photons, carry a specific amount of
energy $E=h\nu$ with $h$ being Planck's constant and $\nu$ the
light's frequency. In 1909, Taylor performed a low-intensity
Young-type experiment, measuring the shadow of a needle with an
exposure time of the photographic plate of 3 months \cite{Tayl1909}.
Despite the feeble light with on average less than one photon at a
time, the interference pattern was observed.

In 1924, de Broglie postulated that also all massive particles
behave as waves \cite{deBr1924}. The wavelength associated with a
particle with momentum $p$ is given by $\lambda=h/p$. This
\textit{wave-particle duality} was confirmed experimentally through
diffraction of an electron beam at a nickel crystal \cite{Davi1927}
and through diffraction of helium atoms at a crystal face of lithium
fluoride \cite{Estermann1930}. In 1961, the first nonphotonic
double-slit-type experiment was performed using electrons
\cite{Joen1961}. A good decade later, neutron interference \cite{Rauch1974} allowed
one to measure the quantum-mechanical phase shift caused by the Earth's
gravitational field \cite{Colella1975}. In modern interferometric
experiments, the wave nature of molecules of approximately 7000
atomic mass units and 1 pm de Broglie wavelength has been
demonstrated \cite{Gerl2011}.

In the language of quantum mechanics, the wave-particle duality is
reflected by the \textit{superposition principle}, i.e.\ the fact
that individual systems are described by quantum states, which can
be superpositions of different states with complex amplitudes. In a
Young-type double-slit experiment, every quantum system is at one
point in time in an equal-weight superposition of being at the left
and the right slit. When detectors are placed directly at
the slits, the system is found only at one of the slits, reflecting
its particle character. At which slit an individual system is found
is completely random. If, however, detectors are not placed at the
slits but at a larger distance, the superposition state will evolve
into a state which gives rise to an interference pattern, reflecting
the wave character of the system. This pattern cannot emerge when
the state at the slit would have been a mere classical mixture of
systems being at the left or the right slit.

To make things more precise, we consider a situation almost
equivalent to the double-slit experiment, namely quantum systems,
e.g.\ photons, electrons, or neutrons, which enter a Mach-Zehnder
interferometer (MZI)~\cite{Zehnder1891,Mach1892} via a
semitransparent mirror (beam splitter). We denote the transmitted
and reflected arm by \textit{b} and \textit{a}, respectively (Fig.~\ref{figMZI}). Let
there be a phase shift $\varphi$ in the reflected arm \textit{a},
additionally to a $\frac{\pi}{2}$ shift due to the reflection. Then
the quantum state of the system is a superposition of the
two path states with in general complex
amplitudes:\begin{figure}[tb]
\begin{center}
\includegraphics{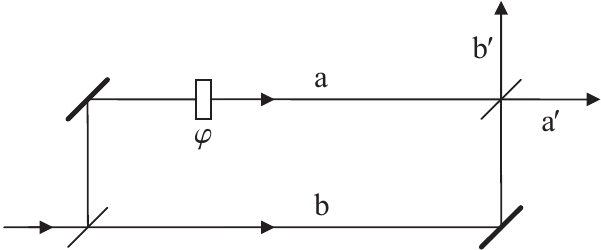}
\caption{Schematic of a Mach-Zehnder interferometer. A quantum
system enters from the left via a semitransparent beam splitter.
When detectors are placed in the paths a and b inside the
interferometer, the system is found in one and only one of the arms
with probability $\tfrac{1}{2}$ each. This reflects the picture that
it traveled one of the two paths as a particle. If, however,
detectors are not placed inside the interferometer but at the exit
ports a$^{\prime}$ and b$^{\prime}$ after the second beam splitter,
the probability of detection depends on the phase $\varphi$. This
reflects the view that the system traveled both paths a and b as a
wave, leading to constructive and destructive interference.}
\label{figMZI}
\end{center}
\end{figure}
\begin{equation}
\left\vert \psi\right\rangle =\tfrac{1}{\sqrt{2}}(\left\vert
\textit{b}\right\rangle
+\text{i}\,\text{e}^{\text{i}\varphi}\left\vert
\textit{a}\right\rangle ).
\end{equation}
Whenever one decides to measure through which path the system is
traveling by putting detectors into the arms a and b, one will find
it in one and only one arm, in agreement with its particle
character. Until the measurement the system is considered to be in a
superposition of both paths. The state $\left\vert
\psi\right\rangle$ determines only the \textit{probabilities} for
the respective outcomes a and b. They are given by the squared
modulus of the amplitudes and are thus
$p_{\textrm{a}}=p_{\textrm{b}}=\tfrac{1}{2}$. If, however, the two
paths are recombined on a second beam splitter with outgoing paths
a$^{\prime}$ and b$^{\prime}$, the quantum state will (up to a
global phase) be transformed into%
\begin{equation}
|\psi^{\prime}\rangle=\cos\tfrac{\varphi}{2}\,|\textrm{a}^{\prime}\rangle-\sin
\tfrac{\varphi}{2}\,|\textrm{b}^{\prime}\rangle.
\end{equation}
This state gives rise to detection probabilities
$p_{\textrm{a}^{\prime}}=\cos^{2}\tfrac{\varphi}{2}$ and
$p_{\textrm{b}^{\prime}}=\sin^{2}\tfrac{\varphi}{2}$. The
$\varphi$-dependent interference fringes indicate that the system
traveled through the interferometer through both arms, reflecting
its wave character. Particle and wave behavior are
\textit{complimentary} \cite{Bohr1928} in the sense that they can
only be revealed in different experimental contexts and not
simultaneously (see section II.C).

When two physical systems 1 and 2 interact with each other, they
will in general end up in an \textit{entangled state}, i.e., a
(non-separable) superposition of joint states. An example would be
two particles, each of them in a separate interferometer:
\begin{equation}
\left\vert \Psi\right\rangle _{12}= \cos\alpha\left\vert
\textrm{a}\right\rangle
_{1}\!\left\vert \bar{\textrm{a}}\right\rangle _{2}+\sin\alpha\,\text{e}%
^{\text{i}\varphi}\left\vert \textrm{b}\right\rangle
_{1}\!\left\vert \bar{\textrm{b}}\right\rangle _{2}.
\end{equation}
Here, with probability $\cos^2\alpha$ the first system is in path a
in interferometer 1, and the second system is in path
$\bar{\textrm{a}}$ of interferometer 2. With probability
$\sin^2\alpha$ they are in paths b and $\bar{\textrm{b}}$,
respectively. Again, the superposition state ``a$\bar{\textrm{a}}$
and b$\bar{\textrm{b}}$" is distinctly different in character from a
classical mixture ``a$\bar{\textrm{a}}$ or b$\bar{\textrm{b}}$".
Entanglement can be studied for multi-partite systems, arbitrary
high-dimensional state spaces, and for mixed states
\cite{Horo2009,Pan2012}. Entanglement also plays a crucial role in
Bell tests of local realism \cite{Bell1964,Brun2013} and it is an
essential resource for modern quantum information applications
\cite{Niel2000,Horo2009}.

Due to the many counter-intuitive features of quantum mechanics, a
still heavily debated question is which \textit{meaning} the quantum
state has, in particular whether it is a real physical property or
whether it is only a mathematical tool for predicting measurement
results. Delayed-choice experiments have particularly highlighted
certain peculiarities and non-classical features. In interferometric
delayed-choice experiments, the choice whether to observe the
particle or wave character of a quantum system can be delayed with
respect to the system entering the interferometer. Moreover, it is
possible to observe a continuous transformation between these two
extreme cases. This rules out the naive classical interpretation
that every quantum system behaves either definitely as a particle or
definitely as a wave by adapting a priori to the specific
experimental situation. Using multi-partite states, one can decide a
posteriori whether two systems were entangled or separable, showing
that, just as ``particle" and ``wave", also ``entanglement" and
``separability" are not realistic physical properties carried by the
systems.

This review is structured as follows: In chapter II, we discuss the
history of delayed-choice gedanken experiments, regarding both
single (wave-particle duality) and multi-partite (entanglement)
scenarios such as delayed-choice quantum erasure and entanglement
swapping. In chapters III, IV, and V, we review their experimental
realizations. Chapter VI contains conclusions and an outlook. Some
fractions of this review are based on Ref.~\cite{Ma2010}

\section{Delayed-choice gedanken experiments}

\subsection{Heisenberg's microscope}
\begin{figure}[tb]
\begin{center}
\includegraphics[width=0.2\textwidth]{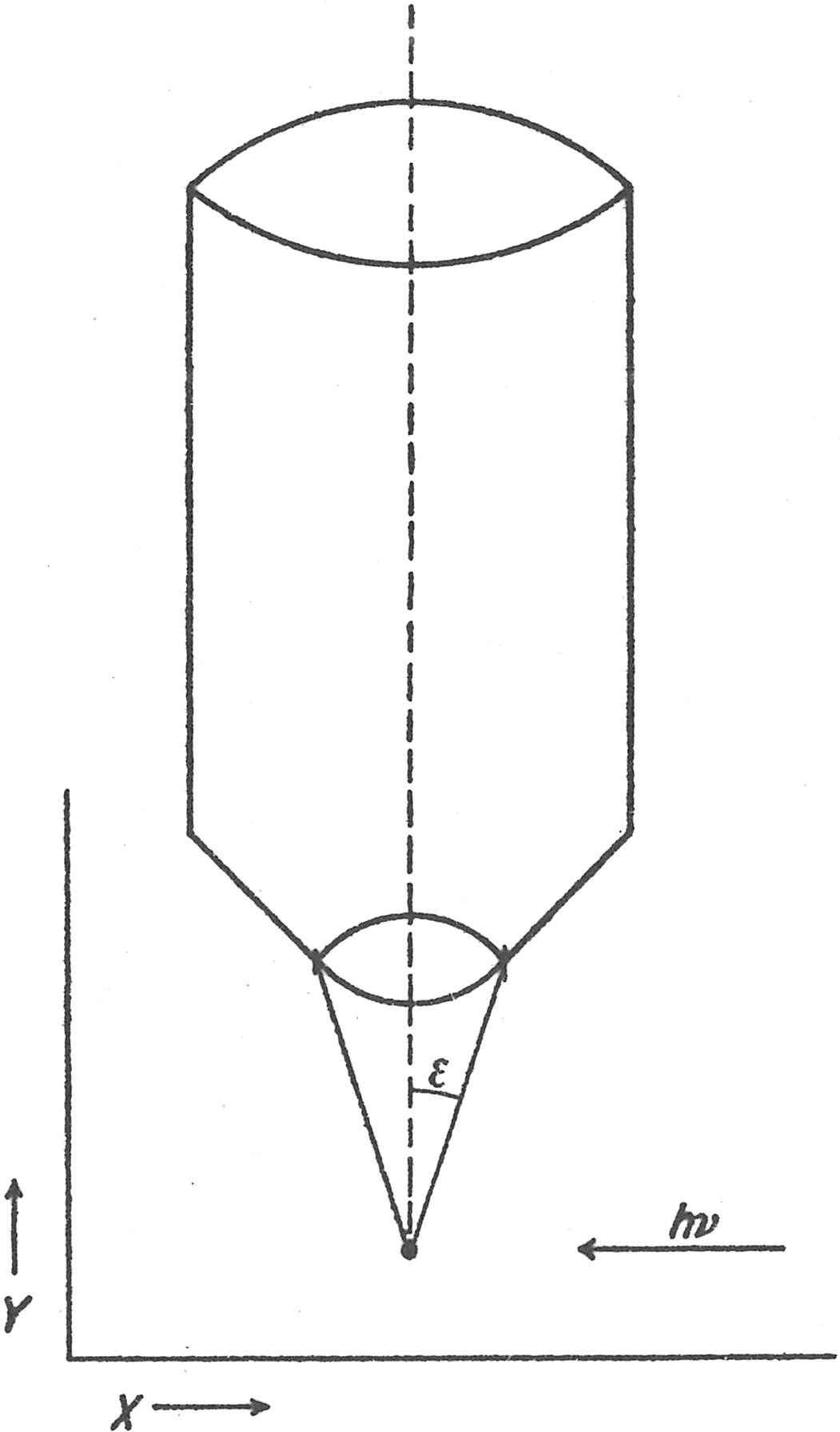}
\caption{Heisenberg's microscope drawing from the notes of his 1929
lectures, printed in 1930. A photon with energy $h\nu$ is scattered
at an electron (represented by a dot) and reaches the observer via a
microscope with opening angle $\varepsilon$. The product of
uncertainty of the position measurement and the momentum disturbance
is of the
order of Planck's constant $h$. Figure taken from Ref.~\cite{Heis1991}.}%
\label{fig_microscope}%
\end{center}
\end{figure}

The history of delayed-choice gedanken experiments can be traced
back to the year 1927, when Heisenberg put forward a rudimentary and
semiclassical version of the uncertainty relation \cite{Heis1927}.
He visualized a microscope which is used to determine the position
of an electron. Due to the Abbe limit the accuracy $\epsilon_{x}$ of
the position measurement is essentially given by the wavelength
$\lambda$ of the light used, and since the resolution gets better
with shorter wavelengths, one also often talks about the ``gamma ray
microscope". Considering the microscope's opening angle
$\varepsilon$ (see Fig.\ \ref{fig_microscope}), the laws of optics
yield the approximate relation $\epsilon_{x}\sim\lambda
/\sin\varepsilon$ for the accuracy. For a position measurement, at
least one photon needs to (Compton) scatter from the electron and
reach the observer through the microscope. The momentum transfer
depends on the angle of the outgoing photon, which is uncertain
within $\varepsilon$. The uncertainty of the momentum transfer in
$x$ direction is thus $\eta_{p}=\sin\varepsilon\cdot h/\lambda$ and
implies the same uncertainty of the electron momentum. The product
of position accuracy and momentum disturbance reads
\begin{equation}
\epsilon_{x}\,\eta_{p}\sim h.
\end{equation}
For Heisenberg, this mathematical relation was a ``direct
illustrative explanation"\ \cite{Heis1927} of the quantum mechanical
commutation relation $[\hat{x},\hat{p}_{x}]=\;$i$\hbar$ for the
position and momentum operator. He noted\ \cite{Heis1991}
(translated from German)

\begin{quote}
that every experiment, which for instance allows a measurement of
the position, necessarily disturbs the knowledge of the velocity to
a certain degree
\end{quote}

In the subsequent years, Heisenberg's uncertainty principle was
derived accurately within the formalism of quantum mechanics\
\cite{Kenn1927,Weyl1928,Robe1929,Schr1930}. However, the resulting
famous inequality
\begin{equation}\label{eq Heisenberg}
\Delta x\,\Delta p\geq\hbar/2
\end{equation}
acquired a different meaning, as it did not involve the notion of
disturbance any longer: $\Delta x$ and $\Delta p$ are the standard
deviations of position and momentum for an ensemble of identically
prepared quantum systems. These quantities can be inferred by
measuring either the position or the momentum of every system. No
sequential or joint measurements are made. Every quantum state
predicts intrinsic uncertainties, which cannot be overcome, that is,
which necessarily fulfill Heisenberg's uncertainty relation (\ref{eq
Heisenberg}). Experimental observations of ensembles of identically
prepared systems confirm these predictions.

More recently, Heisenberg's original derivation in the sense of an
error-disturbance relation has been revisited in fully quantum
mechanical terms\ \cite{Ozaw2004,Bran2013,Busc2013}. In particular,
it is now known that Heisenberg's uncertainty appears in three
manifestations, namely (i) for the widths of the position and
momentum distributions in any quantum state, (ii) for the
inaccuracies of any unsharp joint measurement of both quantities,
and (iii) for the inaccuracy of a measurement of one of the
quantities and the resulting disturbance in the distribution of the
other one\ \cite{Busch2007}. Note that these manifestations are in close connection to wave-particle duality and complementarity, as they provide partial information about complementary observables.

\subsection{von Weizs\"acker, Einstein, Hermann}
In 1931, von Weizs\"acker gave a detailed account of Heisenberg's
thought experiment\ \cite{Weiz1931}. He remarked that one can place
the observer not in the image plane, as originally intended, but in
the focal plane of the microscope. This constitutes a measurement
not of the electron's position but of its momentum. A small but
conceptually very important step was made by Einstein (for a similar
type of experiment) and Hermann (for the Heisenberg microscope), who
made explicit the possibility to delay the choice of measurement
after the relevant physical interaction had already taken place\
\cite{Einstein1931,Hermann1935}. This paved the path for the
paradigm of delayed-choice experiments. In an article on the
interpretation of quantum mechanics, von
Weizs$\ddot{\textrm{a}}$cker wrote\ \cite{Weiz1941} (translated from
German, italics in the original):

\begin{quote}
It is not at all the act of physical interaction between object and
measuring device that defines which quantity is determined and which
remains undetermined, but the act of noticing. If, for example, we
observe an electron with initially known momentum by means of a
single photon, then we are in principle able, after the photon has
traversed the lens, therefore certainly not interacting with the
electron any more, to decide, whether we move a photographic plate
into the \textit{focal} plane or the \textit{image} plane of the
lens and thus determine the \textit{momentum} of the electron after
the observation or its \textit{position}. For here the physical
``disturbance" of the photon determines the description of the state
of the electron, which is related to it not any more physically but
only via the connection of the state probabilities given in the wave
function, the physical influence is apparently merely important as
technical auxiliary means of the intellectual act of constituting a
well-defined observation context.
\end{quote}

\subsection{Bohr's account}
We briefly review Bohr's viewpoint on complementarity, measurement
and temporal order in quantum experiments. Already in 1928, Bohr
said about the requirement of using both a corpuscular and a wave
description for electrons that ``we are not dealing with
contradictory but complementary pictures of the phenomena"
\cite{Bohr1928}. In his ``Discussion with Einstein on
epistemological problems in atom physics" \cite{Bohr1949} he wrote:

\begin{quote}
Consequently, evidence obtained under different experimental conditions cannot
be comprehended within a single picture, but must be regarded as
\textit{complementary} in the sense that only the totality of phenomena
exhausts the possible information about the objects.
\end{quote}

In other words, "in quantum theory the information provided by
different experimental procedures that in principle cannot, because
of the physical character of the needed apparatus, be performed
simultaneously, cannot be represented by any mathematically allowed
quantum state of the system being examined. The elements of
information obtainable from incompatible measurements are said to be
complementary"\ \cite{Stapp2009}.

The term ``phenomenon" is defined by Bohr as follows\
\cite{Bohr1949}:

\begin{quote}
As a more appropriate way of expression I advocated the application
of the word \textit{phenomenon} exclusively to refer to the
observations obtained under specified circumstances, including an
account of the whole experimental arrangement.
\end{quote}

\begin{figure}[tb]
    \includegraphics[width=0.4\textwidth]{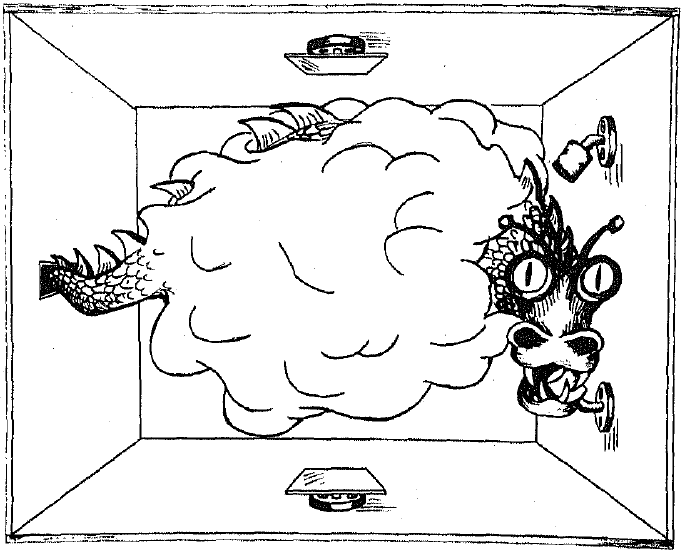}
    \caption{The quantum ``phenomenon" can be viewed as a ``great smoky dragon''.
Figure taken from Ref.~\cite{Miller1983a}.}\label{dragon}
\end{figure}

Miller and Wheeler vividly illustrated the concept of ``elementary
quantum phenomenon" in a cartoon shown in Fig.~\ref{dragon}. The
sharp tail and head of a dragon correspond to Bohr's ``specified
circumstances" (the experimental preparation and arrangement) and
the result of the observation (the outcome of the experiment),
respectively. The body of the dragon, between its head and tail, is
unknown and smoky: ``But about what the dragon does or looks like in
between we have no right to speak, either in this or any
delayed-choice experiment. We get a counter reading but we neither
know nor have the right to say how it came. The elementary quantum
phenomenon is the strangest thing in this strange
world."\cite{Miller1983a}

Already in his response to the Einstein-Podolsky-Rosen (EPR)
argument\ \cite{Eins1935}, Bohr stresses the experimenter's
``freedom of handling the measuring instruments"\ \cite{Bohr1935}.
Later, he wrote \cite{Bohr1949}:

\begin{quote}
It may also be added that it obviously can make no difference as regards
observable effects obtainable by a definite experimental arrangement, whether
our plans for constructing or handling the instruments are fixed beforehand or
whether we prefer to postpone the completion of our planning until a later
moment when the particle is already on its way from one instrument to another.
In the quantum-mechanical description our freedom of constructing or handling
the experimental arrangement finds its proper expression in the possibility of
choosing the classically defined parameters entering in any proper application
of the formalism.
\end{quote}

Therefore, in the language of Heisenberg, von Weizs\"acker, and Bohr
-- the main proponents of the Copenhagen interpretation of quantum
mechanics -- the observer is free to choose at any point in time,
even after physical interactions have been completed, the further
classical conditions of the experiment. This decision, e.g.,\ the
positioning of the detector in the focal or image plane in the
Heisenberg microscope experiment, defines which particular of the
complementary observables is determined.

\subsection{Wheeler's delayed-choice wave-particle duality gedanken experiment}

\begin{figure}[tb]
    \includegraphics[width=0.4\textwidth]{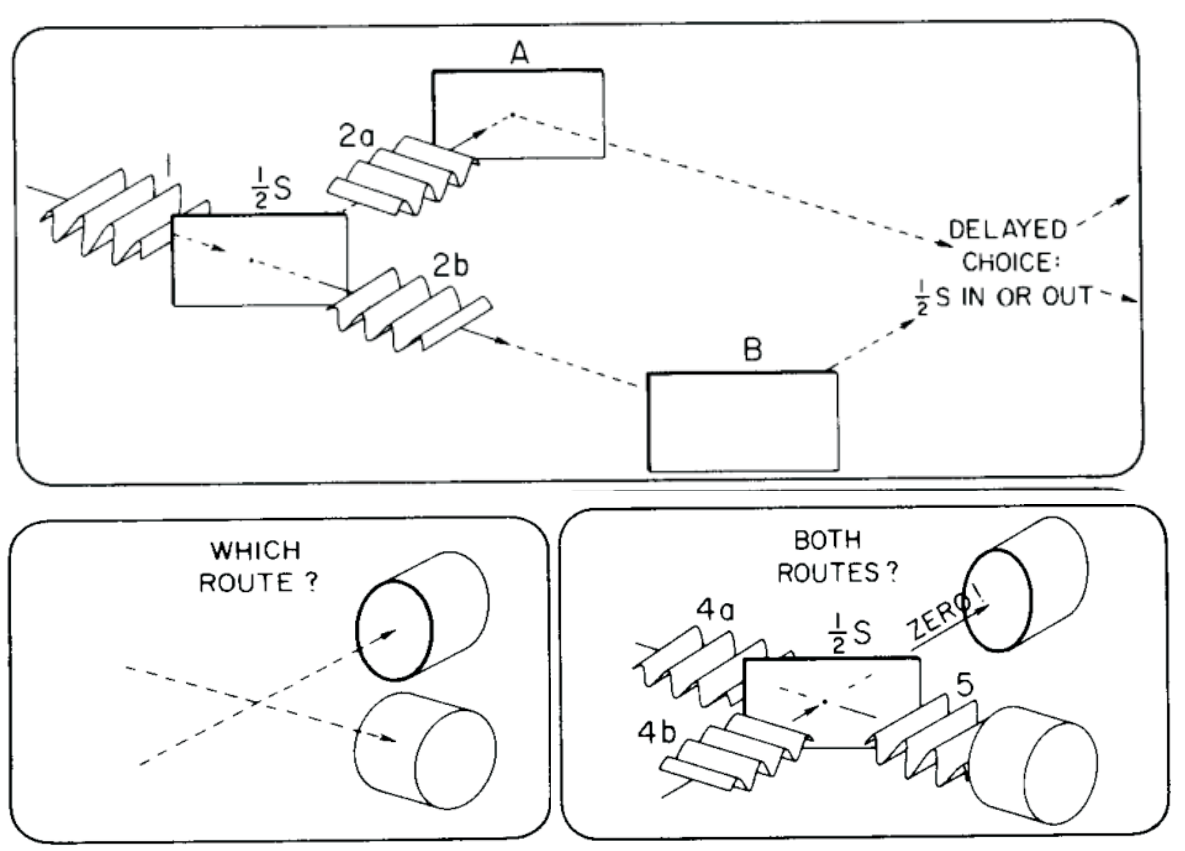}
    \caption{Wheeler's delayed-choice gedanken experiment with a
    single-photon wave packet in a Mach-Zehnder interferometer. Top: The
    second half-silvered mirror ($\frac{1}{2}$S) of the
    interferometer can be inserted or removed at will. Bottom left: When $\frac{1}{2}$S
    is removed, the detectors allow one to determine through which path
    the photon propagated. Which detector fires for an individual photon
    is absolutely random. Bottom right: When $\frac{1}{2}$S is inserted, detection
    probabilities of the two detectors depend on the length difference between
    the two arms. Figure taken from Ref.~\cite{Wheeler1984}.}\label{wheeler1}
\end{figure}
The paradigm of delayed-choice experiments was revived by Wheeler in
Ref.~\cite{Wheeler1978} and a series of works between 1979 and 1981
which were merged in Ref.~\cite{Wheeler1984}. To highlight the
inherently non-classical principle behind wave-particle
complementarity, he proposed a scheme shown at the top in
Fig.~\ref{wheeler1}, where one has a Mach-Zehnder interferometer and
a single-photon wave packet as input. After the first half-silvered
mirror (beam splitter) on the left, there are two possible paths,
indicated by `2a' and `2b'. Depending on the choice made by the
observer, different properties of the photon can be demonstrated. If
the observer chooses to reveal the its particle nature, he should
not insert the second half-silvered mirror ($\frac{1}{2}$S), as
shown at the bottom left in Fig.~\ref{wheeler1}. With perfect
mirrors (A and B) and 100\% detection efficiency, both detectors
will fire with equal probabilities but only one will fire for every
individual photon and that event will be completely random. As
Wheeler pointed out, ``[...]\ one counter goes off, or the other.
Thus the photon has traveled only \textit{one} route"
\cite{Wheeler1984}.

\begin{figure}[tb]
  \begin{center}
    \includegraphics[width=0.4\textwidth]{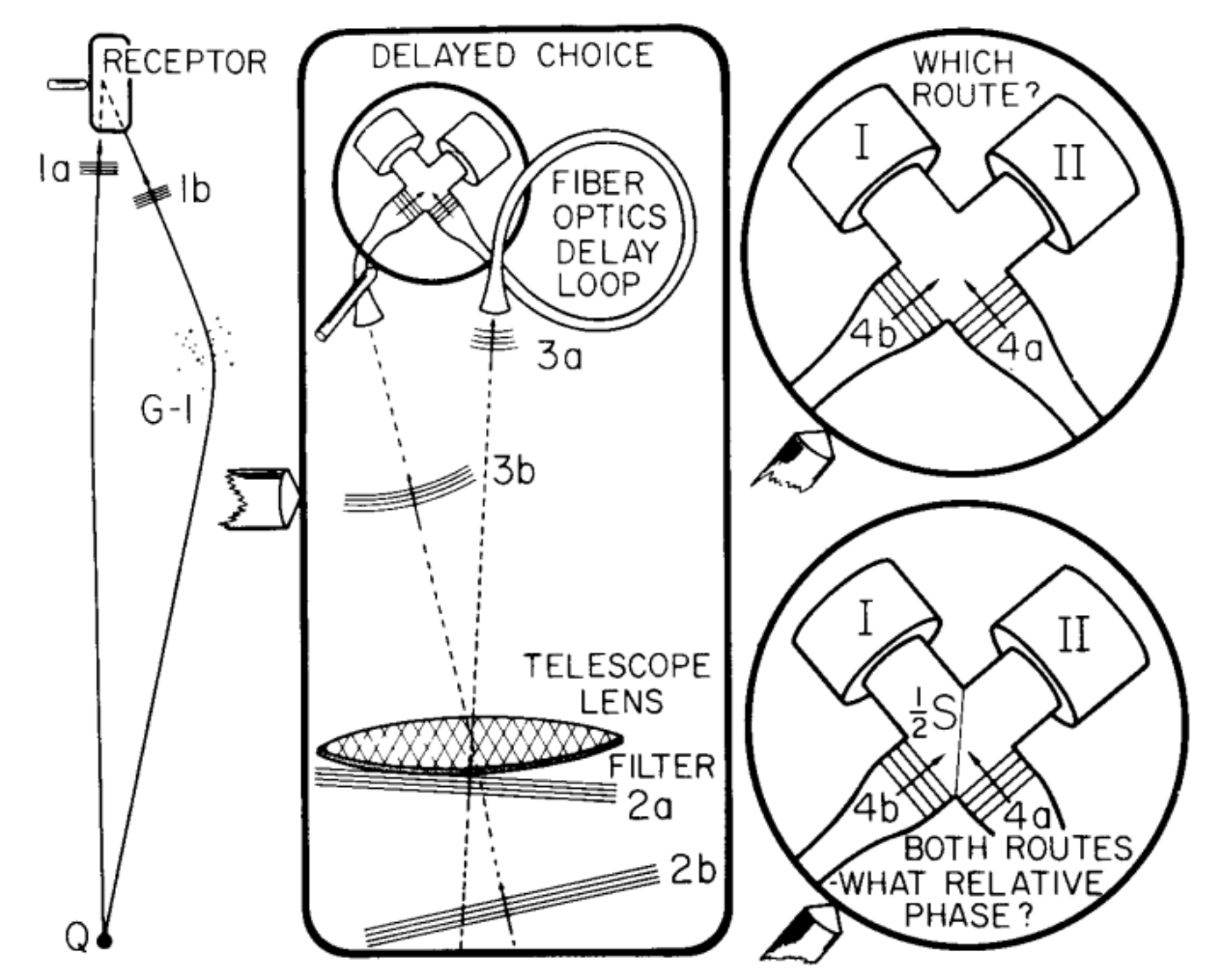}
    \caption{Delayed-choice gedanken experiment at the cosmological scale.
    Left:\ Due to the gravitational lens action of galaxy
    G-1, light generated from a quasar (Q) has two possible paths to reach
    the receptor. This mimics the setup in Fig.~\ref{wheeler1}. Center:\
    The receptor setup. Filters are used to increase the coherence
    length of the light, thus allowing to perform the
    interference experiment. A fiber optics delay loop
    adjusts the phase of the interferometer. Right:\ The
    choice to not insert (top) or insert (bottom) the half-silvered mirror at the final stage of the experiment,
    allows to either measure which particular route the light traveled or what
    the relative phase of the two routes was when it traveled both
    of them. Given the distance between the quasar and the receptor (billions of light
    years), the choice can be made long after the light's entry
    into the interferometer, an extreme example of the delayed-choice gedanken
    experiment. Figure taken from Ref.~\cite{Wheeler1984}.}\label{wheeler2}
  \end{center}
\end{figure}

On the other hand, if the observer chooses to demonstrate the
photon's wave nature, he inserts the beam splitter $\frac{1}{2}$S as
shown at the bottom right in Fig.~\ref{wheeler1}. For identical beam
splitters and zero path difference (or an integer multiple of the
photon wavelength), only the detector on the bottom right will fire.
As Wheeler pointed out:\ ``This is evidence of interference [...],
evidence that each arriving light quantum has arrived by both
routes" \cite{Wheeler1984}.

One might argue that whether the single-photon wave packet traveled
both routes or one route depends on whether the second half-silvered
mirror is inserted or not. In order to rule out naive
interpretations of that kind, Wheeler proposed a ``delayed-choice"
version of this experiment, where the choice of which property will
be observed is made after the photon has passed the first beam
splitter. In Wheeler's words: ``In this sense, we have a strange
inversion of the normal order of time. We, now, by moving the mirror
in or out have an unavoidable effect on what we have a right to say
about the \textit{already} past history of that photon." And:\
``Thus one decides whether the photon `shall have come by one route
or by both routes' after it has \textit{already done} its
travel"~\cite{Wheeler1984}. Very much along the line of the
reasoning of Bohr, one can talk only about a property of the quantum
system, for example, wave or particle, after the quantum phenomenon
has come to a conclusion. In the situation just discussed, this is
only the case after the photon has completely finished its travel
and has been registered.

Illustrated in Fig.~\ref{wheeler2}, Wheeler proposed a most dramatic
``delayed-choice gedanken experiment at the cosmological
scale"~\cite{Wheeler1984}. He explained it as follows:

\begin{quote}
We get up in the morning and spend the day in meditation whether to
observe by ``which route" or to observe interference between ``both
routes." When night comes and the telescope is at last usable we
leave the half-silvered mirror out or put it in, according to our
choice. The monochromatizing filter placed over the telescope makes
the counting rate low. We may have to wait an hour for the first
photon. When it triggers a counter, we discover ``by which route" it
came with one arrangement; or by the other, what the relative phase
is of the waves associated with the passage of the photon from
source to receptor ``by both routes"--perhaps 50,000 light years
apart as they pass the lensing galaxy G-1. But the photon has
already \textit{passed} that galaxy billions of years before we made
our decision. This is the sense in which, in a loose way of
speaking, we decide what the photon ``shall have done" after it has
``already" done it. In actuality it is wrong to talk of the ``route"
of the photon. For a proper way of speaking we recall once more that
it makes no sense to talk of the phenomenon until it has been
brought to a close by an irreversible act of amplification: ``No
elementary phenomenon is a phenomenon until it is a registered
(observed) phenomenon."
\end{quote}

Given the distance between the quasar and the receptor (billions of
light years), the choice is made by the experimenter long after the
photon's entry into the cosmic interferometer (i.e.\ emission by the
quasar). The speed of light of intergalactic space is not exactly
the vacuum speed of light. Therefore, whether the experimenter's
choice is in the time-like future of the emission event or
space-like separated therefrom depends on the size of the
interferometer and the amount of time between the choice event and
the photon arrival at the second beam splitter. Depending on the
specific parameters, Wheeler's delayed choice can thus be thought of
being in the time-like future of, or space-like separated from, the
photon emission.

While Wheeler did not specifically discuss the latter case, it is
particularly appealing because it rules out any causal influence
from the emission to the choice which might instruct the photon to
behave as a particle or as a wave. Note that this resembles the
freedom-of-choice loophole~\cite{Bell2004,Scheidl,Gallicchio2014}
discussed in the context of Bell tests for the falsification of
hidden variable theories using entangled states of at least two
systems. The question in Wheeler's gedanken experiment is about if
and when a single quantum system decides to behave as a particle or
as a wave. Space-like separation excludes unknown communication from
this decision to the choice of the experimenter.

Although Wheeler suggested employing (thermal) light from a quasar,
it is conceptually important to use true single photons rather than
thermal light. This is because the indivisible particle nature of
single photons guarantees that the two detectors will never click at
the same time. Otherwise, one could explain the results by what is
often called a semi-classical theory of light, where light
propagates as a classical wave and is quantized only at the
detection itself~\cite{Paul1982}.

Therefore, important requirements for an ideal delayed-choice
wave-particle duality experiment are (1) a free or random choice of
measurement with space-like separation between the choice and the
entry of the quantum system into the interferometer, and (2) using
single-particle quantum states.

\subsection{Delayed-choice quantum erasure}

\begin{figure*}[tb]
  \begin{center}
    \includegraphics[width=0.65\textwidth]{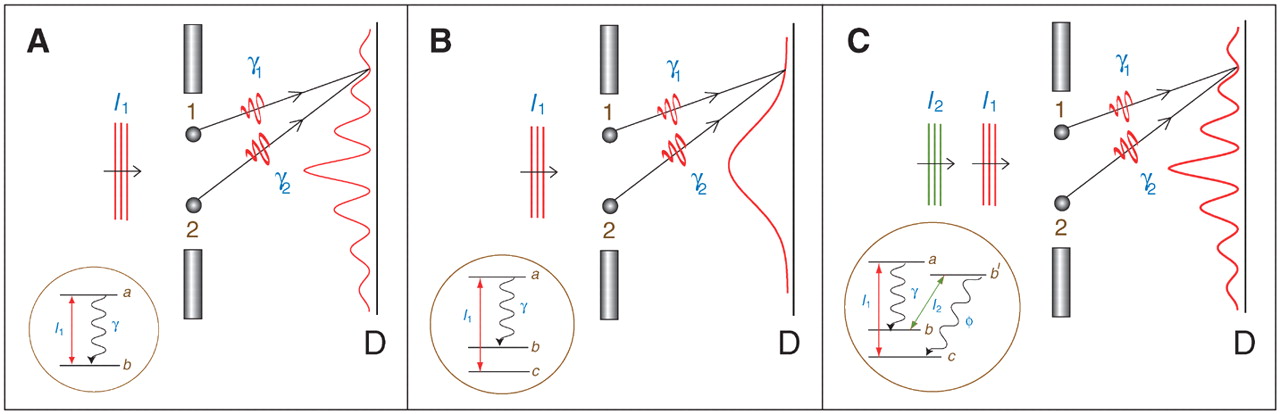}
    \caption{(Color online). The delayed-choice quantum eraser following Ref.~\cite{Scully1982}.
    In \textbf{A}, two two-level atoms are initially in
the state \textit{b}. The incident pulse $l_{1}$ excites one of the
two atoms to state \textit{a} from where it decays to state
\textit{b}, emitting a photon labeled $\gamma$. Because the final
states of both atoms are identical, one can observe interference of
the photons at the detector D. In \textbf{B}, two atoms are
initially in the ground state \textit{c} and one of them is excited
by the pulse $l_{1}$ to state \textit{a} from where it decays to
state \textit{b}. Since the final states of both atoms are
different, one cannot observe interference of the photons. In
\textbf{C}, a fourth level is added. A pulse $l_{2}$ excites the
atom from state \textit{b} to \textit{b'}. The atom in \textit{b'}
emits a photon labeled $\phi$ and ends up in state \textit{c}. If
one can detect $\phi$ in a way that the which-path information is
erased, interference can be recovered for photon $\gamma$. Figure
taken from Ref.~\cite{Aharonov2005}.}\label{Scully1}
  \end{center}
\end{figure*}
Scully and collaborators proposed the so-called quantum eraser
\cite{Scully1982, Scully1991}, in which an entangled atom-photon
system was studied. They considered the scattering of light from two
atoms located at sites 1 and 2 and analyzed three different cases
(Fig.~\ref{Scully1}):

\begin{enumerate}[A.]

  \item A resonant light pulse $l_1$ impinges on two two-level atoms (Fig.~\ref{Scully1}A)
  located at sites 1 and 2. One of the two atoms is
excited to level \textit{a} and emits a photon labeled $\gamma$,
bringing it back to state \textit{b}. As it is impossible to know
which atom emits $\gamma$, because both atoms are finally in the
state \textit{b}, one obtains interference of these photons at the
detector. This is an analog of Young's double-slit experiment.

  \item In the case of three atomic levels (Fig.~\ref{Scully1}B),
  the resonant light $l_1$ excites the atoms from the ground state
\textit{c} to the excited state \textit{a}. The atom in state
\textit{a} can then emit a photon $\gamma$ and end up in state
\textit{b}. The other atom remains in level \textit{c}. This
distinguishability of the atoms' internal states provides which-path
information of the photon and no interference can be observed.

  \item An additional light pulse $l_{2}$ takes the atom from level \textit{b} to
\textit{b'} (Fig.~\ref{Scully1}C). Then a photon labeled $\phi$ is
emitted and the atom ends up in level \textit{c}. Now the final
state of both atoms is \textit{c}, and thus the atoms' internal
states cannot provide any which-path information. If one can detect
photon $\phi$ in a way that its spatial origin (thus which-path
information of $\gamma$) is erased, interference is recovered. Note
that in this case, there are two photons: One is $\gamma$ for
interference, the other one is $\phi$, acting as a which-path
information carrier. (This resembles closely von Weizs\"{a}cker's
account of Heisenberg's microscope.)
\end{enumerate}

\begin{figure}[tb]
  \begin{center}
    \includegraphics{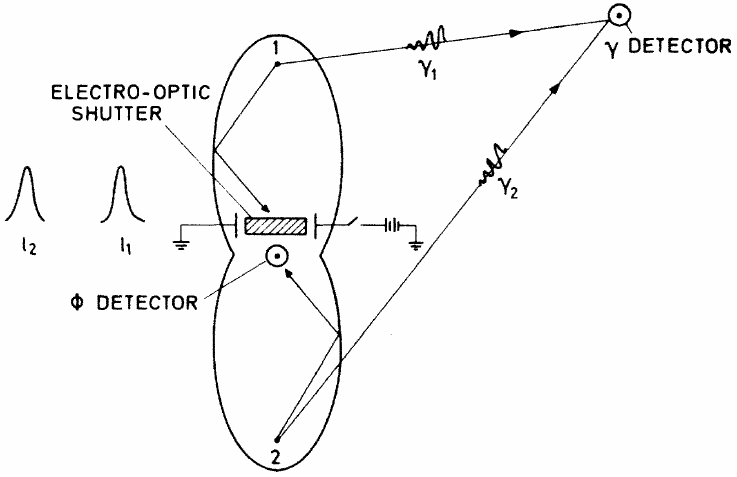}
    \caption{Proposed delayed-choice quantum-eraser setup in Ref.~\cite{Scully1982};
    figure taken therefrom.
    Laser pulses $l_{1}$ and $l_{2}$ are incident on atoms at sites 1 and 2.
    A scattered photon, $\gamma_1$ or $\gamma_2$, is generated by
    $a \rightarrow b$ atomic transition. The atom's decay from
    $b' \rightarrow c$ produces a photon $\phi$. This corresponds to the
    situation depicted in Fig.~\ref{Scully1}\textbf{C}. In order to operate this experiment
    in a delayed-choice mode, two elliptical cavities and an electro-optical
    shutter are employed. The cavities reflect $\phi$ onto a common detector.
    The electro-optical shutter transmits $\phi$ only
    when the switch is open. The choice of open or closed shutter
    determines whether or not the information which atom (1 or 2) emitted the
    photon is erased. This determines whether one can observe the wave
    or particle nature of $\gamma$. The choice can be delayed with
    respect to the generation of $\gamma$.}\label{Scully1p5}
  \end{center}
\end{figure}

Scully and Dr\"{u}hl designed a device based on an electro-optical
shutter, a photon detector, and two elliptical cavities to implement
the above described experimental configuration C in a delayed-choice
arrangement (Fig.~\ref{Scully1p5}). There, one can choose to reveal
or erase the which-path information after the photon $\gamma$ has
been generated.

\begin{figure}[tb]
  \begin{center}
    \includegraphics[width=0.48\textwidth]{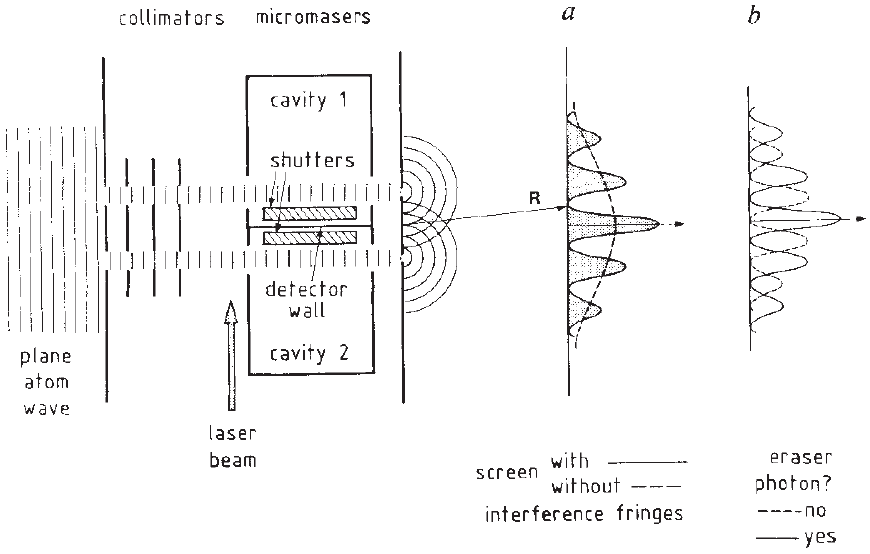}
    \caption{Proposed quantum-eraser setup in Ref.~\cite{Scully1991};
    figure taken therefrom. A detector
wall, separating two cavities for microwave photons, is sandwiched
by two electro-optic shutters. \textit{a}.\ By always opening only
one of the shutters, the photon detections reveal the cavity where
the photon was emitted and thus, which-path information for the
atoms. Consequently, no interference pattern emerges. When, opening
both shutters the photon detections will erase which-path
information of the atoms and interference shows up. \textit{b}.\
Both shutters are open. It is assumed that the detector wall can
only be excited by the symmetric photon state $\ket{+}_{12}$. Hence,
if a photon is being emitted in one of the cavities but not
detected, it was in the antisymmetric state $\ket{-}_{12}$. The
detections of the symmetric and antisymmetric photon state give rise
to oppositely-modulated interference fringes of the atoms (solid and
dashed curve), respectively.}\label{Scully2}
  \end{center}
\end{figure}

In another proposal~\cite{Scully1991}, the interfering system is an
atomic beam propagating through two cavities coherently. The atomic
state is the quantum superposition
$\frac{1}{\sqrt{2}}(\ket{e}_{1}+\ket{e}_{2})$, where $\ket{e}_{i}$
denotes the excited state of the atom passing through cavity $i=1,2$
(see Fig.~\ref{Scully2}). The excited atom can decay to its ground
state $\ket{g}_{i}$ and emit a photon in state $\ket{\gamma}_{i}$.
In conjunction with the perfectly reflective shutters, the two
cavities, separated by a photon detector wall, are used to trap the
photon. Conditional on the emission of one photon $\gamma$ from the
atom in one of the cavities, the state of atom (a) and photon (p)
becomes:
\begin{equation}\label{eraser00}
    \ket{\Phi}_{\textrm{ap}}=\tfrac{1}{\sqrt{2}}(\ket{g}_{1}\ket{\gamma}_{1}+\ket{g}_{2}\ket{\gamma}_{2}).
\end{equation}
If shutter 1 is open and shutter 2 is closed, detection of a photon
(in cavity 1) reveals the atom's position in cavity 1, and vice
versa if shutter 2 is open while 1 is closed. Repeating experiments
with these two configurations (i.e.\ only one shutter open) will not
lead to an interference pattern of the atom detections (dashed curve
in Fig.~\ref{Scully2}\textit{a}). The same pattern will emerge when
both shutters remain closed at all times. The lack of interference
in both cases is because the which-path information is still present
in the universe, independent of whether an observer takes note of it
or not. Ignoring the photon state, which carries which-path
information about the atom, leads to a mixed state of the atom of
the from $\tfrac{1}{2}(\ket{g}_{1}\langle g| + \ket{g}_{2}\langle
g|)$ which cannot show an interference pattern.

However, the state~(\ref{eraser00}) can also be written as
\begin{equation}\label{eraserpm}
    \ket{\Phi}_{\textrm{ap}}=\tfrac{1}{2}(\ket{g}_{1}+\ket{g}_{2})\ket{+}_{12}
    +\tfrac{1}{2}(\ket{g}_{1}-\ket{g}_{2})\ket{-}_{12},
\end{equation}
with the symmetric and antisymmetric photon states
$\ket{+}_{12}=\tfrac{1}{\sqrt{2}}(\ket{\gamma}_{1}+\ket{\gamma}_{2})$
and
$\ket{-}_{12}=\tfrac{1}{\sqrt{2}}(\ket{\gamma}_{1}-\ket{\gamma}_{2})$.
If one opens both shutters and detects the symmetric photon state
$\ket{+}_{12}$, one cannot in principle distinguish which cavity the
atom propagated through as its state is the coherent superposition
$\tfrac{1}{\sqrt{2}}(\ket{g}_{1}+\ket{g}_{2})$. Detection of the
photon in the state $\ket{+}_{12}$ has erased the which-path
information of the atom. Therefore, interference in the atom
detections shows up again (solid curve in
Fig.~\ref{Scully2}\textit{a} and \textit{b}). If one detects the
antisymmetric photon state $\ket{-}_{12}$, the atomic state becomes
a superposition with a different relative phase between the two
paths, $\tfrac{1}{\sqrt{2}}(\ket{g}_{1}-\ket{g}_{2})$, leading to a
shift of the interference pattern (dashed curve in
Fig.~\ref{Scully2}\textit{b}). In Ref.~\cite{Scully1991} it was
assumed that the detector has perfect detection efficiency but
cannot be excited by the antisymmetric photon state, which is why
the shifted interference pattern emerges in the case of both
shutters being open and no eraser photon being detected. The
detector wall used here is sufficiently thin such that it cannot
distinguish which side the photon has impinged on, and hence is able
to collapse the photons' superposition states into the symmetric or
antisymmetric state. It is important to note that the interference
patterns of the atoms can only be seen in coincidence with the
corresponding photon projections into the symmetric or antisymmetric
states.

This gedanken experiment triggered a controversial discussion on
whether complementarity is more fundamental than the uncertainty
principle~\cite{Stoery1994, Englert1995, Stoery1995}. Wiseman and
colleagues reconciled divergent opinions and recognized the novelty
of the quantum eraser concept~\cite{Wiseman1995, Wiseman1997}.
Experimental demonstrations of quantum erasure for atomic systems
have been realized in Refs.~\cite{Eichmann1993} and~\cite{Durr1998},
and will be reviewed in Section IV B.

A delayed-choice configuration can be arranged in this experiment:
one can choose to reveal or erase the which-path information of the
atoms (by not opening or opening both shutters) after the atom
finishes the propagation through the two cavities.

A detailed analysis of the fundamental aspects of single-particle
interference experiments facing decoherence has been reported in
Ref.~\cite{Scully1989}. The authors considered the quantum
(system-apparatus) correlations which are at the root of decoherence
rather than the recoil or collision. This topic will be further
discussed in Chapters III and IV.

\subsection{Delayed-choice entanglement swapping}

When two systems are in an entangled quantum state, the correlations
of the joint system are well defined but not the properties of the
individual systems~\cite{Schroedinger1935, Eins1935}. Peres raised
the question of whether it is possible to produce entanglement
between two systems even after they have been registered by
detectors~\cite{Peres2000}. Remarkably, quantum mechanics allows
this via entanglement swapping~\cite{Zukowski1993}. We note that
Cohen had previously analyzed a similar situation in the context of
counterfactual entanglement generation in separable
states~\cite{Cohen1999}.

\begin{figure}[tb]
    \includegraphics[width=0.35\textwidth]{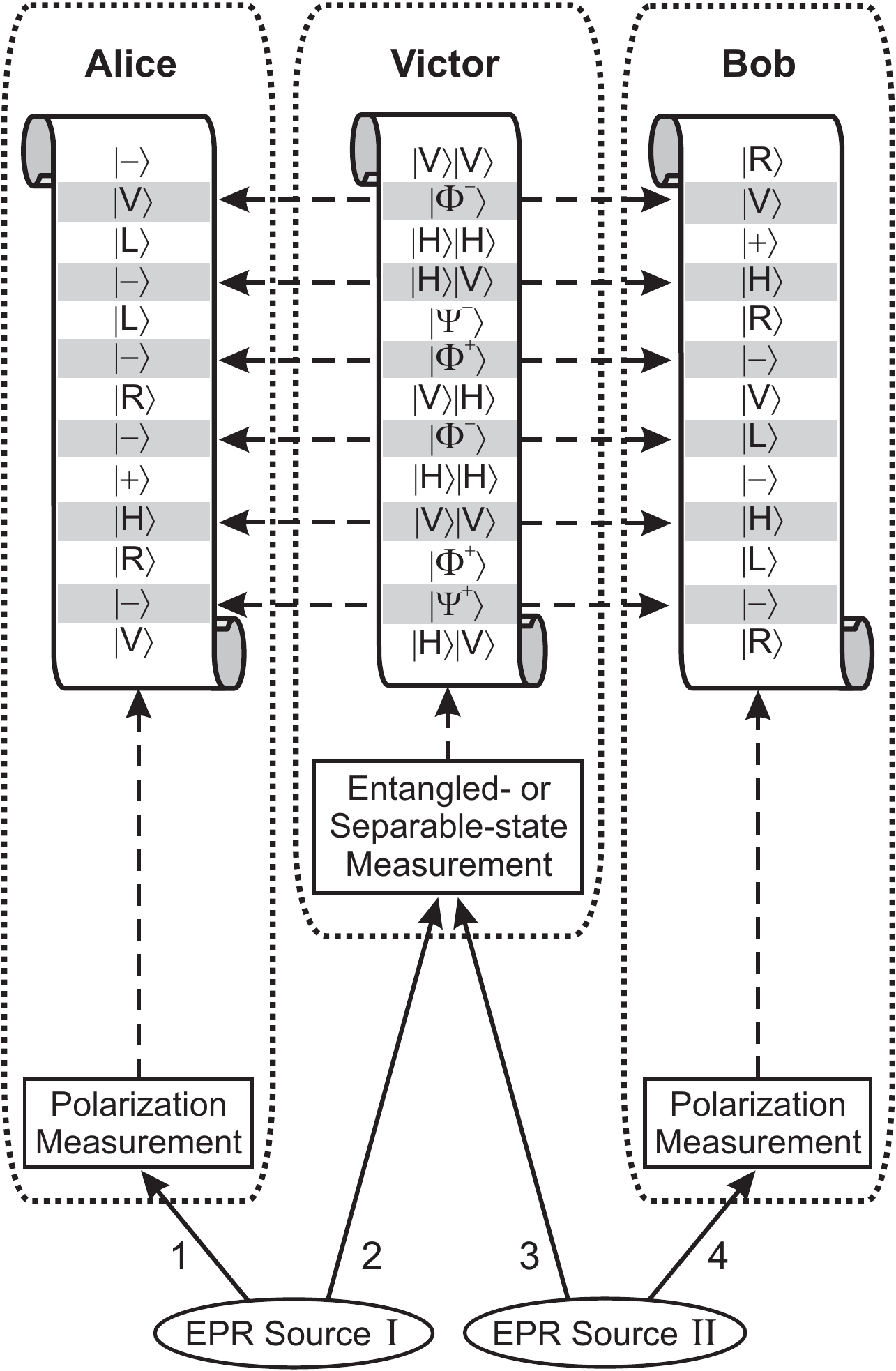}
    \caption{The concept of delayed-choice entanglement swapping. Two entangled
pairs of photons 1\&2 and 3\&4 are produced, e.g., in the joint
state $\ket{\Psi^{-}}_{12} \ket{\Psi^{-}}_{34}$ from the EPR sources
I and II, respectively. Alice and Bob perform polarization
measurements on photons 1 and 4 in any of the three mutually
unbiased bases and record the outcomes. Victor has the freedom of
either performing an entangled- or separable-state measurement on
photons 2 and 3. If Victor decides to perform a separable-state
measurement in the four-dimensional two-particle basis
\{$\ket{\textrm{H}}_2 \ket{\textrm{H}}_3$, $\ket{\textrm{H}}_2
\ket{\textrm{V}}_3$, $\ket{\textrm{V}}_2 \ket{\textrm{H}}_3$,
$\ket{\textrm{V}}_2 \ket{\textrm{V}}_3$\}, then the outcome is
random and one of these four product states. Photons 1 and 4 are
projected into the corresponding product state and remain separable.
On the other hand, if Victor chooses to perform an entangled-state
measurement on photons 2 and 3 in the Bell-state basis
\{$\ket{\Psi^+}_{23}$, $\ket{\Psi^-}_{23}$, $\ket{\Phi^+}_{23}$,
$\ket{\Phi^-}_{23}$\}, then the random result is one of the four
Bell states. Consequently, photons 1 and 4 are also projected into
the corresponding Bell state. Therefore, entanglement is swapped
from pairs 1\&2 and 3\&4 to pairs 2\&3 and 1\&4. Figure adapted from
Ref.~\cite{Ma2012}.}\label{MaSwapping1}
\end{figure}

In a photonic implementation of entanglement swapping, two pairs of
polarization-entangled photons, 1\&2 and 3\&4, are produced from two
different EPR sources (Fig.~\ref{MaSwapping1}). The initial
four-photon entangled state is, e.g., of the form
\begin{equation}\label{psi4}
    \ket{\Psi}_{1234} = \ket{\Psi^-}_{12} \ket{\Psi^-}_{34},
\end{equation}
where $\ket{\Psi^-}_{ij}= \tfrac{1}{\sqrt2} (\ket{\textrm{H}}_i
\ket{\textrm{V}}_j - \ket{\textrm{V}}_i \ket{\textrm{H}}_j)$ are the
entangled antisymmetric Bell (singlet) states of photons $i$ and
$j$. H and V denote horizontal and vertical polarization,
respectively. While photon 1 is sent to Alice and photon 4 is sent
to Bob, photons 2 and 3 propagate to Victor.

Alice and Bob perform polarization measurements on photons 1 and 4,
choosing freely between the three mutually unbiased
bases~\cite{Woot1989} $\ket{\textrm{H}}$/$\ket{\textrm{V}}$,
$\ket{\textrm{R}}$/$\ket{\textrm{L}}$, and $\ket{+}$/$\ket{-}$, with
$\ket{\textrm{R}}=\tfrac{1}{\sqrt2}(\ket{\textrm{H}}+
\textrm{i}\ket{\textrm{V}})$,
$\ket{\textrm{L}}=\tfrac{1}{\sqrt2}(\ket{\textrm{H}}-
\textrm{i}\ket{\textrm{V}})$, and
$\ket{\pm}=\tfrac{1}{\sqrt2}(\ket{\textrm{H}}\pm\ket{\textrm{V}})$.
If Victor chooses to measure his two photons 2 and 3 separately in
the H/V basis, i.e.\ in the basis of separable (product) states
$\ket{\textrm{H}}_2 \ket{\textrm{H}}_3$, $\ket{\textrm{H}}_2
\ket{\textrm{V}}_3$, $\ket{\textrm{V}}_2 \ket{\textrm{H}}_3$, and
$\ket{\textrm{V}}_2 \ket{\textrm{V}}_3$, then the answer of the
experiment is one of the four random results. Upon Victor's
measurement, also photons 1 and 4 will remain separable and be
projected into the corresponding product state $\ket{\textrm{V}}_1
\ket{\textrm{V}}_4$, $\ket{\textrm{V}}_1 \ket{\textrm{H}}_4$,
$\ket{\textrm{H}}_1 \ket{\textrm{V}}_4$, or $\ket{\textrm{H}}_1
\ket{\textrm{H}}_4$, respectively. Alice's and Bob's polarization
measurements are thus only correlated in the
$\ket{\textrm{H}}$/$\ket{\textrm{V}}$ basis.

However, the state~(\ref{psi4}) can also be written in the basis of
Bell states of photons 2 and 3:
\begin{align}\label{psi4Bell}\nonumber
    \ket{\Psi}_{1234} & = \tfrac{1}{2}( \ket{\Psi^+}_{14} \ket{\Psi^+}_{23} - \ket{\Psi^-}_{14} \ket{\Psi^-}_{23} \\
    & \, \, \, \, \, \; - \ket{\Phi^+}_{14} \ket{\Phi^+}_{23} - \ket{\Phi^-}_{14}
    \ket{\Phi^-}_{23},
\end{align}
with the entangled symmetric Bell (triplet) states
$\ket{\Psi^+}_{ij} = \tfrac{1}{\sqrt2} (\ket{\textrm{H}}_i
\ket{\textrm{V}}_j + \ket{\textrm{V}}_i \ket{\textrm{H}}_j$,
$\ket{\Phi^\pm}_{ij} = \tfrac{1}{\sqrt2} (\ket{\textrm{H}}_i
\ket{\textrm{H}}_j \pm \ket{\textrm{V}}_i \ket{\textrm{V}}_j$. When
Victor decides to perform a Bell-state measurement, i.e.\ when he
measures in the basis of entangled states $\ket{\Psi^+}_{23}$,
$\ket{\Psi^-}_{23}$, $\ket{\Phi^+}_{23}$, and $\ket{\Phi^-}_{23}$,
then the answer of the experiment is one of the four random results.
Alice's and Bob's photons are then projected into the corresponding
entangled state $\ket{\Psi^+}_{14}$, $\ket{\Psi^-}_{14}$, $\ket{\Phi^+}_{14}$,
or $\ket{\Phi^-}_{14}$, respectively. Alice's and Bob's polarization
measurements are thus correlated in all possible bases. This implies
that Victor can establish entanglement between photons 1 and 4,
although they have never interacted nor share any common past. After
entanglement swapping, pairs 1\&2 and 3\&4 are no longer entangled,
obeying the monogamy of entanglement \cite{Coffman2000}.

Peres suggested an addition to the entanglement-swapping protocol,
thereby combining it with Wheeler's delayed-choice paradigm. He
proposed that the correlations of photons 1 and 4 can be defined
even after they have been detected via a later projection of photons
2 and 3 into an entangled state. According to Victor's choice and
his results, Alice and Bob can sort their already recorded data into
subsets and can verify that each subset behaves as if it consisted
of either entangled or separable pairs of distant photons, which
have neither communicated nor interacted in the past. Such an
experiment leads to the seemingly paradoxical situation, that
``entanglement is produced \textit{a posteriori}, after the
entangled particles have been measured and may even no longer
exist"~\cite{Peres2000}.

Since the property whether the quantum state of photons 1 and 4 is
separable or entangled, can be freely decided by Victor's choice of
applying a separable-state or Bell-state measurement on photons 2
and 3 after photons 1 and 4 have been already measured, the
delayed-choice wave-particle duality of a single particle is brought
to an \textit{entanglement-separability duality} of two particles.

\subsection{Quantum delayed-choice}

\begin{figure}[tb]
\begin{center}
\includegraphics{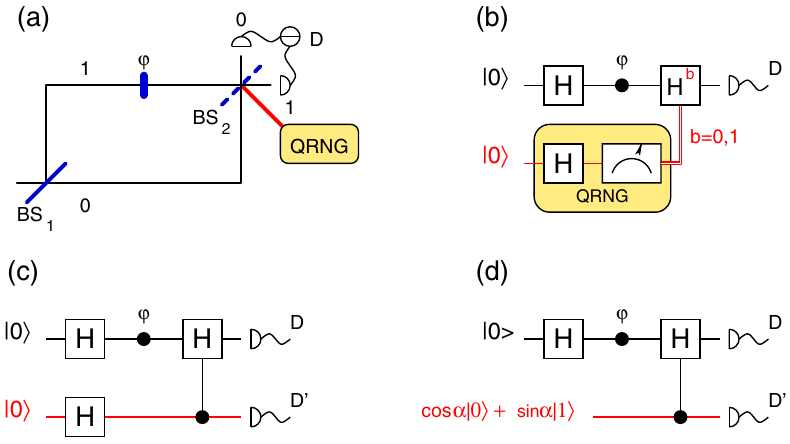}
\end{center}
\caption{(Color online). (a) The `classical' delayed-choice experiment: The second
beam splitter BS$_{2}$ is inserted or not after the photon has
already entered the interferometer. (b) An equivalent quantum
network: An ancilla (lower input, red) acts as a quantum random
number generator (QRNG). Its initial state $\left\vert
0\right\rangle$ is transformed by a Hadamard gate H into
$(\left\vert 0\right\rangle +\left\vert 1\right\rangle )/\sqrt{2}$.
A measurement in the computational $\left\vert 0\right\rangle
/\left\vert 1\right\rangle$ basis gives a random outcome, which
determines whether or not the second Hadamard gate is applied with
the system qubit (equivalent to BS$_{2}$) is applied. (c) Delayed
choice with a quantum beam splitter: The second beam splitter
BS$_{2}$ (represented by a controlled Hadamard gate) is coherently
controlled by the state of the ancilla qubit. It is now in
superposition of present and absent. (d) The QRNG can be biased by
preparing the ancilla in the state $\cos\alpha\left\vert
0\right\rangle +\sin\alpha\left\vert 1\right\rangle $. Figure taken
from
Ref.~\cite{Ioni2011}.}%
\label{fig_quantum_delayed_choice}%
\end{figure}

Wheeler's delayed choice experiment [redrawn in Fig.\
\ref{fig_quantum_delayed_choice}(a)] of a photon in an
interferometer with phase $\varphi$ can be translated into the
language of quantum circuits\ \cite{Niel2000}, where Hadamard gates
represent the beam splitters and an ancilla is used in a quantum
random number generator (QRNG) for making the choice\ [Fig.\
\ref{fig_quantum_delayed_choice}(b)]. A quantum version of this
experiment was suggested~\cite{Ioni2011}, where the ancilla can
coherently control the second beam splitter of the interferometer
[Fig.\ \ref{fig_quantum_delayed_choice}(c)]. Bias can be achieved by
more general ancilla states $\cos\alpha\left\vert 0\right\rangle
+\sin\alpha\left\vert 1\right\rangle $ with amplitudes depending on
a parameter $\alpha$ [Fig.\ \ref{fig_quantum_delayed_choice}(d)]. By
this, the second beam splitter can be in a superposition of being
present and absent. Following the language of Wheeler, the photon
must consequently be in a superposition of particle and wave at the
same time. Moreover, one can arbitrarily choose the temporal order
of the measurements. In particular, if one measures the ancilla
\textit{after} the photon, the latter can be described as having
behaved as a particle or as a wave after it has been already
detected. From the experimental point of view, it is advantageous
that no fast switching of any devices is required.

With an appropriate alignment of the interferometer, before the
detectors in
Fig.\ \ref{fig_microscope}(d) the state of the photon and the ancilla reads%
\begin{equation}
\left\vert \Psi\right\rangle
=\cos\alpha\,|\text{particle}\rangle\left\vert 0\right\rangle
+\sin\alpha\,|\text{wave}\rangle\left\vert 1\right\rangle
,\label{eq particle-wave}%
\end{equation}
with the photon states%
\begin{align}
|\text{particle}\rangle &  =\tfrac{1}{\sqrt{2}}(\left\vert
0\right\rangle
+\text{e}^{\text{i}\varphi}\left\vert 1\right\rangle ),\label{eq particle}\\
|\text{wave}\rangle &  =\text{e}^{\text{i}\varphi/2}\,(\cos\tfrac{\varphi}%
{2}\left\vert 0\right\rangle
+\text{i}\sin\tfrac{\varphi}{2}\left\vert
1\right\rangle ).\label{eq wave}%
\end{align}
The overlap between the latter states is
$\langle$particle$|$wave$\rangle =2^{-1/2}\cos\varphi$. As $\varphi$
varies, the probability to find the photon in state 0 is
$I_{\text{p}}(\varphi)=\tfrac{1}{2}$ (visibility $V=0$) for the
particle state and
$I_{\text{w}}(\varphi)=\cos^{2}\tfrac{\varphi}{2}$ (visibility
$V=1$) for the wave state. Eq.\ (\ref{eq particle-wave}) is a
quantitative expression of complementarity, and the question whether
a system behaves as a wave can now be seen in the language of
mutually unbiased bases. If the photon data is analyzed in the
respective subensembles of the ancilla outcomes, it shows either
perfect particle-like (ancilla in $\left\vert 0\right\rangle $,
photon visibility $V=0$) or wave-like behavior (ancilla in
$\left\vert 1\right\rangle $, photon visibility $V=1$).

For an equal-weight superposition ($\alpha=\frac{\pi}{4}$),
analyzing only the photon data itself as a function of $\varphi$
leads to an interference pattern with a reduced visibility of
$V=\frac{1}{2}$. Changing $\alpha$ from 0 (photon certainly in state
$\left\vert \text{particle}\right\rangle $) to $\frac{\pi }{2}$
($\left\vert \text{wave}\right\rangle $) allows to continuously
morph into particle and wave properties. Ignoring the ancilla
outcome, the
detector for the photon state 0 fires with probability $I_{\text{p}}%
(\varphi)\cos^{2}\alpha+I_{\text{w}}(\varphi)\sin^{2}\alpha$, i.e.%
\begin{equation}
\tfrac{1}{2}\cos^{2}\alpha+\cos^{2}\tfrac{\varphi}{2}\sin^{2}\alpha,
\label{eq intensity}%
\end{equation}
corresponding to a visibility $V=\sin^{2}\alpha$.

A hidden-variable based analysis of quantum delayed-choice
experiments needs to describe the entire (entangled) system of
photon and ancilla. It was argued that quantum delayed-choice
experiments without space-like separation between system photon and
ancilla are equivalent to classical delayed-choice experiments with
space-like separation\ \cite{Cele2014}. The continuous morphing
behavior predicted by quantum mechanics in quantum delayed-choice
experiments cannot be described by hidden-variable theories for the
system photon and the ancilla, which obey objectivity (``particle''
and ``wave'' are intrinsic attributes of the system photon during
its lifetime), determinism (the hidden variables determine the
individual outcomes), and independence (the hidden variables do not
depend on the experimental setting, i.e.\ the choice of $\alpha$)
\cite{Ioni2014}. Moreover, these three assumptions are indeed
incompatible with any theory, not only quantum mechanics\
\cite{Ioni2015}.

\section{Realizations of delayed-choice Wave-particle duality experiments}

\subsection{First realizations of Wheeler's delayed-choice experiment}

Inspired by Wheeler's gedanken experiment, there have been several
concrete experimental proposals and analyses for different physical
systems, including neutron interferometers~\cite{Miller1983a,
Miller1983, Greenberger1983} and photon
interferometers~\cite{Alley1983,Mittelstaedt1986}. Pioneering
endeavors in realizing these experiments have been reported in
Refs.~\cite{Alley1986, Schleich1986, Hellmuth1987}.

\begin{figure}[tb]
  \begin{center}
    \includegraphics{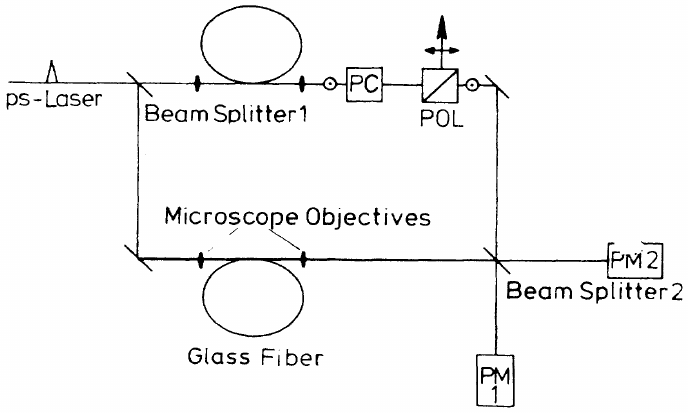}
    \caption{Setup of the delayed-choice experiment reported in
    Ref.~\cite{Hellmuth1987}; figure taken therefrom. The
combination of a Pockels cell (PC) and a polarizer (POL) in the
upper arm of the interferometer was used as a
shutter.}\label{Hellmuth1}
  \end{center}
\end{figure}

Hellmuth and collaborators performed delayed-choice experiments with
a low-intensity Mach-Zehnder interferometer (MZI) in the spatial
domain as well as time-resolved atomic fluorescence in the time
domain~\cite{Hellmuth1987}. The layout of the delayed-choice
experiment in the spatial domain is shown in Fig.~\ref{Hellmuth1}.
An attenuated picosecond laser (on average less than 0.2 photons per
pulse) was used as the light source for the MZI. Two 5~m (20 ns)
glass fibers were used to delay the input photon. The transit time
of the photon through the whole interferometer was about 24~ns. The
combination of a Pockels cell (PC) and a polarizer (POL) was placed
in the upper arm of the MZI as a shutter.

When a half-wave voltage was applied on the Pockels cell, it rotated
the polarization of the photons propagating through it, such that
they were reflected out of the interferometer. In this case the
shutter was closed and interference vanished as the upper path of
the interferometer was interrupted and only photons from the lower
arm could reach the photomultipliers (PM 1 and PM 2). This provided
which-path information, as the photon arrived at beam splitter~2
because it could only have come via the other, open, path. On the
other hand, if the shutter was open upon the photon's arrival, one
could observe the interference pattern, because then no information
was present about the path the photon took.

The temporal structure in the ``delayed-choice mode" of this
experiment was as follows. The input photon met beam splitter~1
first, where its amplitude was split between two paths through the
interferometer. It then was kept in a fiber, one in each path, for
20~ns. During the photon propagation in the fiber, the shutter
opened after 4~ns PC rise time. Then the photon exited from the
fibers, and met the opened shutter and beam splitter~2 sequentially.
Therefore, in this case, opening of the shutter was delayed until
after the input photon met beam splitter~1 and was well inside of
the interferometer. With this experimental arrangement, the photon's
entry into the MZI was clearly located in the past light cone of
opening the shutter.

\begin{figure}[tb]
  \begin{center}
    \includegraphics{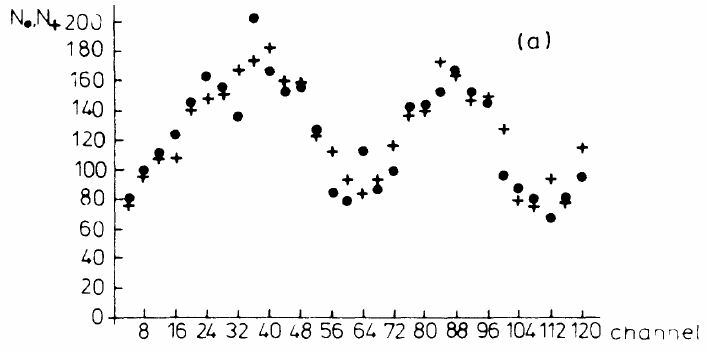}
    \caption{Experimental results of the delayed-choice
experiment in Ref.~\cite{Hellmuth1987}; figure taken therefrom.
Interference patterns for normal mode (dots) and delayed-choice mode
(crosses) measured by PM 1 are similar and consistent with quantum
mechanical predictions.}\label{Hellmuth2}
  \end{center}
\end{figure}

In the ``normal mode", opening the shutter was prior to the input
photon meeting beam splitter~1. The authors alternated the
experimental arrangement from the normal mode (opening the shutter
\textit{before} the photon reaches beam splitter~1) to the
delayed-choice mode (opening the shutter \textit{after} the photon
reaches beam splitter~1) for each successive light pulse, while they
kept all the other experimental configurations to be the same, in
particular the phase of the MZI. The photon counts detected by PM 1
as a function of the phase variation are presented in
Fig.~\ref{Hellmuth2}. The results measured by PM 2 showed
complementary behavior, i.e.\ the pattern was shifted by a phase
$\pi$ with respect to the one recorded by PM~1.

This experiment was one of the pioneering realizations of Wheeler's
gedanken experiment, although no true single photons were used and
no real active choices were implemented. The switch-on time of the
Pockels cell was delayed, but eventually it was turned on such that
always the light's wave character was tested.

\begin{figure}[tb]
  \begin{center}
    \includegraphics[width=0.48\textwidth]{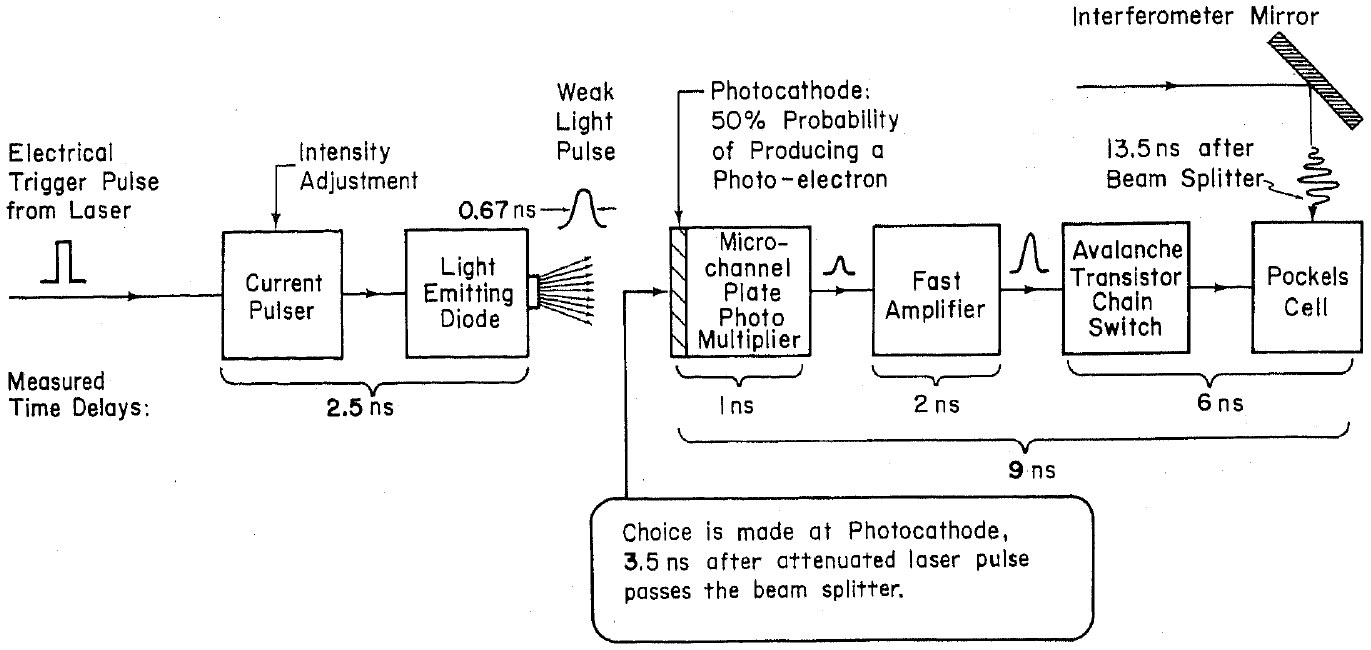}
    \caption{Schematic diagram of the device generating the random choices
proposed in~\cite{Alley1983} and used in~\cite{Alley1986}; figure
taken therefrom. A weak light pulse emitted from a light emitting
diode has a pulse duration of 0.67~ns. The detection event of this
light pulse makes the random choice which determines the setting of
the Pockels cell. To realize that, a photocathode with 50\%
probability of producing a photo-electron within 1~ns is amplified
by a fast amplifier within 2~ns. This electric pulse then triggers
the avalanche transistor chain switch and hence the Pockels cell.
The time of the choice can be tuned with respect to the photon's
entry into the MZI.}\label{Alley1}
  \end{center}
\end{figure}

\begin{figure}[tb]
  \begin{center}
    \includegraphics[width=0.35\textwidth]{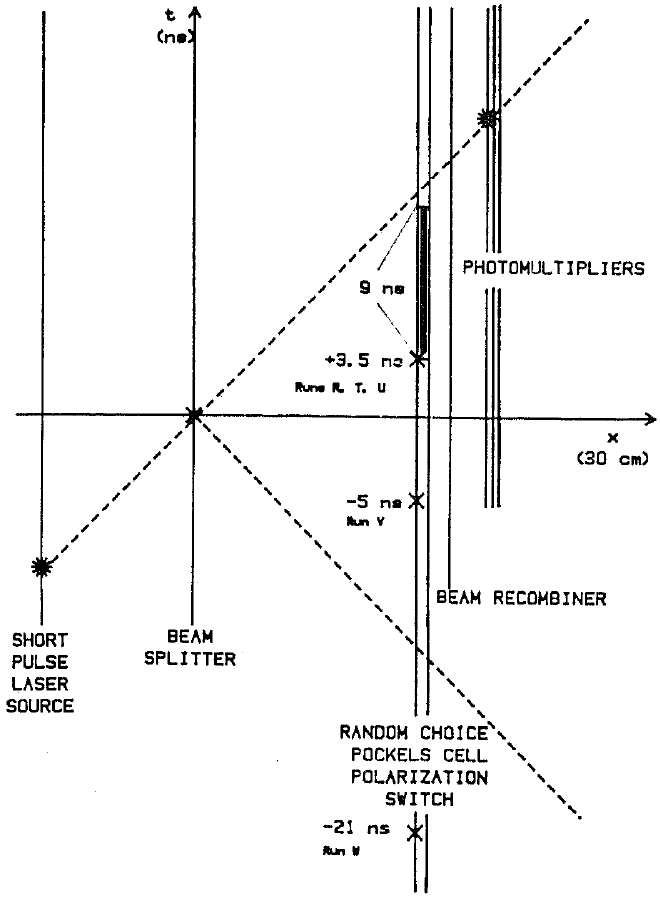}
    \caption{Space-time diagram reported in Ref.~\cite{Alley1986}; figure
taken therefrom. It shows the locations of the random-choice events
for different runs with respect to the photons meeting the beam
splitter and hence entering into the MZI in the laboratory reference
frame. In runs R, T, U, and Y, the choice events were space-like
separated from the photon's entry into the interferometer (origin of
the diagram).}\label{Alley2}
  \end{center}
\end{figure}

Alley and co-workers have put forward a concrete scheme for
realizing Wheeler's gedanken experiment with a delayed and random
choice for the configuration (open or closed) of a
MZI~\cite{Alley1983}. Three years later, they reported successful
experimental demonstration in Ref.~\cite{Alley1986}. The full
details of this work are described in~\cite{Jakubowicz1984}. The
experiment was conceptually similar to that in ~\cite{Hellmuth1987}
with some important differences. It was realized with a 4~m by 0.3~m
free space interferometer, where delayed random choices were
implemented. An additional photomultiplier was used to detect the
photons which were reflected out of the interferometer by the
combination of the Pockels cell and the polarizer. The random choice
was made at a photocathode which had a 50\% probability of producing
a photo-electron upon the strike of a laser pulse. This
photo-electron was then amplified and used to switch the EOMs in the
MZI. Detailed information of the random-choice generation is shown
in Fig.~\ref{Alley1}.

Five experimental runs with different space-time configurations were
implemented. The equivalent one-space-dimensional space-time diagram
is shown in Fig.~\ref{Alley2}. In runs R, T and U the choice events
were not only 3.5~ns delayed with respect to the entry of the
photons into the MZI in the laboratory reference frame (event E, the
origin point in Fig.~\ref{Alley1}) but also space-like separated
from E. In run Y, the choices were also space-like separated from E
but took place 5~ns earlier. In run W, the choices were in the
time-like past of E. About 90\% interference visibility was obtained
when the wave property of the input photons was measured, and no
observable interference was obtained when the particle property was
measured. The authors conclude:\ ``The predictions of quantum
mechanics are confirmed even with the choice of the final
configuration being made randomly during the course of the
`elementary quantum phenomenon"~\cite{Alley1986}.

\subsection{Wheeler's delayed-choice experiment with single particles: Photons and atoms}

\begin{figure}[tb]
  \begin{center}
    \includegraphics[width=0.4\textwidth]{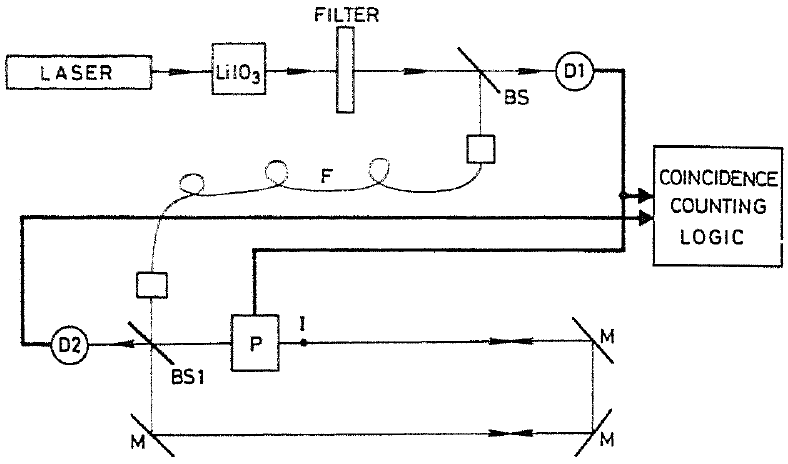}
    \caption{Setup of the delayed-choice experiment reported in
Ref.~\cite{Baldzuhn1989}; figure taken therefrom. Photon pairs were
produced by parametric down-conversion in the LiIO$_3$ crystal.
Detection of the (trigger) photon 1 in D1 heralded (signal) photon 2
propagating through fiber F to a Sagnac interferometer. The
detection at D1 triggered a Pockels cell P in the interferometer
through which the signal photons propagated in a clockwise or
anti-clockwise path before reaching detector D2. The signal photons
showed wave behavior, if the Pockels cell was continuously left on
or off. Particle behavior was revealed if the Pockels cell was
switched on at the moment when the signal photons reached the
reference point I in the interferometer.}\label{Baldzuhn1}
  \end{center}
\end{figure}

To meet the requirement of using a single-particle quantum state,
Baldzuhn and collaborators used heralded single photons generated
from spontaneous parametric down-conversion (SPDC)
~\cite{Friberg1985} to perform a delayed-choice wave-particle
experiment~\cite{Baldzuhn1989}. The layout of the setup is shown in
Fig.~\ref{Baldzuhn1}\textbf{A}. The detection of one (trigger)
photon was used to trigger a Pockels cell (P) in a Sagnac
interferometer~\cite{Sagnac1913} through which the other (signal)
photon propagated.

In the clockwise path, the signal photon first passed the Pockels
cell P and then the reference point I. In the anti-clockwise path,
however, the situation is reverse. (a) If the Pockels cell was off
during the photon's propagation through the whole interferometer,
the polarization of the signal photon was not rotated and remained
the same for both the clockwise and the anti-clockwise path. (b)
Similarly, if the Pockels cell was continuously on, the polarization
was rotated in both paths. In both cases (a) and (b) the final
polarization state was the same for both paths, leading to
interference. If, however, the Pockels cell was switched on at the
time when the signal photon arrives at the reference point I and was
kept on until after the photon met the beam splitter again, no
interference was observed. This is because the polarization of the
clockwise path remained unchanged, while the polarization of the
counter-clockwise path was rotated. The polarization degree of
freedom introduced a distinguishability between the two paths and
hence destroyed the possibility of interference.
\begin{figure}[tb]
  \begin{center}
    \includegraphics[width=0.48\textwidth]{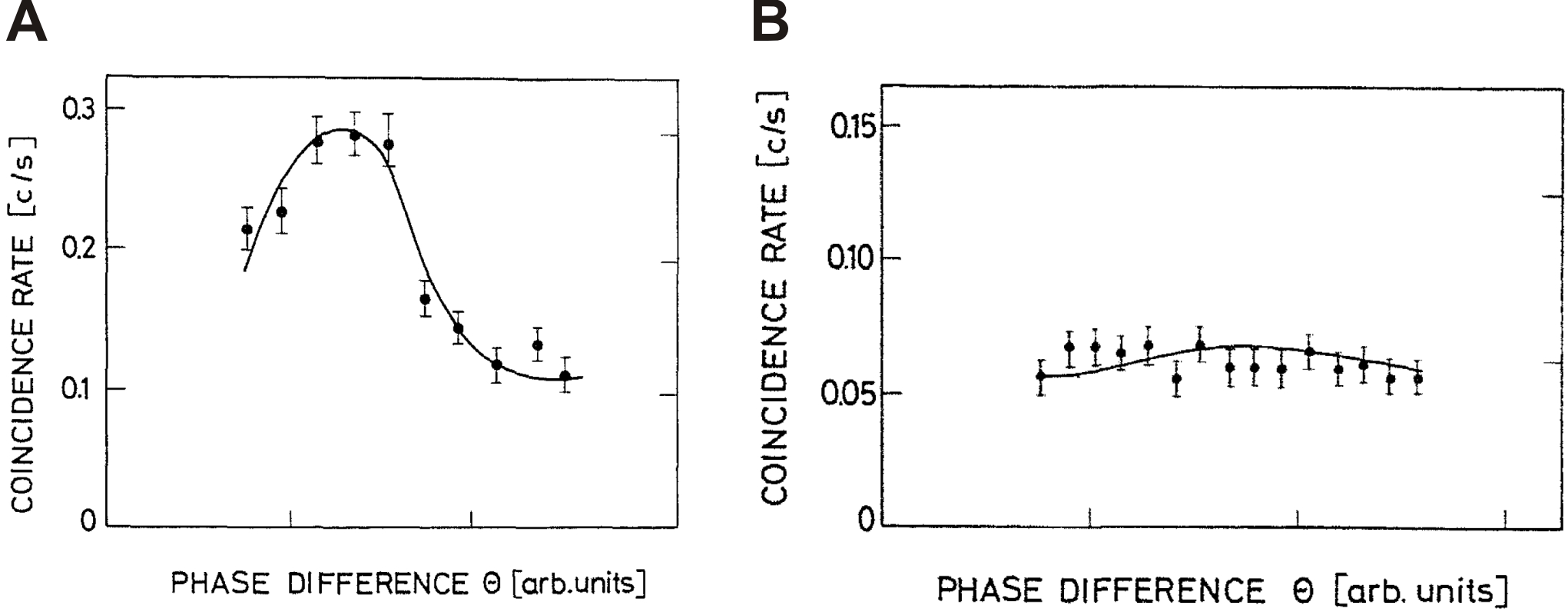}
    \caption{Experimental results of the delayed-choice
experiment in Ref.~\cite{Baldzuhn1989}; figures taken therefrom.
\textbf{A}:\ If the Pockels cell was continuously on or off, an
interference pattern was observed. \textbf{B}:\ If the Pockels cell
was switched on when the signal photon reached the reference point
I, indicated in Fig.~\ref{Baldzuhn1}, no interference showed
up.}\label{Baldzuhn2}
  \end{center}
\end{figure}

The experimental results are presented in Fig.~\ref{Baldzuhn2}. If
the Pockels cell was continuously on or off, one observed an
interference pattern (Fig.~\ref{Baldzuhn2}\textbf{A}). This
corresponds to the photon's wave-like behavior. On the other hand,
if the Pockels cell was switched on at the time when photon passed
the reference point I, no interference pattern was observed
(Fig.~\ref{Baldzuhn2}\textbf{B}). This corresponds to the
particle-like behavior of the photon.

The delayed-choice aspect of this experiment was realized by
delaying the signal photon by an optical fiber (labeled `F' in
Fig.~\ref{Baldzuhn2}\textbf{A}) and varying the time of the
application of the voltage on the Pockels cell via electronic
delays. This allowed to switch the Pockels cell at the time when the
photon was at the reference point, i.e.\ already within the
interferometer. Space-like separation between the choice of the
performed measurement and the entering of the photon into the
interferometer was not implemented in this experiment.

Very recently, a realization of Wheelers' delayed-choice gedanken experiment with single atoms has been reported~\cite{Manning2015}. The physical beam splitters and mirrors were replaced with optical Bragg pulses. The choice of either applying the last beam splitting pulse or not was controlled by an external quantum random number generator. This choice event occurred after the entry of the atoms into the interferometer.

\subsection{Wheeler's delayed-choice experiment with single photons and spacelike separation}

\begin{figure}[tb]
  \begin{center}
    \includegraphics[width=0.48\textwidth]{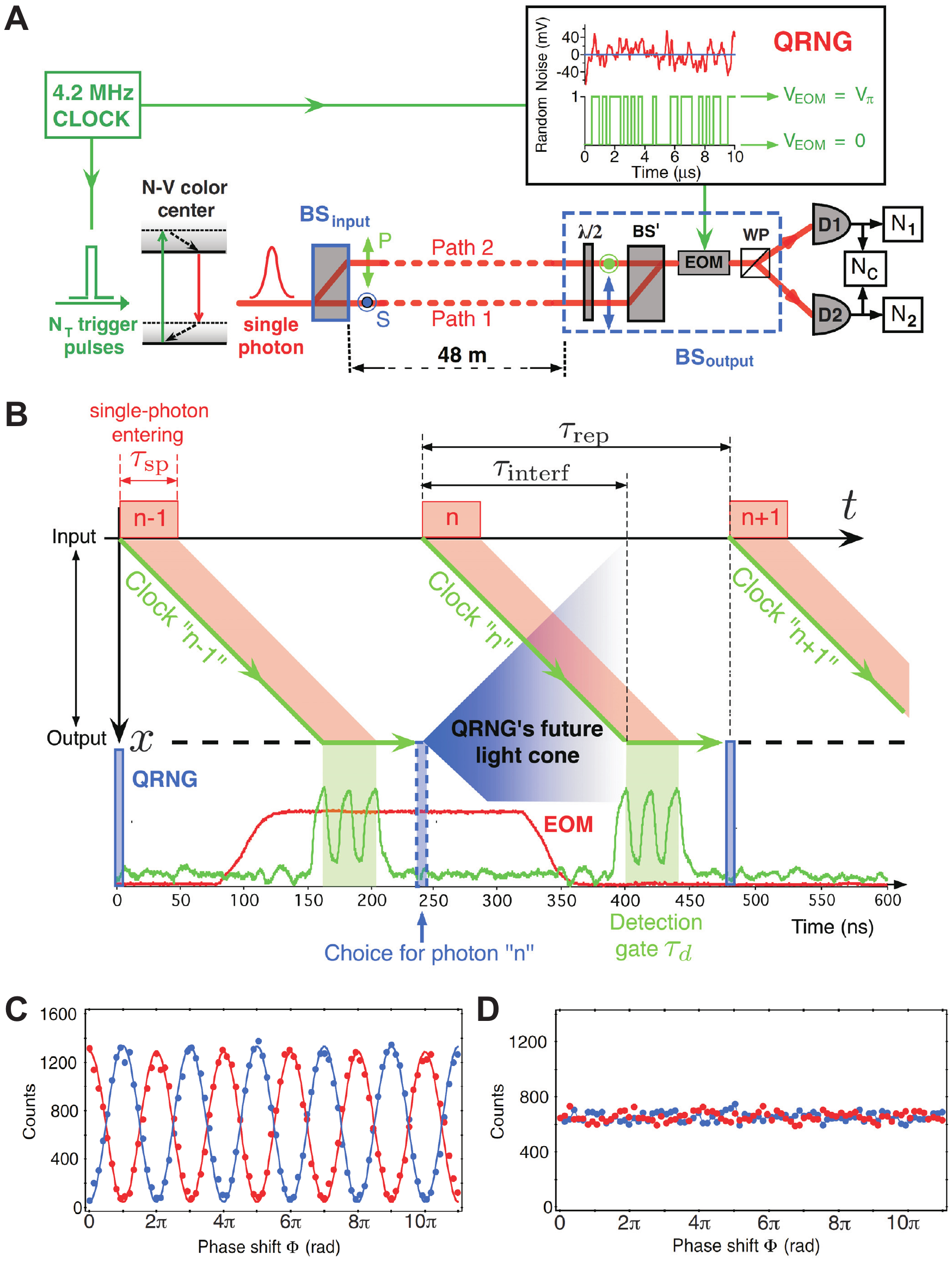}
    \caption{(Color online). The delayed-choice experiment realized in Ref.~\cite{Jacques2007};
    figures taken therefrom.
    \textbf{A}:\ Layout of the setup. Single photons were generated by NV color
    centers in diamond. A 48-meter-long
polarization interferometer and a fast electro-optical modulator
(EOM), controlled by a quantum random number generator (QRNG), were
used to fulfil the relativistic space-like separation condition. The
space-time diagram is shown in \textbf{B}. The choice whether to
open or close the interferometer was space-like separated from the
entry of the photon into the interferometer. If the EOM was on, the
polarization distinguishability of the two paths was erased and thus
an interference pattern emerged (\textbf{C}). If, however, the EOM
was switched off, no interference showed up due to the polarization
distinguishability of the two paths (\textbf{D}).}\label{Jacques1}
  \end{center}
\end{figure}

Two important requirements of an ideal realization of delayed-choice
wave-particle duality gedanken experiment -- namely, use of
single-particle quantum states as well as space-like separation
between the choice of the measurement and the entry of the particle
into the interferometer -- have been fulfilled simultaneously in
Refs.~\cite{Jacques2007, Jacques2008}. NV color centers in diamonds
were employed as single-photon sources~\cite{Kurt2000}. As shown in
Fig.~\ref{Jacques1}\textbf{A}, a 48-meter-long polarization
interferometer and a fast electro-optical modulator (EOM) controlled
by a quantum random number generator (QRNG) were used to fulfill
relativistic space-like separation. The random numbers were
generated from the amplified shot noise of a white light beam.

The space-time diagram of this experiment is illustrated in
Fig.~\ref{Jacques1}\textbf{B}. The sequence for the measurement
applied to the $n$-th photon constituted of three steps. First, the
choice was made by the QRNG, creating a binary random number (blue),
which determined the interferometer configuration. This choice
happened simultaneously with the trigger pulse of the $n$-th
photon's emission. Second, the random number (bit values 1, 0, 1 for
photons $n-1$, $n$, $n+1$ in Fig.~\ref{Jacques1}\textbf{B}) drove
the EOM voltage to $V = 0$ (bit value 0) or $V = V_{\pi}$ (bit value
1) within rise time 40 ns (red), which determined the state of the
second beam splitter (BS$_\textrm{output}$). Finally, the photon was
recorded by detectors D1 or D2, after its time of flight
$\tau_{\textrm{interf}}$ in the interferometer. The blue zone in
Fig.~\ref{Jacques1}\textbf{B} represents the future light cone of
the choice. The event ``entry of photon into the interferometer" was
space-like separated from the respective choice. If the EOM was on
with voltage $V = V_{\pi}$, one erased the polarization
distinguishability of the two paths and thus observed an
interference pattern when tuning the phase of the interferometer by
tilting BS$_\textrm{intput}$ (Fig.~\ref{Jacques1}\textbf{C}). If,
however, the EOM was switched off, due to the polarization
distinguishability of the two paths, no interference showed up
(Fig.~\ref{Jacques1}\textbf{D}).
\begin{figure}[tb]
  \begin{center}
    \includegraphics[width=0.35\textwidth]{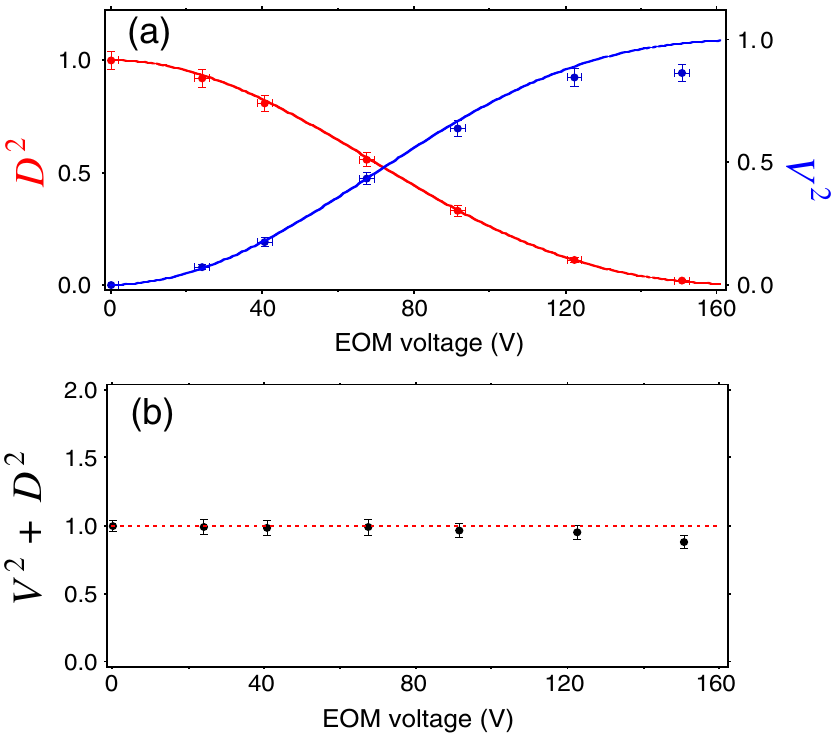}
    \caption{(Color online). Experimental visibility ($V^2$, starting at 0) and distinguishability ($D^2$, starting at 1) results from
Ref.~\cite{Jacques2008}; figure taken therefrom. (a) $V^2$ and $D^2$
as functions of the EOM voltage (corresponding to the reflectivity
of the second beam splitter). Solid lines are the theoretical
expectations. (b) $V^2+D^2$ as a function of the EOM voltage, in
agreement with Ineq.~(\ref{eq V-D}).}\label{Jacques2}
  \end{center}
\end{figure}

Furthermore, Jacques and co-workers varied the driving voltages
applied to the EOM and thus realized a fast switchable beam splitter
with an adjustable reflection coefficient
\textit{R}~\cite{Jacques2008}. The QRNG switched this beam splitter
on and off randomly. Each randomly set value of \textit{R} allowed
them to obtain partial interference with visibility $V$ and partial
which-path information. The which-path information was parameterized
by the distinguishability $D$. The authors confirmed that $V$ and
$D$ fulfilled the complementary relation~\cite{Wootters1979,
Greenberger1988,Jaeger1995,Englert1996}
\begin{equation}\label{eq V-D}
V^2+D^2 \leq 1,
\end{equation}
where equality holds for pure states (see Fig.~\ref{Jacques2}). The
visibility is defined as $V =
(p_{\textrm{max}}-p_{\textrm{min}})/(p_{\textrm{max}}+p_{\textrm{min}})$,
with $p_{\textrm{max}}$ and $p_{\textrm{min}}$ the maximal and
minimal probability for recording a photon in a chosen detector when
scanning through the phase of the interferometer. The
distinguishability (or which-path information) is defined as $D =
D_1+D_2$ with $D_i = |p(i,1)-p(i,2)|$ and $p(i,j)$ the probability
that the photon traveled path $i=1,2$ and is recorded by detector
$j=1,2$. The quantity $D_1$ is measured by blocking path 2, and vice
versa.

\section{Realizations of delayed-choice quantum-eraser experiments}

Delayed-choice experiments with two particles offer more
possibilities than those with single particles. Especially in the
experiments performed with entangled particles in the context of
quantum erasure, the choice of measurement setting for one particle
can be made even after the other particle has been registered. This
has been shown in delayed-choice quantum eraser experiments, where
the which-path information of one particle was erased by a later
suitable measurement on the other particle. This allowed to a
posteriori decide a single-particle characteristic, namely whether
the already measured photon behaved as a wave or as a particle. We
will discuss the experimental realizations along this line in the
following sections.

\subsection{Photonic quantum erasure}

Energy-time~\cite{Friberg1985,Joobeur1994},
momentum~\cite{Rarity1990} and
polarization~\cite{Shih1988,Kwiat1995} entanglement of photon pairs
generated from SPDC have been widely used in experiments realizing
photonic quantum erasure. Herzog and co-workers used photon pairs
generated from type-I SPDC and demonstrated the quantum eraser
concept via various experiments~\cite{Herzog1995}. Polarization as
well as time delay were used as quantum markers, and wave plates as
well as narrow-bandwidth interference filters as quantum erasers.
They harnessed the momentum entanglement and polarization
correlation between photon pairs, and performed remote measurements
on one photon either revealing or erasing which-path information of
the other one.

\begin{figure}[tb]
  \begin{center}
    \includegraphics[width=0.375\textwidth]{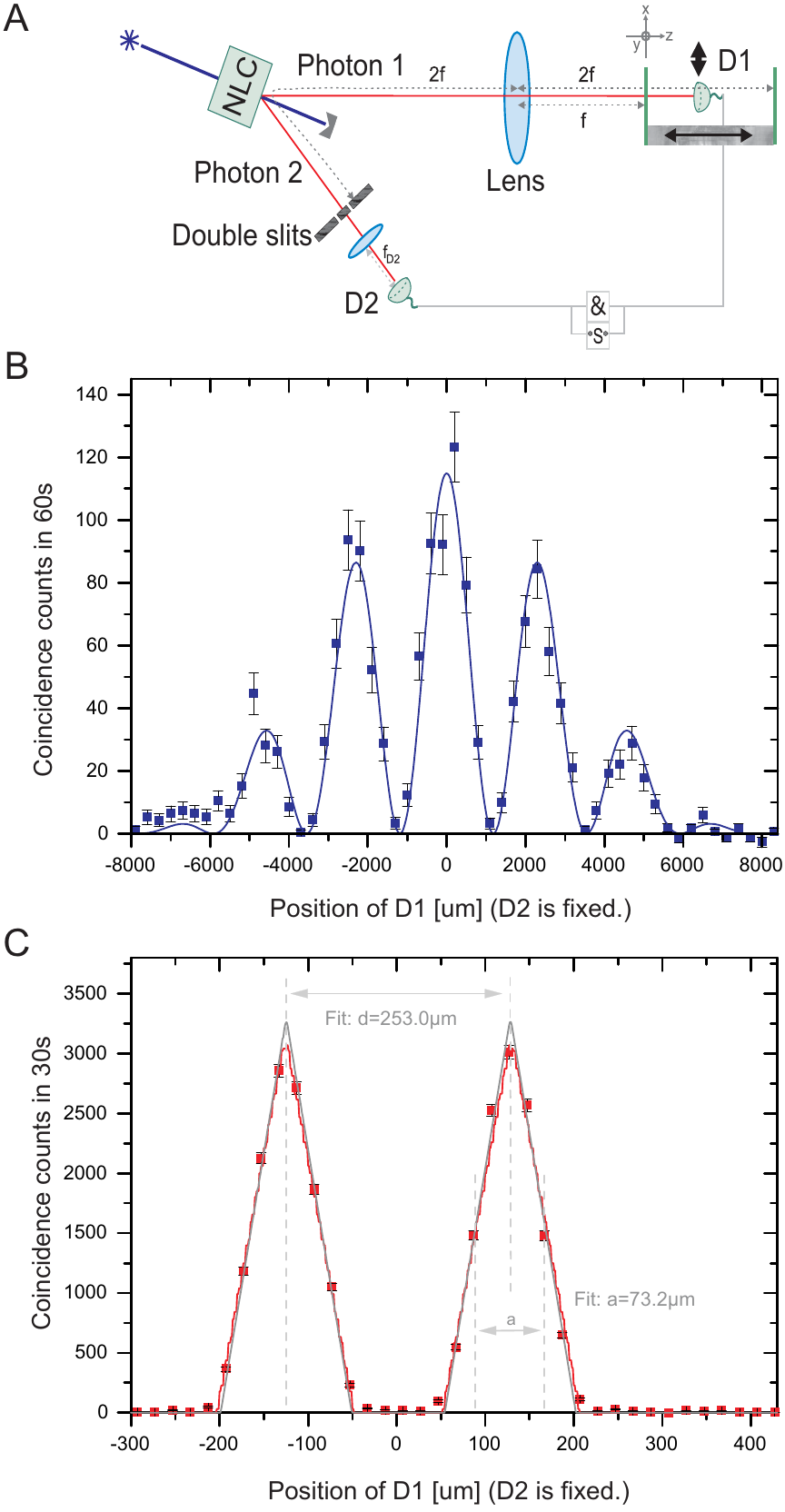}
    \caption{(Color online). \textbf{A}:\ Experimental scheme of the experiment,
    using a momentum entangled state of two photons. See text for
    details. \textbf{B}:\ A high-visibility interference pattern in the
conditional photon counts was obtained when D1 was positioned in the
focal plane of the lens, thus erasing all path information.
\textbf{C}:\ The profile of the double slit was resolved when D1 was
positioned in the image plane of the lens, revealing path
information and therefore no interference pattern arose. Figures
taken from~\cite{Dopfer1998}.}\label{dopfer}
  \end{center}
\end{figure}

An arrangement consisting of a double slit and two entangled
particles allows a combination of the gedanken experiments of
Heisenberg's microscope and the quantum eraser. Dopfer and
collaborators employed photon pairs generated from type-I
SPDC~\cite{Dopfer1998,Zeil1999,Zeil2005}. Due to the phase matching
condition, photons 1 and 2 were entangled in their linear momentum
states. Fig.~\ref{dopfer}\textbf{A} shows one of their experimental
configurations. Photon 2 passed a double-slit and a lens and was
measured by a static detector D2 in the focal plane. Photon 1 was
sent through another lens with focal length \textit{f} and was
measured by detector D1, which was mounted on translation stages
capable of moving along both axes \textit{x} and \textit{z}. This
allowed an implementation of the idea of von Weizs\"{a}cker
switching from the focal plane to the image plane.

If D1 was placed in the focal plane of the lens (i.e.\ at distance
$f$ from the lens), one measured photon 1's momentum state and hence
lost the information about its position. Due to the momentum
entanglement, the measurement of photon 1's momentum state projected
the state of photon 2 into a momentum eigenstate which could not
reveal any position information. One therefore had no information
whatsoever about which slit photon 2 went through. When both photons
were detected, neither photon 1 nor photon 2 revealed any path
information. Therefore, when coincidence counts between D1 and D2
were measured as a function of D1's position along the $x$-axis, an
interference pattern showed up with a visibility as high as
$97.22\%$ (Fig.~\ref{dopfer}\textbf{B}).

On the other hand, when D1 was placed in the image plane (i.e.\
distance $2f$ from the lens), the detection events of photon 1
revealed the path photon 2 took through the double slit. In
Fig.~\ref{dopfer}\textbf{C}, two prominent peaks indicate the
profile of the double-slit assembly with no interference pattern.

In the experiment of Walborn and collaborators~\cite{Walborn2002},
one photon of a polarization-entangled pair impinged on a special
double-slit device, where two quarter-wave plates, oriented such
that their fast axes are orthogonal, were placed in front of each
slit to serve as which-path markers. The quarter-wave plates rotated
the polarization states of the photons passing through them and
hence the subsequent slits. This rotation introduced a
distinguishability of the two possible paths and thus destroyed the
interference pattern. To recover interference, polarization
entanglement was used and the polarization of the other entangled
photon was measured in a proper basis. This experiment was also
performed under delayed erasure conditions, in which the interfering
photon is detected before its entangled twin. The experimental data
was in agreement with the predictions of quantum mechanics.

\subsection{Matter-wave quantum erasure}

Light scattered from laser-cooled atoms provides information on the
localization of atoms and can be used to realize quantum eraser
experiments, if the atomic separation is large enough and the wave
length of the scattered light short enough to allow in principle
identification of the atom's position by imaging conditions. In
Ref.~\cite{Eichmann1993}, an experiment with light scattered from
two ions was performed. By employing the polarization of the
scattered light, the authors realized the above mentioned cases A
and B in Section II.E and observed polarization-detection dependent
interference patterns.
\begin{figure}[tb]
  \begin{center}
    \includegraphics[width=0.48\textwidth]{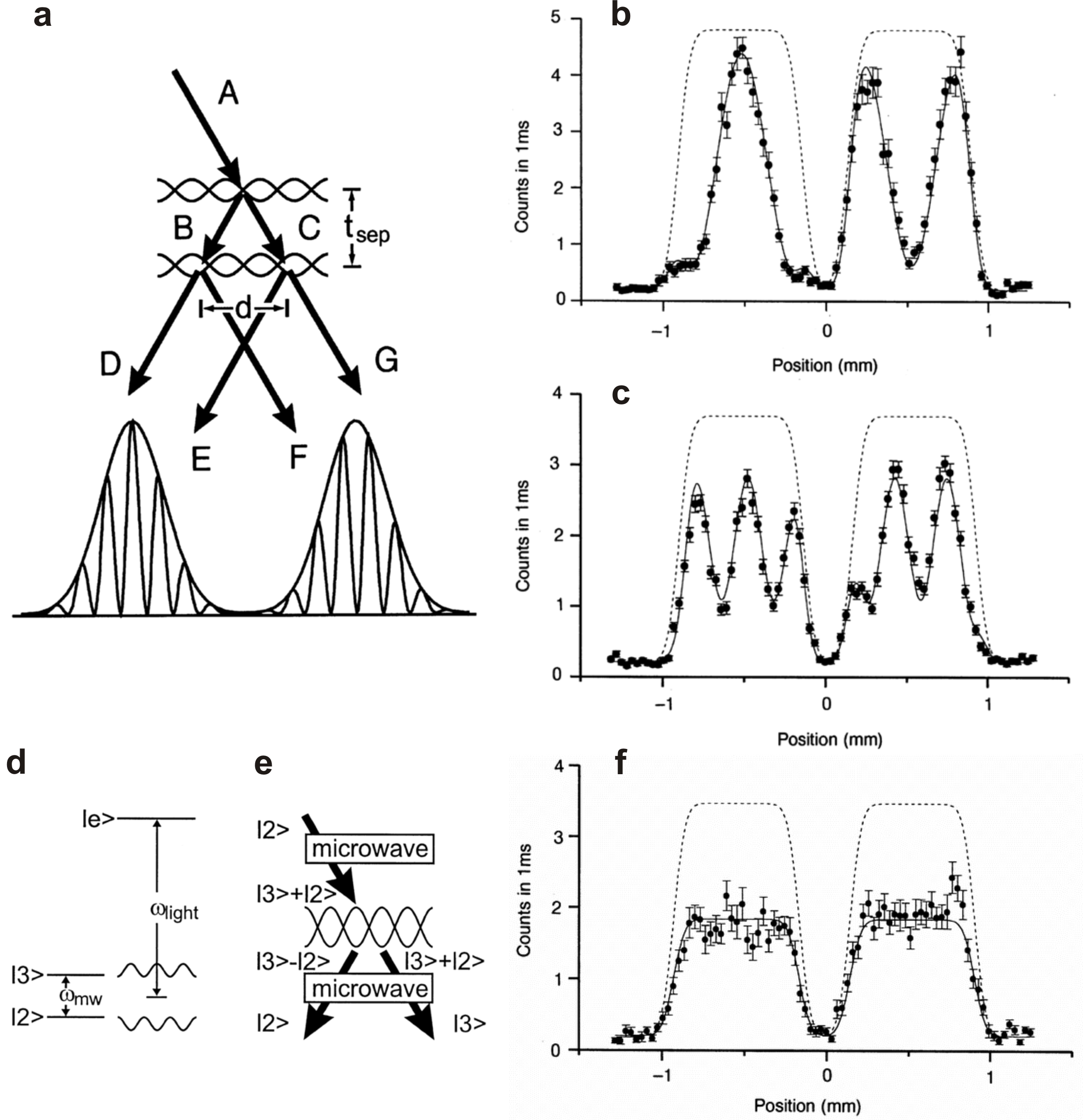}
    \caption{Quantum-eraser experiment from Ref.~\cite{Durr1998} based on an atomic
interferometer; figures taken therefrom. \textbf{a}:\ The atomic
beam A was split into two beams; beam B was reflected by the first
Bragg grating formed by a standing wave, and beam C was transmitted.
The atomic beams freely propagated for a time duration of
$t_{\textrm{sep}}$ and acquired a lateral separation $d$. The beams
B and C were then split again by a second standing light wave
grating. In the far field, complementary spatial interference
patterns were observed in two regions. These interference patterns
were due to superpositions of beams D and E (F and G). \textbf{b}
and \textbf{c} show the spatial fringe patterns in the far field of
the interferometer for $t_{\textrm{sep}} = 105\, \mu$s with $d =
1.3\, \mu$m and $t_{\textrm{sep}} = 255\, \mu$s with $d = 3.1\,
\mu$m, respectively. The left and right complementary interference
patterns were respectively generated by the atomic beams D and E,
and F and G (shown in \textbf{a}). The dashed lines indicate the sum
of the intensities of two interference patterns obtained with a
relative phase shift of $\pi$. \textbf{d} illustrates the simplified
scheme of the internal atomic states, which were addressed using
microwave (mw) radiation and light. \textbf{e} illustrates the
principle of correlating the path the atoms took with their internal
electronic states. The standing-wave grating produced a relative
$\pi$ phase shift of state $\ket{2}$ relative to $\ket{3}$
conditional on its path. A Ramsey interferometer employed two
microwave $\frac{\pi}{2}$ pulses and converted different relative
phases into different final internal states $\ket{2}$ and $\ket{3}$,
respectively. \textbf{f}:\ When the which-path information was
stored in the internal atomic state, the interference patterns
vanished.}\label{Durr}
  \end{center}
\end{figure}

D\"{u}rr and collaborators carried out an atomic interferometric
experiment showing that the disturbance of path detection on an
atom's momentum is too small to destroy the interference
pattern~\cite{Durr1998}. The principle of this experiment is shown
in Fig.~\ref{Durr}\textbf{a}. By using a standing-wave grating
formed by off-resonant laser light, the collimated atomic beam A was
split into two beams: beam B was reflected and beam C was
transmitted. After free propagation for a time duration of
$t_{\textrm{sep}}$, they were separated by a lateral distance $d$.
The beams B and C were then split again by a second standing light
wave grating. In the far field, complementary spatial interference
patterns were observed in two regions. Experimentally, the authors
varied the phase of the atomic interferometer by setting different
separation durations $t_{\textrm{sep}}$ between the first and the
second standing-wave gratings. Interference patterns with
visibilities of $(75\pm1)\%$ and $(44\pm1)\%$ for $t_{\textrm{sep}}
= 105\, \mu$s and $255\, \mu$s, shown in Fig.~\ref{Durr}\textbf{b}
and \textbf{c}, have been observed which were in good agreement with
the theoretical expectations.

The internal electronic states $\ket{2}$ and $\ket{3}$ were used as
a which-path detector for the paths B and C (shown in
Fig.~\ref{Durr}\textbf{d}). These two states were addressed and
manipulated with microwave pulses. Fig.~\ref{Durr}\textbf{e} shows
how the atomic internal electronic states were employed in
controlling the paths the atoms took. The authors converted the
input state $\ket{2}$ to a superposition state $\ket{2}+\ket{3}$ by
a $\pi/2$ microwave pulse with frequency
$\omega_{\textrm{mw}}=\omega_{3}-\omega_{2}$, where $\omega_{2}$ and
$\omega_{3}$ are the frequencies of states $\ket{2}$ and $\ket{3}$.
(We omit the normalization to be consistent with the original
notation). Then a standing-wave grating was formed by a laser with
frequency $\omega_{\textrm{light}}$, which was tuned to be halfway
between the $\ket{2}\rightarrow\ket{e}$ and
$\ket{3}\rightarrow\ket{e}$ transitions to the excited state
$\ket{e}$, i.e.\
$\omega_{\textrm{light}}=\omega_{e}-\frac{\omega_{3}-\omega_{2}}{2}$.
Due to these detunings an internal-state dependent phase shift was
implemented. In the reflected arm (B), the light grating induced a
$\pi$ phase shift on state $\ket{2}$ with respect to $\ket{3}$
resulting in state $\ket{3}-\ket{2}$. In the transmitted arm (C), no
phase shift was induced and hence the state remained
$\ket{3}+\ket{2}$. A subsequent $\pi/2$ microwave pulse converted
the superposition states in the reflected and transmitted arm to
$\ket{2}$ and $\ket{3}$, respectively. Therefore, the atom path in
the interferometer was correlated with its internal electronic
states. Consequently, no interference patterns did arise, as shown
in Fig.~\ref{Durr}\textbf{f}.

In this experiment, the disturbance of the path, which was induced
by using microwave pulses, was four orders of magnitude smaller than
the fringe period and hence was not able to explain the
disappearance of the interference patterns. Instead, ``the mere fact
that which-path information is stored in the detector and
\textit{could} be read out already destroys the interference
pattern"~\cite{Durr1998}.

Recently, an experimental realization of quantum erasure in a
mesoscopic electronic device has been reported in
Ref.~\cite{Weisz2014}. Interacting electrons have been used to
extract which-path information and a smooth variation of the degree
of quantum erasure has been demonstrated.

We also remark here that neutral kaon systems have been
theoretically suggested to be suitable for a demonstration of
quantum erasure as shown in Ref.~\cite{Bramon2004}. There,
strangeness oscillations would represent the interference pattern
linked to wave-like behavior.

\subsection{Quantum erasure with delayed choice}

\begin{figure}[tb]
  \begin{center}
    \includegraphics[width=0.48\textwidth]{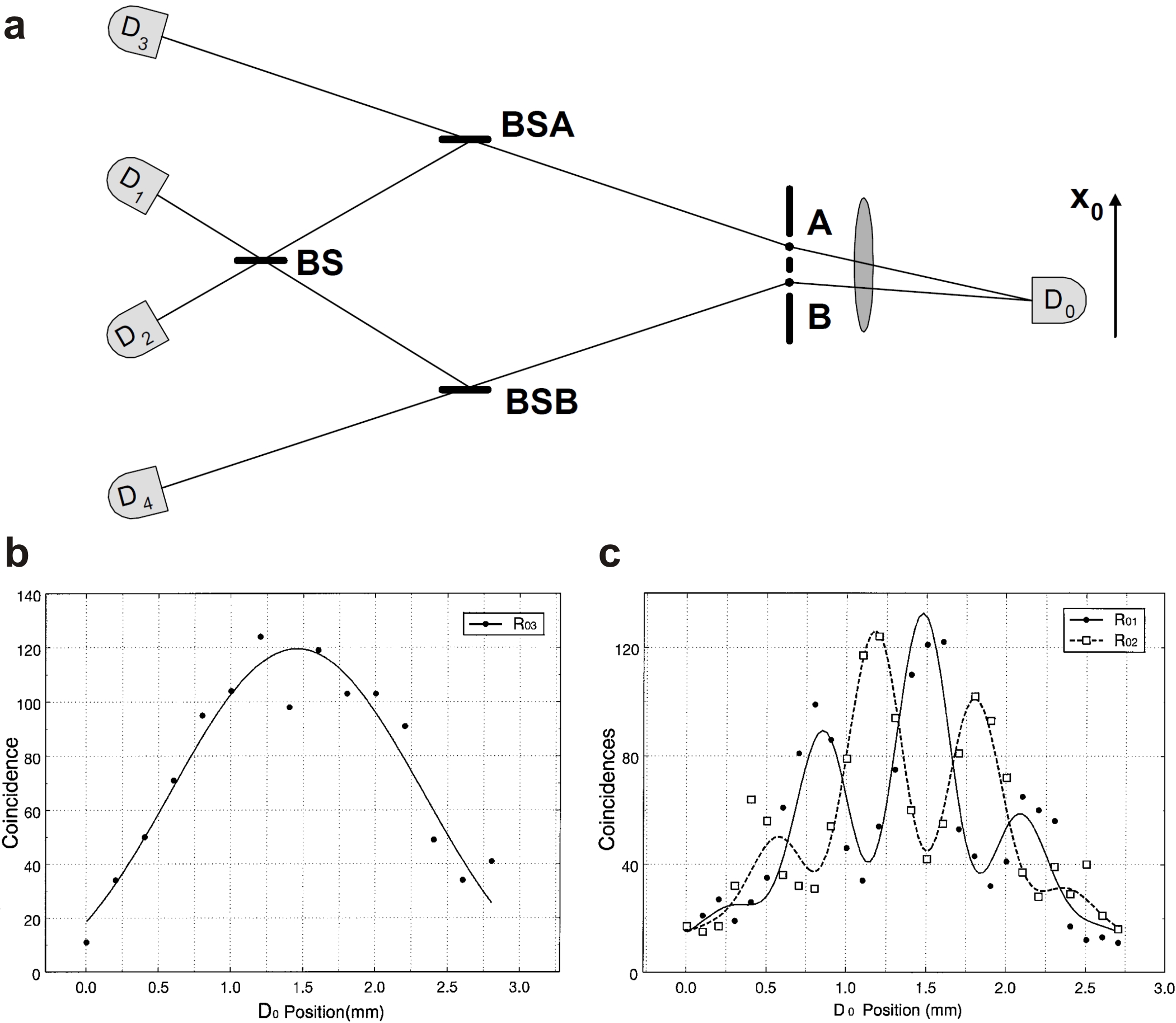}
    \caption{Delayed-choice quantum-eraser experiment realized in
    Ref.~\cite{Kim2000}; figures taken therefrom. \textbf{a}:\ Experimental
    scheme. Pairs of entangled photons were emitted
    from either region A or region B of a BBO crystal via spontaneous
    parametric down-conversion. These two emission processes were coherent.
    Detections at D$_3$ or D$_4$ provided which-path information and
    detections at D$_1$ or D$_2$ erased it.
    \textbf{b}:\ Coincidence counts between D$_0$ and D$_3$, as a function
    of the lateral position $x_0$ of D$_0$. Absence of
    interference was demonstrated. \textbf{c}:\ Coincidence counts
    between D$_0$ and D$_1$ as well as between D$_0$ and D$_2$ are plotted as a function of $x_0$.
    Interference fringes were obtained. See text for details.}\label{Kim1}
  \end{center}
\end{figure}

In~\cite{Kim2000}, pairs of entangled photons were used to mimic the
entangled atom-photon system proposed in~\cite{Scully1991}. The
layout of the experimental setup is shown in
Figure~\ref{Kim1}\textbf{a}. Photon pairs were generated
noncollinearly either from region A or region B of a $\beta$-Barium
borate (BBO) crystal via type-I SPDC. From each pair, photon 1,
simulating the atom, propagated to the right and was focused by a
lens. It was then detected by $D_0$, which was mounted on a step
motor capable of changing the lateral position $x_0$.

Photon 2, propagating to the left, passed through one or two of the
three beam splitters. If the pair was generated in region A, photon
2 would follow path a and meet beam splitter BSA, where it had a
50\% chance of being reflected or transmitted. If the pair was
generated in region B, photon 2 would propagate path b and meet beam
splitter BSB, again with a 50\% chance of being reflected or
transmitted.

In the case that photon 2 was transmitted at BSA or BSB, it would be
detected by detector D$_3$ or D$_4$, respectively. The detection of
D$_3$ or D$_4$ provided which-path information (path a or path b)
for photon 2, thus also providing the which-path information for
photon 1 due to the linear momentum entanglement of the photon pair.
Therefore, there was no interference, as verified by the results
shown in Figure~\ref{Kim1}\textbf{b}.

On the other hand, given a reflection at BSA or BSB, photon 2
continued its path to meet another 50:50 beam splitter BS and was
then detected by either D$_1$ or D$_2$. The detection by D$_1$ or
D$_2$ erased the which-path information carried by photon 2 and
therefore an interference pattern showed up for photon 1
(Figure~\ref{Kim1}\textbf{c}). This confirmed the theoretical
prediction.

The ``choice'' of observing interference or not was made randomly by
photon 2 by being either reflected or transmitted at BSA or BSB. In
the actual experiment, the photons traveled almost collinearly, but
the distance from the BBO to BSA and BSB was about 2.3~m (7.7~ns)
longer than the distance from the BBO to D$_0$. Thus, after D$_0$
was triggered by photon 1, photon 2 was still be on its way to BSA
or BSB, i.e., the which-path or the both-path choice was ``delayed"
compared to the detection of photon 1.

As an extension, a delayed-choice quantum eraser experiment based on
a two-photon imaging scheme using entangled photon pairs (signal and
idler photons) was reported in Ref.~\cite{Scarcelli2007}. The
complete which-path information of the signal photon was transferred
to the distant idler photon through a ``ghost" image. By setting
different sizes of the apertures, the authors could either obtain or
erase which-path information. In the case of which-path information
erasure, interference with a visibility of about 95\% was obtained.
When not erasing which-path information, no interference was
observed.

\subsection{Quantum erasure with active and causally disconnected choice}
Quantum erasure with an active and causally disconnected choice was
experimentally demonstrated in Ref.~\cite{Ma2013a}. To this end, the
erasure event of which-path information had to be space-like
separated from the passage of the interfering system through the
interferometer as well as its detection event. Based on the special
theory of relativity, the event of quantum erasure was therefore
causally disconnected from all relevant interference events.
\begin{figure}[tb]
    \includegraphics[width=0.48\textwidth]{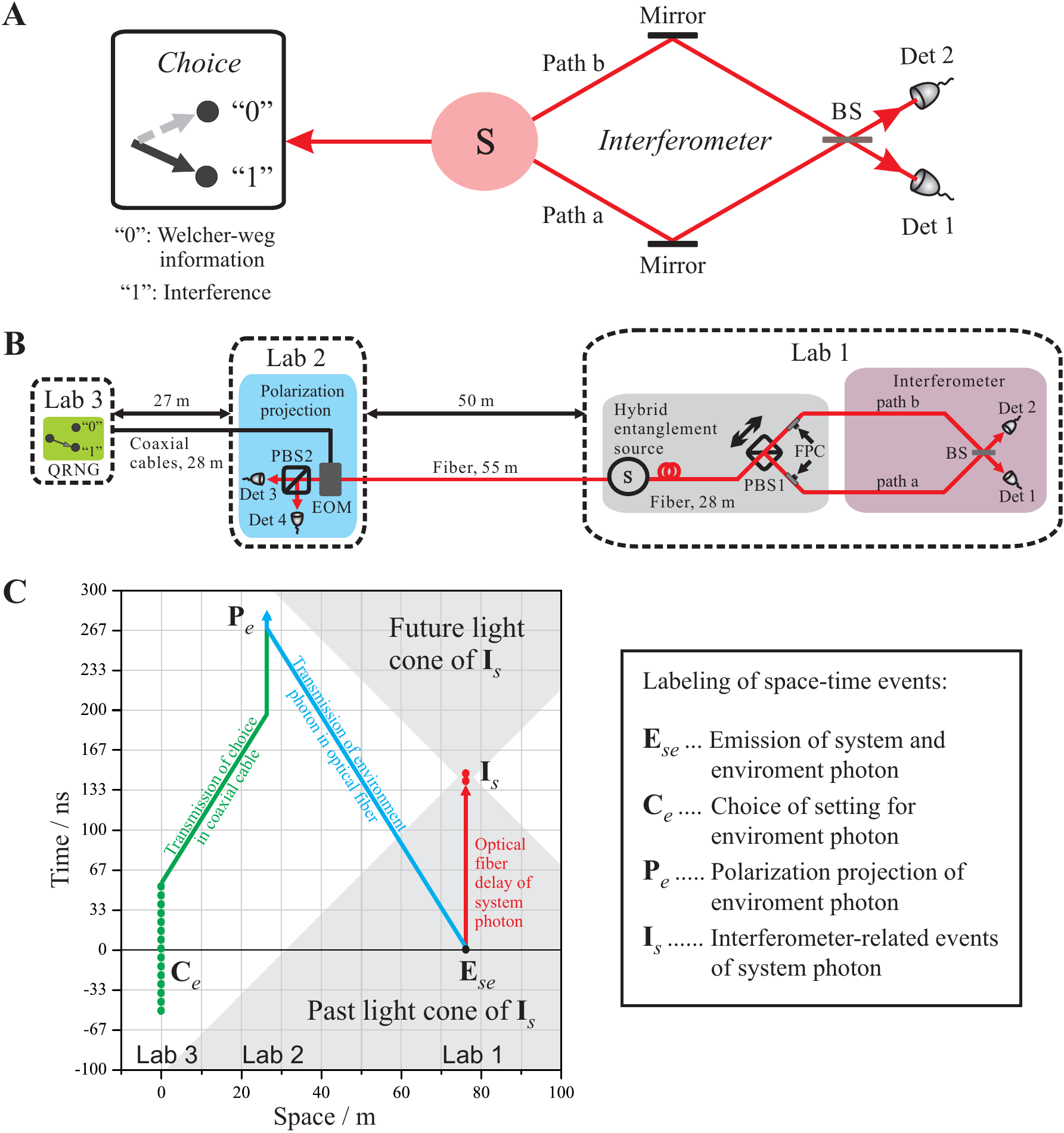}
    \caption{(Color online). Quantum erasure with causally disconnected
choice. \textbf{A}:\ Principle:\ The source S emitted
path-polarization entangled photon pairs. The system photons
propagated through an interferometer (right side), and the
environment photons were subject to polarization measurements
(left). \textbf{B}:\ Scheme of the Vienna experiment: In Lab~1, the
polarization entangled state generated via type-II spontaneous
parametric down-conversion, was converted into a hybrid entangled
state with a polarizing beam splitter (PBS1) and two fiber
polarization controllers (FPC). In Lab~2, the polarization
projection setup of the environment photon consisted of an
electro-optical modulator (EOM) and another polarizing beam splitter
(PBS2). In Lab~3, the choice was made with a quantum random number
generator (QRNG)~\cite{Jennewein2000}. \textbf{C}:\ Space-time
diagram. The choice-related events \textbf{C}$_{e}$ and the
polarization projection of the environment photon \textbf{P}$_{e}$
were space-like separated from all events of the interferometric
measurement of the system photon \textbf{I}$_{s}$. Additionally, the
events \textbf{C}$_{e}$ were also space-like separated from the
emission of the entangled photon pair from the source
\textbf{E}$_{se}$. The shaded areas are the past and the future
light cones of events \textbf{I}$_{s}$. This ensured that Einstein
locality was fulfilled. Figures taken
from~\cite{Ma2013a}.}\label{MaNQE1}
\end{figure}

The concept of the experiment is illustrated in Fig.\
\ref{MaNQE1}\textbf{A}. Hybrid entangled photon pairs~\cite{Ma2009}
were produced, with entanglement between the path a or b of one
photon (the system photon $s$) in an interferometer, and the
polarization H or V of the other photon (the environment photon
$e$):
\begin{equation} \label{hybrid1}
{\left| \Psi \right\rangle} _{se} ={\tfrac{1}{\sqrt{2} }} ({\left|
\textrm{a} \right\rangle} _{s} {\left| \textrm{H} \right\rangle}
_{e} +{\left| \textrm{b} \right\rangle} _{s} {\left| \textrm{V}
\right\rangle} _{e} ).
\end{equation}
Analogous to the original proposal of the quantum eraser, the
environment photon's polarization carried which-path information of
the system photon due to the entanglement between the two photons.
Depending on which polarization basis the environment photon was
measured in, one was able to either acquire which-path information
of the system photon and observe no interference, or erase
which-path information and observe interference. In the latter case,
it depended on the specific outcome of the environment photon which
one out of two different interference patterns the system photon was
showing.

The quantum eraser concept under Einstein locality was tested on two
different length scales. In a first experiment performed in Vienna
in 2007, the environment photon was sent away from the system photon
via a 55~m long optical fiber (Fig.\ \ref{MaNQE1}\textbf{B} and
\textbf{C}). In a second experiment performed on the Canary Islands
in 2008, the photons were separated by 144 km via a free-space link.
See the caption of Fig.\ \ref{MaNQE1} for details on the first
experiment and its space-time diagram.
\begin{figure}[tb]
    \includegraphics[width=0.325\textwidth]{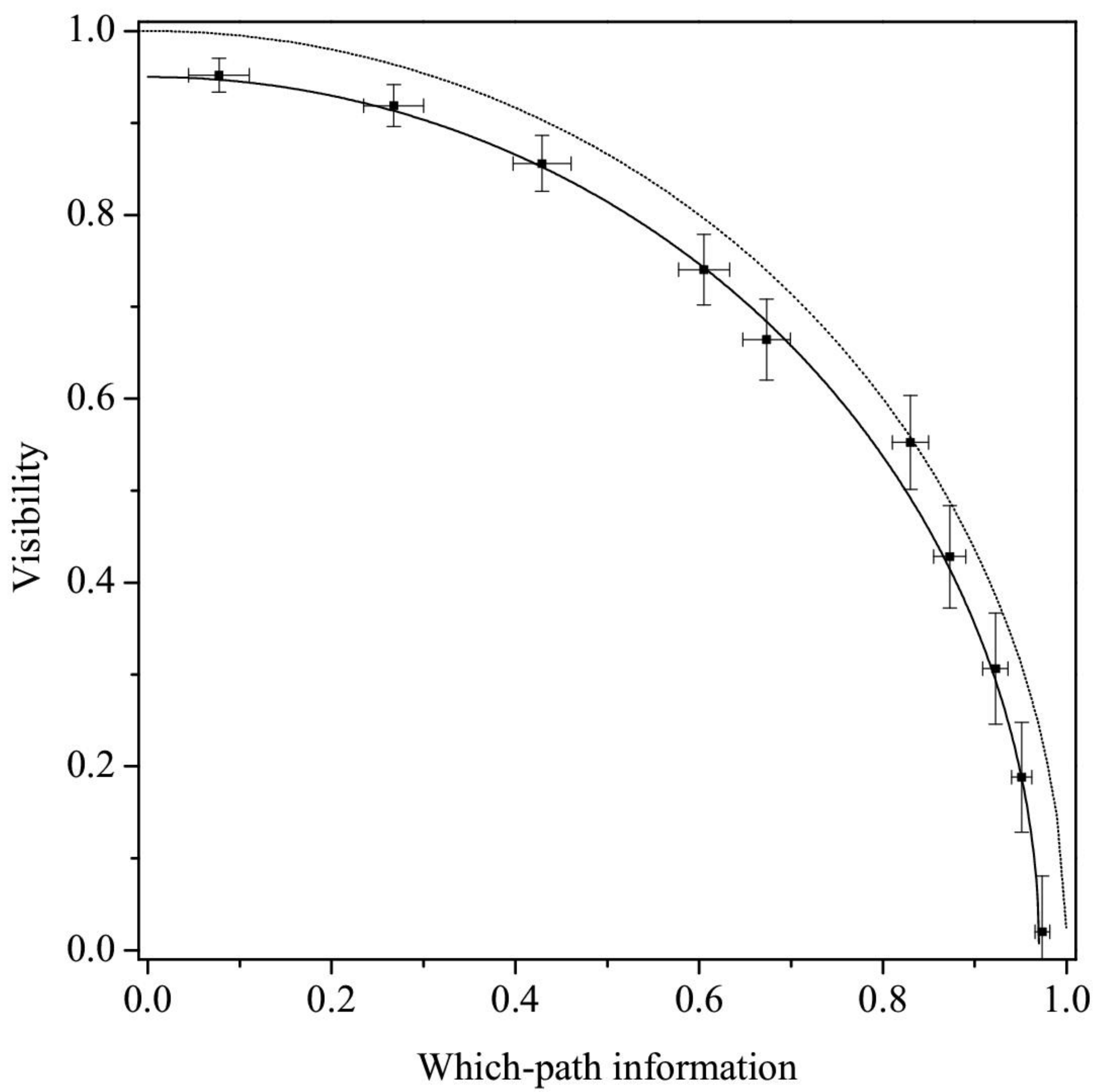}
    \caption{Experimental test of the complementarity inequality under Einstein
locality, manifested by a trade-off of the which-path information
parameter $D$ and the interference visibility $V$. The dotted line
is the ideal curve from the saturation of the complementary
inequality. The solid line, $V=0.95\,
[1-\left(D/0.97\right)^{2}]^{1/2}$, is the estimation from
experimental imperfections. Figure taken
from~\cite{Ma2013a}.}\label{MaNQE2}
\end{figure}

In order to quantitatively demonstrate quantum erasure under
Einstein locality, the authors employed a bipartite complementarity
inequality of the form (\ref{eq
V-D})~\cite{Wootters1979,Greenberger1988,Jaeger1995,Englert1996}, in
which $D$ and $V$ stand for conditional which-path information
(distinguishability) and interference visibilities respectively. It
is an extension of the single-particle complementarity inequality
(experimentally verified in Ref.\ \cite{Jacques2008} and discussed
in Section III.C). Under Einstein locality, $D$ and $V$ were
measured in sequential experimental runs as a function of the
applied voltage of the EOM, which changed the polarization
projection basis of the environment photon. Hence, a continuous
transition between measurements of particle nature and wave nature
was acquired. The results are shown in Fig.\ \ref{MaNQE2}.

\begin{figure}[tb]
    \includegraphics[width=0.35\textwidth]{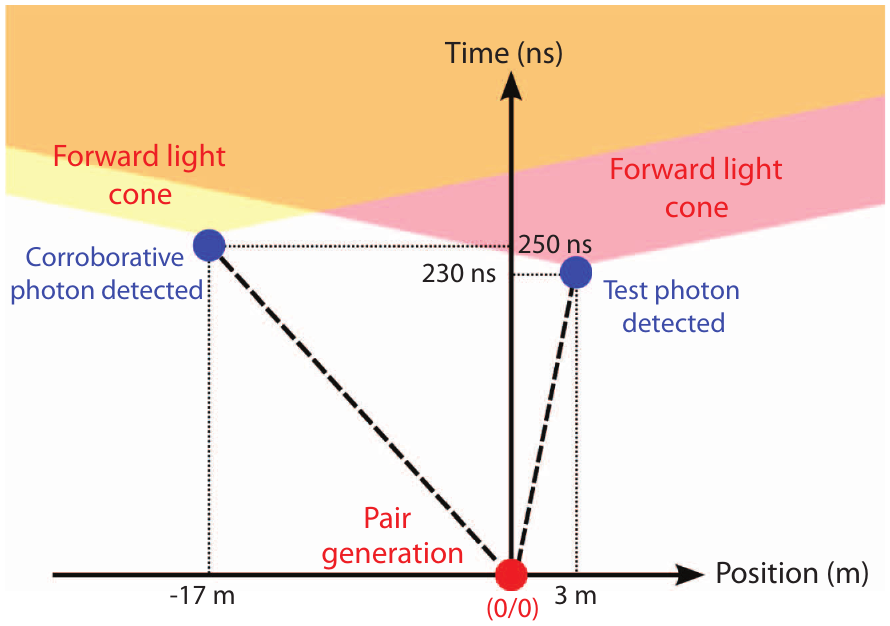}
    \caption{(Color online). Space-time diagram of the experiment reported in
Ref.~\cite{Kaiser2012}; figure taken therefrom. The detections of
the corroborative photon and the test photon were space-like
separated.}\label{kaiser0}
\end{figure}

Note that similar setups have been proposed in Refs.\
\cite{Grangier1986a, Ballentine1998, Kwiat2004}. Another successful
experiment along this line was reported in~\cite{Kaiser2012}. Kaiser
and collaborators used polarization-entangled photon pairs at the
telecom wavelength. Every `test' photon was sent into an
interferometer (with phase $\theta \equiv \varphi$), while the
corresponding `corroborative' photon was subject to a polarization
measurement. While no active random choices were implemented in
their experiment, the detection events of the corroborative and test
photon were space-like separated
(Fig.~\ref{kaiser0}).

The interferometer employed a polarization-dependent beam splitter
(PDBS) with bulk optics which was able to reflect horizontally
polarized test photons with close to 100\% probability and
reflect/transmit vertically polarized photons with 50\%/50\%
probability. Then polarizing beam splitters oriented at 45° to the
H/V basis erased all polarization information that potentially
existed at the PDBS output. The corroborative photon passed an EOM
which rotated its polarization state by an angle $\alpha$ before it
was measured. The total quantum state of the test (\textit{t}) and
corroborative photon (\textit{c}) was
\begin{align} \label{hybrid2}
\left\vert \Psi\right\rangle_{tc}\nonumber
&=\tfrac{1}{\sqrt{2}}[(\cos\alpha\,|\text{particle}\rangle_{t}-\sin\alpha\,|\text{wave}\rangle_{t})\left\vert\text{H}\right\rangle_{c}\\
&\,\,\,\,\,\,\,\,\,\,+(\cos\alpha\,|\text{wave}\rangle_{t}+\sin\alpha\,|\text{particle}\rangle_{t})\left\vert\text{V}\right\rangle_{c}].
\end{align}
Here, $|$particle$\rangle$ and $|$wave$\rangle$ are defined similar
to (\ref{eq particle}) and (\ref{eq wave}). This allowed for a
continuous transition between wave and particle properties,
verifying the predicted intensity pattern of Eq.\ (\ref{eq
intensity}).
\begin{figure}[tb]
    \includegraphics[width=0.40\textwidth]{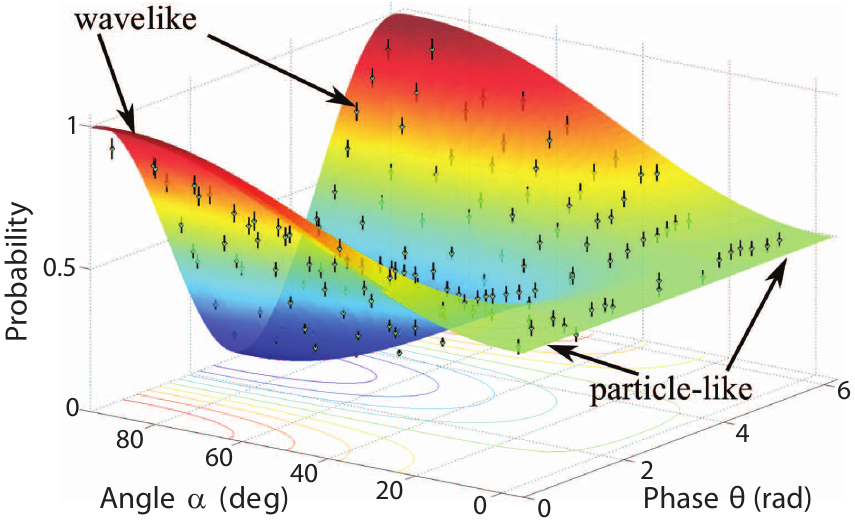}
    \caption{(Color online). Experimental results reported in
Ref.~\cite{Kaiser2012}; figure taken therefrom. When the
corroborative photon was found to be horizontally polarized, the
test photons produced an intensity pattern following expression
(\ref{eq intensity}), with $\theta \equiv \varphi$. The same pattern
emerged when the corroborative photon was measured in the
$\ket{+}$/$\ket{-}$ basis, verifying the
entanglement.}\label{kaiser2}
\end{figure}

\subsection{Quantum delayed-choice}

Quantum delayed choice shares a few features with quantum erasure.
An experiment following the proposal described in chapter II.G has
been realized using \textit{single} photons in an interferometer
\cite{Tang2012}. This was achieved by taking the polarization state
of the photon itself as the ancilla. Only the horizontally polarized
photons $\left\vert \text{H}\right\rangle $ passed through a second
beam splitter, while for vertical polarization $\left\vert \text{V}%
\right\rangle $ the interferometer was open. Similar to Eq.\
(\ref{eq particle-wave}), with initial polarization state $\sin
\alpha\left\vert \text{V}\right\rangle +\cos\alpha\left\vert \text{H}%
\right\rangle $ the total one-photon state was transformed into:%
\begin{equation}
\left\vert \psi\right\rangle
=\sin\alpha\,|\text{particle}\rangle\left\vert \text{V}\right\rangle
+\cos\alpha\,|\text{wave}\rangle\left\vert
\text{H}\right\rangle ,\label{eq p-w}%
\end{equation}
where $|$particle$\rangle=(\left\vert 0\right\rangle
+\,$e$^{\text{i}\varphi
}\left\vert 1\right\rangle )/\sqrt{2}$ and $|$wave$\rangle=\;$e$^{\text{i}%
\varphi/2}\times(\cos\tfrac{\varphi}{2}\left\vert 0\right\rangle -\,$%
i$\,\sin\tfrac{\varphi}{2}\left\vert 1\right\rangle )$, similar to
(\ref{eq particle}) and (\ref{eq wave}), and $\left\vert
0\right\rangle $ and $\left\vert 1\right\rangle $ are the path
states in the interferometer. (Note the different conventions for
the ancilla bias parameter $\alpha$ in states (\ref{eq particle-wave}) and (\ref{eq
p-w}).)\begin{figure}[tb]
\begin{center}
\includegraphics[width=0.48\textwidth]{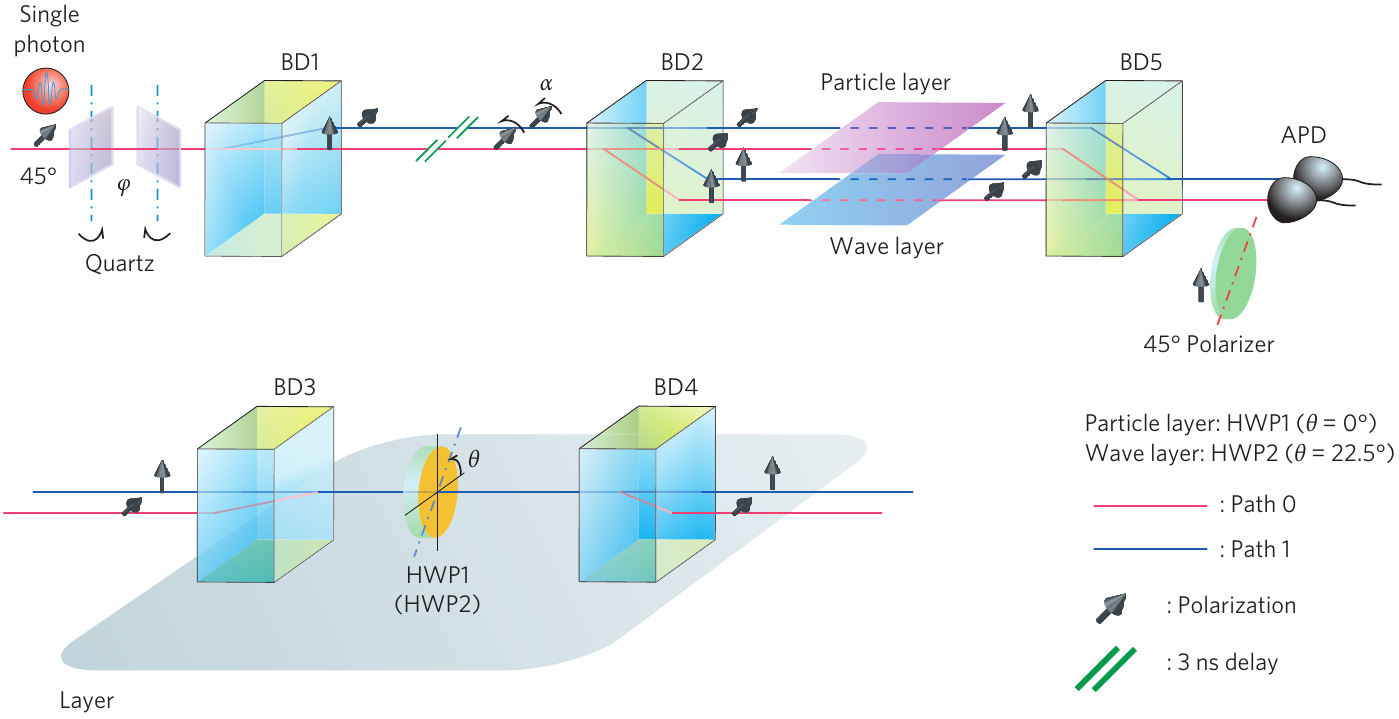}
\end{center}
\caption{(Color online). Experimental quantum delayed-choice with single photons.
Single photons entered the interferometer at beam displacer BD1
which split the light into horizontal and vertical polarization. The
phase $\varphi$ was scanned by the quartz plates before BD1. The
``second beam splitter" in the closed (open) interferometer was
provided by the combination of BD3, BD4 and half-wave plates HWP2
(HWP1). For HWP2 in the wave layer, the optical-axis direction
$\theta$ was set to 22.5${^{\circ}}$ and interference appears
(closed interferometer). For HWP1 in the particle layer $\theta$ was
set to 0${^{\circ}}$ (open interferometer), showing the particle
properties. Depending on the polarization (parameter $\alpha$), BD2
controlled whether the photons passed through the particle or wave
layer. The two layers were combined by BD5. A 45${^{\circ}}$
polarizer could be inserted to post-select on the polarization state
$(\left\vert \text{H}\right\rangle +\left\vert \text{V}\right\rangle
)/\sqrt{2}$. Finally, two detectors counted the photons in
paths 0 and 1. Figure taken from Ref.~\cite{Tang2012}.}%
\label{fig_Tang}%
\end{figure}

\begin{figure}[tb]
\begin{center}
\includegraphics{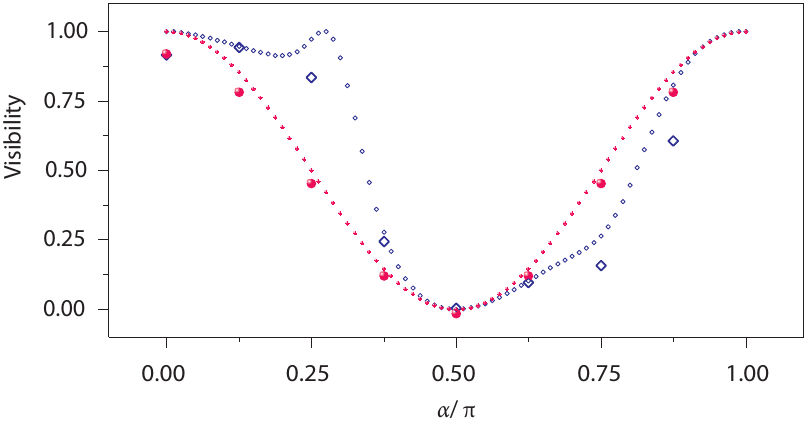}
\end{center}
\caption{(Color online). Visibility as a function of $\alpha$ for the mixed state
(\ref{eq mix}) in red (filled circles) and for the superposition state (\ref{eq sup}) in blue (unfilled diamonds). The red curve has the form $\cos^{2}\alpha$, while the blue one also reflects interference between wave and particle properties. The larger symbols are experimental data, while the smaller symbols are theoretical simulation results. Figure taken from Ref.~\cite{Tang2012}.}%
\label{fig_Tang2}%
\end{figure}

The experimental setup is shown and explained in Fig.\
\ref{fig_Tang}. If the final path measurement was not sensitive to
the polarization (i.e.\ no polarizer at the end), the detection
results were described by a mixed state
(density matrix) of the form%
\begin{equation}
\sin^{2}\alpha\,|\text{particle}\rangle\langle\text{particle}|+\cos^{2}%
\alpha\,|\text{wave}\rangle\langle\text{wave}|.\label{eq mix}%
\end{equation}
This corresponded to ignoring the ancilla outcome in chapter II.G and
lead to a visibility pattern of the form $\cos^{2}\alpha$. If,
however, the photon was post-selected in the polarization state
$(\left\vert \text{H}\right\rangle +\left\vert \text{V}\right\rangle
)/\sqrt{2}$ (i.e.\ polarizer at 45${^{\circ}}$), its path state was
left in the ``wave-particle superposition"
\begin{equation}
\sin\alpha\,|\text{particle}\rangle+\cos\alpha\,|\text{wave}\rangle
.\label{eq sup}%
\end{equation}
The experimental results for these two states were very different.
Fig.\ \ref{fig_Tang2} shows the visibility as a function of $\alpha$
for state (\ref{eq mix}) in red and for state (\ref{eq sup}) in
blue. The red curve follows the expected form $\cos^{2}\alpha$, as
only the wave-part in Eq.\ (\ref{eq mix}) leads to fringes. The blue
curve is more complicated and reflects the fact that there was also
quantum interference between the wave and particle
properties.

\begin{figure}[tb]
\begin{center}
\includegraphics[width=0.48\textwidth]{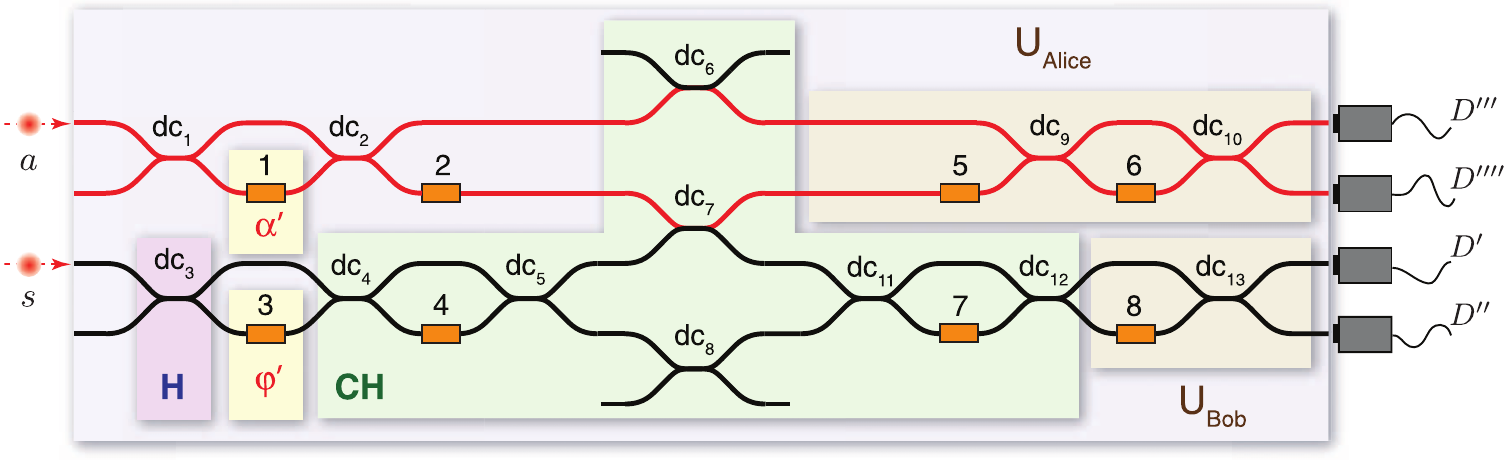}
\end{center}
\caption{(Color online). Two-photon experimental quantum delayed-choice experiment.
Non-entangled photon pairs were injected into an integrated photonic
device. The system photon ($s$, black optical path) passed a Hadamard gate (H) and a phase shifter $\varphi$. The ``second beam splitter" was a
controlled-Hadamard gate (CH), implemented with additional
Mach-Zehnder interferometers. The ancilla photon ($a$, red optical path) passed a phase shifter $\alpha$, allowing a superposition of present and
absent beam splitter for the system photon. For the Bell test,
single qubit rotations ($U_{\text{Alice}}$ and $U_{\text{Bob}}$)
were performed before the photon detectors. Directional couplers are
abbreviated by `dc', and resistive heaters are shown by orange
rectangles. Figure taken from Ref.~\cite{Peru2012}.}%
\label{fig_Peruzzo}%
\end{figure}

Also a two-photon experiment was performed realizing the proposal of
Ref.~\cite{Ioni2011} has been performed \cite{Peru2012}. The setup,
which used an integrated photonic device, is explained in Fig.\
\ref{fig_Peruzzo}.

\begin{figure}[tb]
\begin{center}
\includegraphics{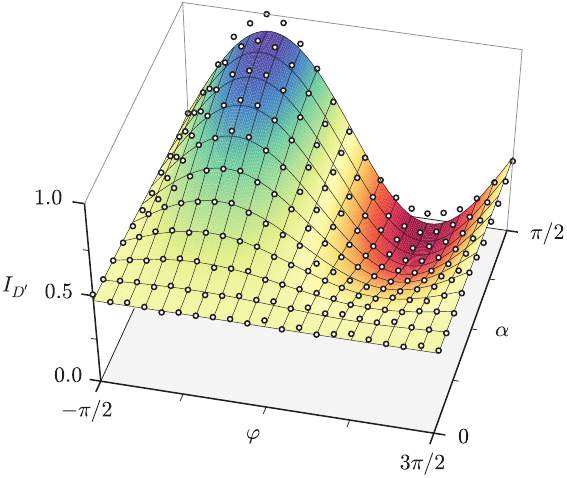}
\end{center}
\caption{(Color online). Continuous transition between wave and particle behavior.
The experimental data are shown by white dots and were
fitted (colored surface) based on Eq.\ (\ref{eq
intensity}). Figure taken from Ref.~\cite{Peru2012}.}%
\label{fig_Peruzzo2}%
\end{figure}

The measured intensity at detector $D^{\prime}$ was in excellent
agreement with the theoretical prediction given by Eq.\ (\ref{eq
intensity}), as shown in Fig.\ \ref{fig_Peruzzo2}. Since the ancilla
photon was finally measured in its computational basis, the system
photon data could be explained by a classical model in which the
ancilla photon was prepared in a mixture of the form
$\cos^{2}\alpha\left\vert 0\right\rangle \! \left\langle
0\right\vert +\sin^{2}\alpha\left\vert 1\right\rangle \!
\left\langle 1\right\vert.$ The particular state in every run would
be known to the system photons beforehand, deciding whether their
particle or wave behavior is measured.

To ensure that the choice cannot have been a classical variable
known in advance, the entanglement of state (\ref{eq particle-wave})
needed to be shown. This was done using unitary transformations at
the final stage of the setup (Fig.\ \ref{fig_Peruzzo}) and
performing a test of the Clauser-Horne-Shimony-Holt \cite{Clau1969}
inequality. Maximal entanglement of the state (\ref{eq
particle-wave}) is reached for $\alpha=\frac{\pi}{4}$ (ancilla
initially in equal weight superposition) and $\varphi=\frac{\pi}{2}$
(for which $\left\langle \text{particle}\right.  \!\left\vert \text{wave}%
\right\rangle =0$). For this parameter choice, a Bell value of
$S=2.45\pm0.03$ was reported, a significant violation of the local
realistic bound 2 \cite{Peru2012}. However, the authors acknowledged
correctly that the claim to have ruled out a classical description
of the wave-particle duality without further assumptions would
require a loophole-free Bell test, which has been demonstrated recently by three groups (Hensen2015; Giustina2015; Shalm2015).

Two other successful realizations of the quantum delayed-choice
scenario were achieved in nuclear magnetic resonance (NMR)
experiments with $^{13}$CHCl$_3$ molecules. In Ref.~\cite{Roy2012}
the system qubits (i.e.\ path in the interferometer) were encoded in
the hydrogen nuclear spins, while the ancilla qubits (control of the
interferometer) were encoded in carbon nuclear spins. In
Ref.~\cite{Auccaise2012} it was exactly the opposite. Both
experiments showed excellent agreement with the quantum predictions.

\subsection{Delayed-choice quantum random walk}

\begin{figure}[tb]
    \includegraphics[width=0.48\textwidth]{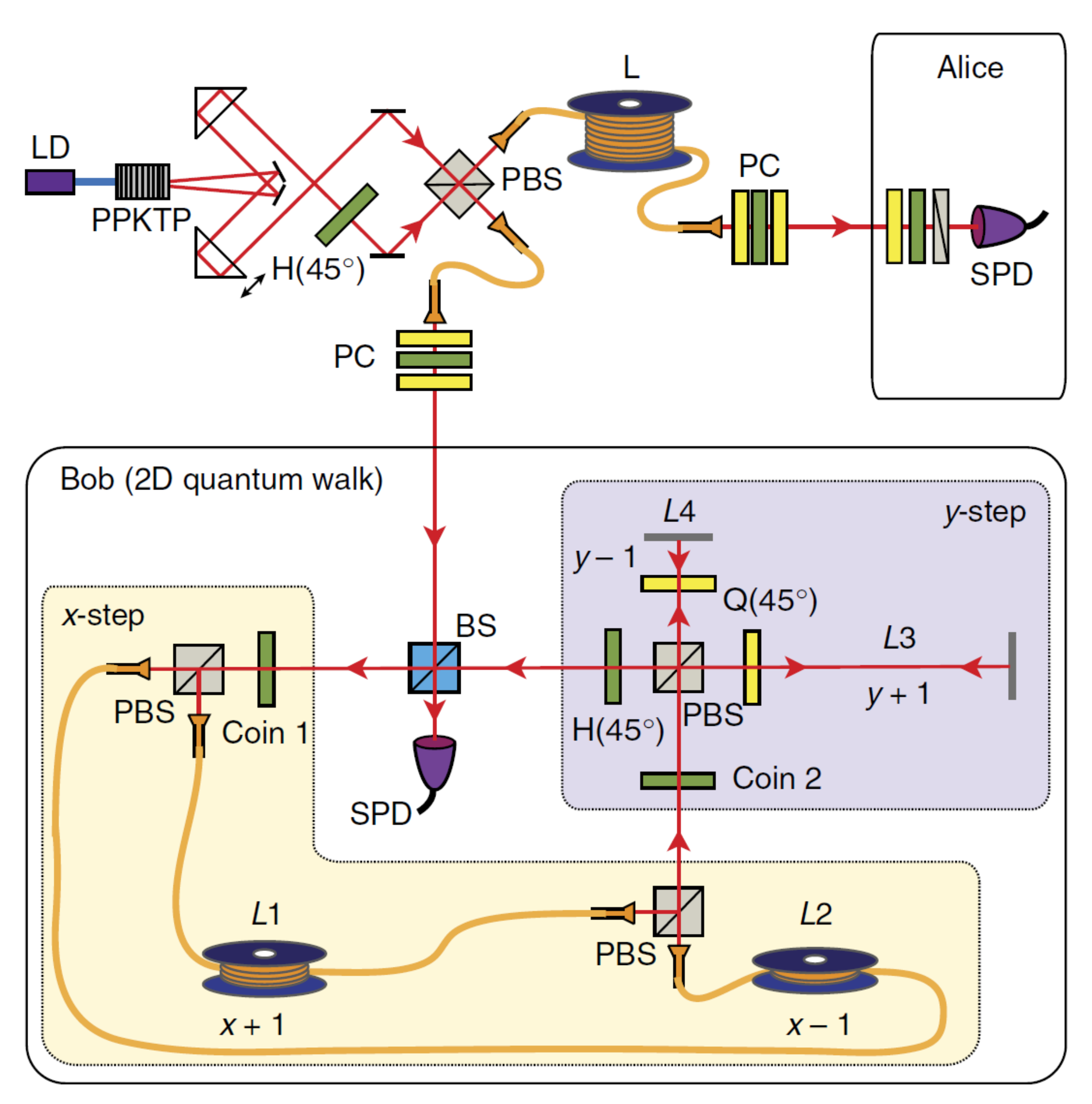}
    \caption{(Color online). Experimental scheme of the delayed-choice quantum walk reported in
Ref.~\cite{Jeong2013}; figure taken therefrom. Entangled photon
pairs were generated in a PPKTP crystal. One photon of every pair
delayed in a 340~m optical fiber and then sent to Alice, who was
able to perform polarization measurements in any basis. This
constituted a delayed-choice projection of the initial coin
(polarization) state of the other photon, which was already sent to
Bob without any fiber delay. In an optical loop, Bob's photons
performed a 2D ($x$ and $y$ steps) quantum walk in the time domain.
Before each step operation is taken, the coin operation (`Coin 1'
and `Coin 2' are Hadamard gates) was applied. In order to map the 2D
quantum walk lattice uniquely onto the photon arrival times, the
lengths of the optical fibers (L1-L4) were chosen
appropriately.}\label{quantumwalk}
\end{figure}

An experimental realization of a delayed-choice two-dimensional (2D)
quantum walk has been reported in~\cite{Jeong2013}. There, the
standard single-photon interferometer was replaced by a 2D quantum
walk lattice, which was mapped to a temporal grid for the arrival
times of a single photon by using polarization optical elements and
fibers. In a quantum walk, a coin and a shift operator are applied
repeatedly. The experimental scheme is shown in
Fig.~\ref{quantumwalk}.

The essence of the experiment is similar to the quantum eraser
concept. The way in which a photon interfered in the 2D quantum walk
circuitry depended on its polarization, which was determined by the
(delayed) polarization measurement of its distant twin. This was the
first experiment realizing a 2D quantum walk with a single photon
source and in a delayed-choice fashion. Additionally, the authors
also showed the first experimental simulation of a Grover walk, a
model that can be used to implement the Grover quantum search
algorithm~\cite{Grover1997}. The similarities between the
theoretical and experimental probability distributions in the Grover
walk were above 0.95.

\section{Realizations of delayed-choice entanglement-swapping experiments}

Entanglement swapping~\cite{Zukowski1993} is a generalization of
quantum teleportation~\cite{Bennett1993} and can teleport entangled
states. It is of crucial importance in quantum information
processing because it is one of the basic building blocks of quantum
repeaters \cite{Briegel1998,Duan2001}, third-man quantum
cryptography~\cite{Chen2005} and other protocols. On the other hand,
entanglement swapping also allows experiments on the foundations of
quantum physics, including loophole-free Bell tests~\cite{Simon2003}
and other fundamental tests of quantum
mechanics~\cite{Greenberger2008, Greenberger2008a}. The entanglement
swapping protocol itself has been experimentally demonstrated with
various physical
systems~\cite{Pan1998,Riebe2004,Barrett2004,Halder2007,Matsukevich2008,Yuan2008,Kaltenbaek2009}.

In the light of finding which kind of physical interactions and
processes are needed for the production of quantum entanglement,
Peres has put forward the radical idea of delayed-choice
entanglement swapping~\cite{Peres2000}. Realizations of this
proposal are discussed in the following.

\subsection{Delayed entanglement swapping}

\begin{figure}[t]
    \includegraphics[width=0.48\textwidth]{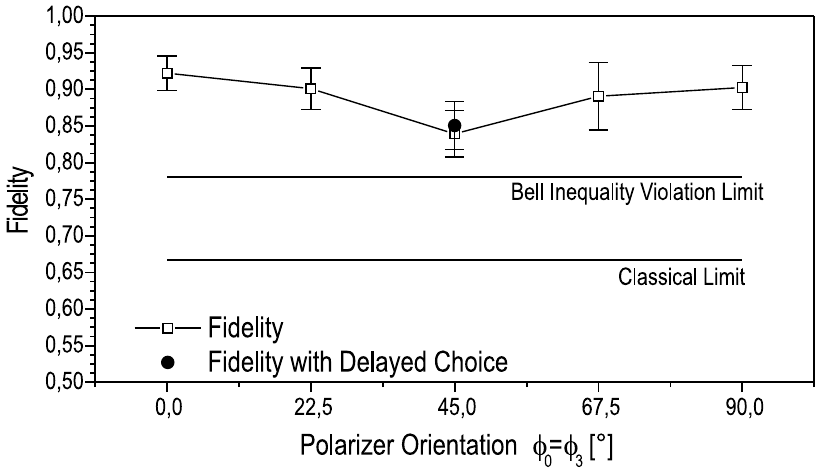}
    \caption{Experimental results of delayed entanglement swapping reported
in Ref.~\cite{Jennewein2001}; figure taken therefrom. Data points show the entanglement fidelity obtained through correlation measurements between photons 1 and 4. Data shown with white open squares (a black filled circle) was obtained when Victor's Bell state measurement was space-like separated from (in the time-like future of) Alice's and Bob's measurements. The angles $\phi_0$/$\phi_3$ are the setting of the polarization analyzer for photons 1/4 (Fig.~\ref{MaSwapping1}), which were aligned to be equal. The minimum fidelity is above the limit achievable with classical swapping protocols as well as above the limit necessary for violating a Bell inequality with the swapped entangled state.}\label{Jennewein}
\end{figure}
In~\cite{Jennewein2001}, a delayed entanglement swapping experiment
was performed. For the conceptual setup see Fig.~\ref{MaSwapping1}.
Detection of photons 2 and 3 by Victor was delayed by two 10~m
(about 50~ns) optical fiber delays after the outputs of the
Bell-state analyzer. Alice's and Bob's detectors were located next
to each other. The traveling time of photons 1 and 4 from the source
to these detectors was about 20~ns. Victor was separated from Alice
and Bob by about 2.5 m, corresponding to luminal traveling time of
approximately 8~ns between them. Therefore, Victor's measurements
were in the time-like future of Alice's and Bob's measurements. The observed
fidelity of the measured state $\rho_{14}$ of photons 1 and 4 with
the ideal singlet state, defined as $_{14}\langle \Psi^- | \rho_{14}
| \Psi^- \rangle_{14}$ was around 0.84, both above the classical
limit of 2/3 and the limit of approximately 0.78 necessary to violate Bell's
inequality, as shown in Fig.~\ref{Jennewein}. This was the first attempt of the realization of delayed-choice entanglement swapping, although a switchable Bell-state analyzer has not been implemented.

We note that in Ref.~\cite{Sciarrino2002} a successful experiment on
delayed entanglement swapping was performed with two singlet
entangled states comprised by the vacuum and the one-photon states.
This allowed to use a pair of entangled photons rather than four
photons. The obtained correlation visibility was $(91 \pm 2)\%$.

\subsection{Delayed-choice entanglement swapping}

\begin{figure}[t]
    \includegraphics[width=0.35\textwidth]{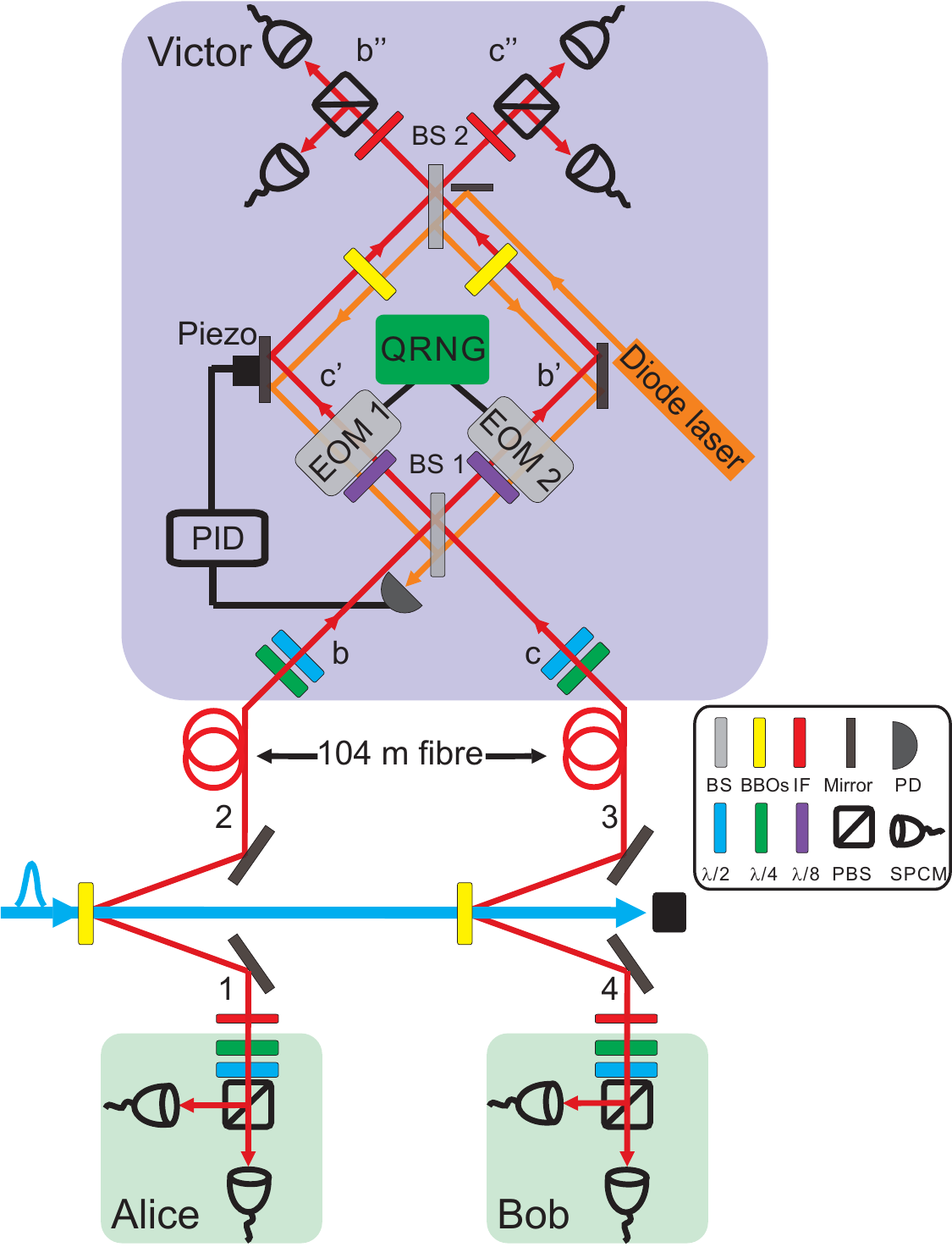}
    \caption{(Color online). Experimental setup of delayed-choice entanglement swapping reported
in Ref.~\cite{Ma2012}; figure taken therefrom. Two
polarization-entangled photon pairs (photons 1\&2 and photons 3\&4)
were generated from BBO crystals. Alice and Bob measured the
polarization of photons 1 and 4 in whatever basis they chose.
Photons 2 and 3 were each delayed with 104 m fiber and then
overlapped on the tunable bipartite state
analyzer (BiSA) (purple block). The BiSA either performed a Bell-state
measurement (BSM) or a separable-state measurement (SSM),
depending on the outcome of a QRNG. An
active phase stabilization system was employed in order to compensate
the phase noise in the tunable BiSA.}\label{MaSwapping2}
\end{figure}
\begin{figure}[t]
    \includegraphics[width=0.48\textwidth]{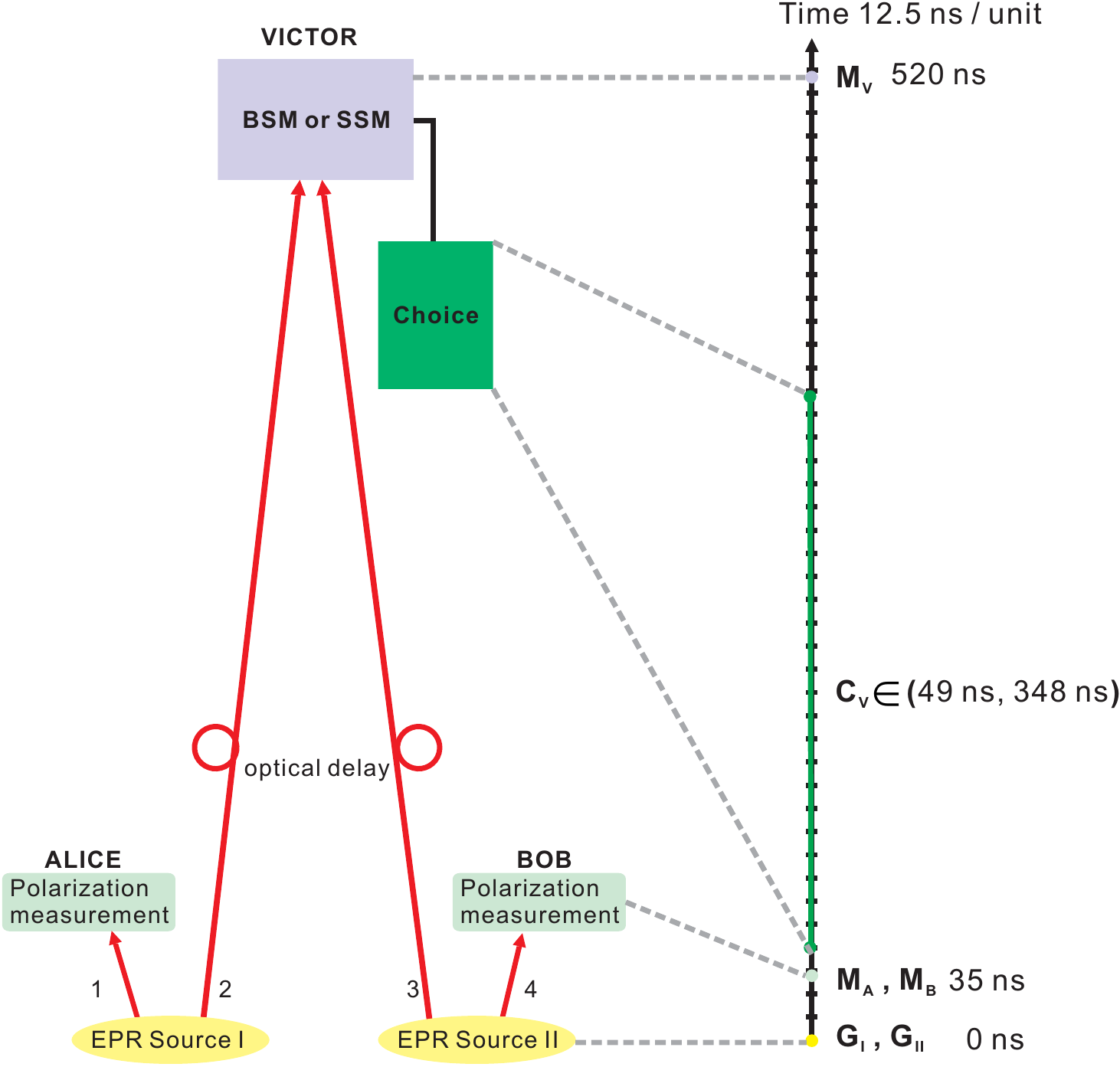}
    \caption{(Color online). Time diagram of the delayed-choice entanglement swapping experiment
reported in Ref.~\cite{Ma2012}; figure taken therefrom. Two
entangled photon pairs (1\&2 and 3\&4) were generated by EPR
sources I and II (events \textbf{G}$_\textrm{I}$ and
\textbf{G}$_\textrm{II}$) at 0 ns. Alice and Bob measured the
polarization of photons 1 and 4 at 35~ns (events
\textbf{M}$_\textrm{A}$ and \textbf{M}$_\textrm{B}$). Photons 2 and
3 were delayed and sent to Victor who chose (event
\textbf{C}$_\textrm{V}$) to perform a Bell-state measurement (BSM)
or a separable-state measurement (SSM) (event
\textbf{M}$_\textrm{V}$). Victor's choice and measurement were made
after Alice's and Bob's polarization
measurements.}\label{MaSwapping3}
\end{figure}
\begin{figure}[tb]
    \includegraphics[width=0.48\textwidth]{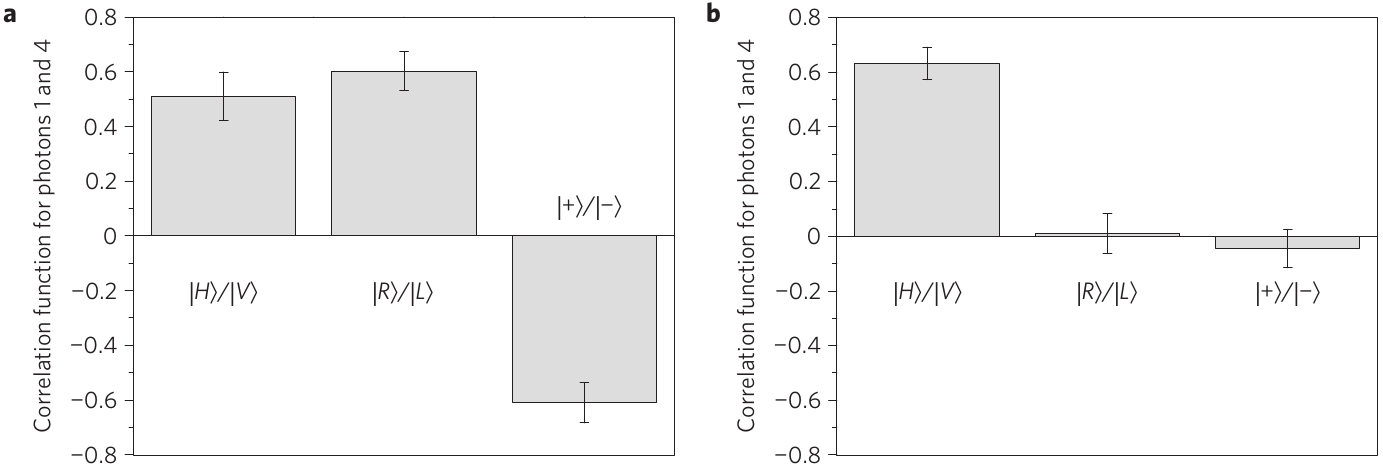}
    \caption{Correlation functions from the experiment~\cite{Ma2012}; figure
taken therefrom. \textbf{a}:\ Victor subjected photons 2 and 3 to a
Bell-state measurement and observed the result $\ket{\Phi^-}_{23}$.
Alice's and Bob's photons 1 and 4 were projected into the
corresponding entangled state $\ket{\Phi^-_{14}}$, showing
correlations in all three mutually unbiased bases
$\ket{\textrm{H}}$/$\ket{\textrm{V}}$,
$\ket{\textrm{R}}$/$\ket{\textrm{L}}$, and $\ket{+}$/$\ket{-}$.
Entanglement between photons 1 and 4 is witnessed by the absolute
sum of the correlation values exceeding 1. \textbf{b}:\ When Victor
performed a separable-state measurement in the
$\ket{\textrm{H}}$/$\ket{\textrm{V}}$ basis, also photons 1 and 4
ended up in the corresponding separable state and hence showed
correlations only in that basis but not the other two.}\label{Ma4}
\end{figure}
A refined and conclusive realization of Peres' gedanken experiment
was reported~\cite{Ma2012}. The layout of this experiment is
illustrated in Fig.\ \ref{MaSwapping2}. The essential point was the
implementation of bipartite state projections based on the random
and delayed choice. The choice was to either perform a Bell-state
measurement (BSM) or a separable-state measurement (SSM) on photons
2 and 3. In order to realize this, a bipartite state
analyzer (BiSA) with two-photon interference on a
high-speed tunable beam splitter combined with photon detections was
used.

The initial four-photon entangled state was of the form
(\ref{psi4}). Alice and Bob measured the polarization of photons 1
and 4 without any delay. Photons 2 and 3 were sent through 104~m
single-mode fibers, corresponding to a delay time of 520~ns. Victor
actively chose and implemented the measurements on photons 2 and 3
(either BSM or SSM) by using a high-speed tunable bipartite state
analyzer (BiSA). A quantum random number generator (QRNG) was used
to make the random choice. Both the choice and the measurement of photons 2
and 3 were in the time-like future of the registration of photons 1
and 4. This projected the state of the two already registered photons, 1 and 4,
onto either an entangled or a separable state.

The diagram of the temporal order of the relevant events is shown in
Fig.\ \ref{MaSwapping3}. For each successful run (a 4-fold
coincidence count), both Victor's measurement event and his choice
were in the time-like future of Alice's and Bob's
measurements.

In that experiment, the existence of entanglement was verified by measuring the
state fidelities and the expectation values of entanglement witness
operators~\cite{Gueh2009}. It was found that whether photons 1 and 4
were entangled or separable only depended on the type of the
measurements Victor implement, not on the temporal order
(Fig.~\ref{Ma4}).

\section{Conclusion and outlook}

Delayed-choice gedanken experiments and their realizations play an
important role in the foundations of quantum physics, because they
serve as striking illustrations of the counter-intuitive and
inherently non-classical features of quantum mechanics. A summary of
the photonic delayed-choice experiments discussed in this review is
presented in Table~\ref{tab_comp}.\begin{table*}[t!]
  \centering
  \caption{A summary of delayed-choice experiments realized with
photons. `\textbf{C} and \textbf{I}', and `\textbf{M} and
\textbf{I}' stand for the space-time relations between events
\textbf{C} (choice), \textbf{I} (entry into interferometer), and
\textbf{M} (measurement of the photon). Note that in the experiments involving more than one photon, \textbf{M} stands for the measurement of ancillary photon(s). Other abbreviations: `sep.'
stands for `space-like separated', `after' and `before' stand for
`time-like after' and `time-like before', `ext.' and `int.' stand
for `external' and `internal', `QRNG' stands for `quantum random number generator', `BS' for `beam splitter'. For example, the entry `before' in the `\textbf{C} and \textbf{I}' column means that \textbf{C} happens time-like before \textbf{I}.}

    \begin{tabular}{lcrrr}
    \hline
    Experiment / Ref. & Number of photons & Nature of the Choice & \textbf{C} and \textbf{I} & \textbf{M} and \textbf{I} \\
    \hline
    \cite{Alley1983} & 1   & ext.\ choice, photon detection & sep. & after \\
    \cite{Hellmuth1987} & 1   & fixed setting & before & after \\
    \cite{Baldzuhn1989} & 1   & fixed setting & before & after \\
    \cite{Jacques2007, Jacques2008} & 2   & ext. choice, shot noise & sep.\ & after \\
    \cite{Dopfer1998} & 2   & fixed setting & before & after \\
    \cite{Walborn2002} & 2   & fixed setting & before & after \\
    \cite{Kim2000} & 2   & int.\ choice, 50/50 BS & after & after \\
    \cite{Ma2013a} & 2   & ext.\ choice, QRNG with 50/50 BS  & sep.\ \& after & sep.\ \& after \\
    \cite{Tang2012} & 1   & quantum delayed choice, fixed setting & before & after \\
    \cite{Kaiser2012} & 2   & quantum delayed choice, fixed setting & before & sep. \\
    \cite{Peru2012} & 2   & quantum delayed choice, fixed setting & before & after \\
    \cite{Jeong2013} & 1   & fixed setting & before & after \\
    \cite{Jennewein2001} & 4  & fixed setting & before & after \\
    \cite{Sciarrino2002} & 2   & fixed setting & before & n/a \\
    \cite{Ma2012} & 4  & ext.\ choice, QRNG with 50/50 BS  & after & after \\
    \hline
    \end{tabular}%
  \label{tab_comp}%
\end{table*}%

Wheeler's delayed-choice experiments challenge a realistic
explanation of the wave-particle duality. In such an explanation
every photon is assumed to behave either definitely as a wave
(traveling both paths in an interferometer) or definitely as a
particle (traveling only one of the paths), by adapting a priori on
the experimental situation. Especially when the choice of whether or
not to insert the second beam splitter into an interferometer is
made space-like separated from the photon's entry into the
interferometer, this picture becomes untenable.

In delayed-choice experiments with two entangled quantum systems
such as the delayed-choice quantum eraser, one can choose that one
system exhibits wave or particle behavior by choosing different
measurements for the other one. These choices and measurements can
be made even after the former system has already been detected.

In delayed-choice entanglement swapping experiments, one can
demonstrate that whether two quantum systems are entangled or
separable can be decided even after they have been measured. This
generalizes the wave-particle duality for single systems to an
entanglement-separability duality for two (and more) systems.

It is a general feature of delayed-choice experiments that quantum
effects can mimic an influence of future actions on past events.
However, there never emerges any paradox if the quantum state is
viewed only as `catalogue of our knowledge'~\cite{Schroedinger1935}
without any underlying hidden variable description. Then the state
is a probability list for all possible measurement outcomes and not
a real physical object. The relative temporal order of measurement
events is not relevant, and no physical interactions or signals, let
alone into the past, are necessary to explain the experimental
results. To interpret quantum experiments, any attempt in explaining
what happens in an individual observation of one system has to
include the whole experimental configuration and also the complete
quantum state, potentially describing joint properties with other
systems. According to Bohr and Wheeler, no elementary phenomenon is
a phenomenon until it is a registered
phenomenon~\cite{Bohr1949,Wheeler1984}. In light of quantum erasure
and entanglement swapping, one might like to even say that some
registered phenomena do not have a meaning unless they are put in
relationship with other registered phenomena~\cite{Ma2012}.

Delayed-choice gedanken experiments and their realizations have
played important roles in the development of quantum physics. The
applicability of the delayed-choice paradigm for practical quantum
information processing is yet to be explored. For example, the
authors in~\cite{Lee2014} introduced and experimentally demonstrated
a delayed-choice decoherence suppression protocol. In their
experiment, the decision to suppress decoherence on an entangled
two-qubit state is delayed until after the decoherence and even
after the detection of the qubit. This result suggests a new way to
tackle Markovian decoherence in a delayed-choice way, which could be
useful for practical entanglement distribution over a dissipative
channel.

The concept of delayed-choice entanglement swapping is of importance
for the security of quantum communication schemes such as third-man
quantum cryptography~\cite{Chen2005} and could also be employed in
probabilistic instantaneous quantum computing. In the latter case,
quantum state teleportation and entanglement swapping imply a
computational speed-up in time over classical
procedures~\cite{Brukner2003, Jennewein2002a}. This can be realized
by sending one photon of a Bell state into the input of a quantum
computer and performing a quantum computation with it. Since this
photon is a part of a Bell state, its individual property is not
well defined. Therefore, also the output of the quantum computation
will not be defined. However, as soon as the required input is
known, it can be teleported onto the state of the photon which had
been fed into the quantum computer. If the Bell-state measurement
(BSM) results in one specific Bell state which requires no
corrective unitary transformation, then immediately the output of
the quantum computer will be projected into the correct result. By
this means the computation is performed quasi instantaneously. Note
that this instantaneous quantum computation is intrinsically
probabilistic because the BSM results in all four Bell states with
equal probability of 1/4.

Finally, we observe that the development of quantum mechanics has
been accompanied initially by a series of ingenious gedanken
experiments, which have -- with the advance of technology -- found
more and more realizations over time. This again has opened up
avenues for new experiments and even applications. Likewise, while
the history of the delayed-choice paradigm dates back to the early
days of quantum mechanics, only in the past decades have many
remarkable experiments demonstrated its counter-intuitive aspects in
different scenarios and with different physical systems. It can be
expected that delayed-choice gedanken experiments will continue to
lead to novel foundational tests as well as further practical
implementations.

\section*{Acknowledgments}
We thank Guido Bacciagaluppi, Daniel Greenberger, Radu Ionicioiu, Thomas Jennewein, Florian Kaiser, Marlan
Scully, S\'{e}bastien Tanzilli, Daniel R. Terno, and Marek Zukowski
for helpful comments and remarks. X.S.M. is supported by a Marie
Curie International Outgoing Fellowship within the $7^{th}$ European
Community Framework Programme. A.Z. acknowledges support by the
Austrian Science Fund through the SFB program and by the European
Commission.

\bibliographystyle{apsrmp}

\begin{thebibliography}{88}
\expandafter\ifx\csname natexlab\endcsname\relax\def\natexlab#1{#1}\fi
\expandafter\ifx\csname bibnamefont\endcsname\relax
  \def\bibnamefont#1{#1}\fi
\expandafter\ifx\csname bibfnamefont\endcsname\relax
  \def\bibfnamefont#1{#1}\fi
\expandafter\ifx\csname citenamefont\endcsname\relax
  \def\citenamefont#1{#1}\fi
\expandafter\ifx\csname url\endcsname\relax
  \def\url#1{\texttt{#1}}\fi
\expandafter\ifx\csname urlprefix\endcsname\relax\def\urlprefix{URL }\fi
\providecommand{\bibinfo}[2]{#2}
\providecommand{\eprint}[2][]{\url{#2}}

\bibitem[{\citenamefont{Aharonov and Zubairy}(2005)}]{Aharonov2005}
\bibinfo{author}{\bibnamefont{Aharonov}, \bibfnamefont{Y.}} and
  \bibinfo{author}{\bibfnamefont{M.~S.} \bibnamefont{Zubairy}},
  \bibinfo{year}{2005}, \bibinfo{journal}{Science}
  \textbf{\bibinfo{volume}{307}}, \bibinfo{pages}{875}.

\bibitem[{\citenamefont{Alley} \emph{et~al.}(1983)\citenamefont{Alley,
  Jakubowicz, Steggerda, and Wickes}}]{Alley1983}
\bibinfo{author}{\bibnamefont{Alley}, \bibfnamefont{C.~O.}},
  \bibinfo{author}{\bibfnamefont{O.}~\bibnamefont{Jakubowicz}},
  \bibinfo{author}{\bibfnamefont{C.~A.} \bibnamefont{Steggerda}}, and
  \bibinfo{author}{\bibfnamefont{W.~C.} \bibnamefont{Wickes}},
  \bibinfo{year}{1983}, in \bibinfo{booktitle}{\emph{International Symposium on
  foundations of Quantum Mechanics in the light of New Technology} \textrm{(Physical Society of Japan, Tokyo)}.}

\bibitem[{\citenamefont{Alley} \emph{et~al.}(1986)\citenamefont{Alley,
  Jakubowicz, and Wickes}}]{Alley1986}
\bibinfo{author}{\bibnamefont{Alley}, \bibfnamefont{C.~O.}},
  \bibinfo{author}{\bibfnamefont{O.}~\bibnamefont{Jakubowicz}}, and
  \bibinfo{author}{\bibfnamefont{W.~C.} \bibnamefont{Wickes}},
  \bibinfo{year}{1986}, in \bibinfo{booktitle}{\emph{2nd International
  Symposium on foundations of Quantum Mechanics in the light of New Technology} \textrm{(Physical Society of Japan, Tokyo)}.}

\bibitem[{\citenamefont{Auccaise} \emph{et~al.}(2012)\citenamefont{Auccaise,
  Serra, Filgueiras, Sarthour, Oliveira, and C\'eleri}}]{Auccaise2012}
\bibinfo{author}{\bibnamefont{Auccaise}, \bibfnamefont{R.}},
  \bibinfo{author}{\bibfnamefont{R.~M.} \bibnamefont{Serra}},
  \bibinfo{author}{\bibfnamefont{J.~G.} \bibnamefont{Filgueiras}},
  \bibinfo{author}{\bibfnamefont{R.~S.} \bibnamefont{Sarthour}},
  \bibinfo{author}{\bibfnamefont{I.~S.} \bibnamefont{Oliveira}}, and
  \bibinfo{author}{\bibfnamefont{L.~C.} \bibnamefont{C\'eleri}},
  \bibinfo{year}{2012}, \bibinfo{journal}{Phys. Rev. A}
  \textbf{\bibinfo{volume}{85}}, \bibinfo{pages}{032121}.

\bibitem[{\citenamefont{Baldzuhn} \emph{et~al.}(1989)\citenamefont{Baldzuhn,
  Mohler, and Martienssen}}]{Baldzuhn1989}
\bibinfo{author}{\bibnamefont{Baldzuhn}, \bibfnamefont{J.}},
  \bibinfo{author}{\bibfnamefont{E.}~\bibnamefont{Mohler}}, and
  \bibinfo{author}{\bibfnamefont{W.}~\bibnamefont{Martienssen}},
  \bibinfo{year}{1989}, \bibinfo{journal}{Z. Phys. B: Cond. Mat.}
  \textbf{\bibinfo{volume}{77}}, \bibinfo{pages}{347}.

\bibitem[{\citenamefont{Ballentine}(1998)}]{Ballentine1998}
\bibinfo{author}{\bibnamefont{Ballentine}, \bibfnamefont{L.~E.}},
  \bibinfo{year}{1998}, \emph{\bibinfo{title}{Quantum Mechanics: A Modern
  Development}} (\bibinfo{publisher}{World Scientific Publishing Company}).

\bibitem[{\citenamefont{Barrett} \emph{et~al.}(2004)\citenamefont{Barrett,
  Chiaverini, Schaetz, Britton, Itano, Jost, Knill, Langer, Leibfried, Ozeri,
  and Wineland}}]{Barrett2004}
\bibinfo{author}{\bibnamefont{Barrett}, \bibfnamefont{M.~D.}},
  \bibinfo{author}{\bibfnamefont{J.}~\bibnamefont{Chiaverini}},
  \bibinfo{author}{\bibfnamefont{T.}~\bibnamefont{Schaetz}},
  \bibinfo{author}{\bibfnamefont{J.}~\bibnamefont{Britton}},
  \bibinfo{author}{\bibfnamefont{W.~M.} \bibnamefont{Itano}},
  \bibinfo{author}{\bibfnamefont{J.~D.} \bibnamefont{Jost}},
  \bibinfo{author}{\bibfnamefont{E.}~\bibnamefont{Knill}},
  \bibinfo{author}{\bibfnamefont{C.}~\bibnamefont{Langer}},
  \bibinfo{author}{\bibfnamefont{D.}~\bibnamefont{Leibfried}},
  \bibinfo{author}{\bibfnamefont{R.}~\bibnamefont{Ozeri}}, and
  \bibinfo{author}{\bibfnamefont{D.~J.} \bibnamefont{Wineland}},
  \bibinfo{year}{2004}, \bibinfo{journal}{Nature}
  \textbf{\bibinfo{volume}{429}}, \bibinfo{pages}{737}.

\bibitem[{\citenamefont{Bell}(1964)}]{Bell1964}
\bibinfo{author}{\bibnamefont{Bell}, \bibfnamefont{J. S.}},
  \bibinfo{year}{1964}, \bibinfo{journal}{Physics (NY)}
  \textbf{\bibinfo{volume}{1}}, \bibinfo{pages}{195}.

\bibitem[{\citenamefont{Bell}(2004)}]{Bell2004}
\bibinfo{author}{\bibnamefont{Bell}, \bibfnamefont{J. S.}},
  \bibinfo{year}{2004}, \emph{\bibinfo{title}{Speakable and Unspeakable in Quantum
Mechanics}}, Rev.\ Ed., pp 243-244 (Cambridge Univ. Press,
Cambridge, UK).

\bibitem[{\citenamefont{Bennett} \emph{et~al.}(1993)\citenamefont{Bennett,
  Brassard, Cr\'epeau, Jozsa, Peres, and Wootters}}]{Bennett1993}
\bibinfo{author}{\bibnamefont{Bennett}, \bibfnamefont{C.~H.}},
  \bibinfo{author}{\bibfnamefont{G.}~\bibnamefont{Brassard}},
  \bibinfo{author}{\bibfnamefont{C.}~\bibnamefont{Cr\'epeau}},
  \bibinfo{author}{\bibfnamefont{R.}~\bibnamefont{Jozsa}},
  \bibinfo{author}{\bibfnamefont{A.}~\bibnamefont{Peres}}, and
  \bibinfo{author}{\bibfnamefont{W.~K.} \bibnamefont{Wootters}},
  \bibinfo{year}{1993}, \bibinfo{journal}{Phys. Rev. Lett.}
  \textbf{\bibinfo{volume}{70}}, \bibinfo{pages}{1895}.

\bibitem[{\citenamefont{Bohr}(1928)}]{Bohr1928}
\bibinfo{author}{\bibnamefont{Bohr}, \bibfnamefont{N.}}, \bibinfo{year}{1928},
  \bibinfo{journal}{Nature} \textbf{\bibinfo{volume}{121}},
  \bibinfo{pages}{580}.

\bibitem[{\citenamefont{Bohr}(1935)}]{Bohr1935}
\bibinfo{author}{\bibnamefont{Bohr}, \bibfnamefont{N.}}, \bibinfo{year}{1935},
  \bibinfo{journal}{Phys. Rev.}
  \textbf{\bibinfo{volume}{48}}, \bibinfo{pages}{696}.

\bibitem[{\citenamefont{Bohr}(1949)}]{Bohr1949}
\bibinfo{author}{\bibnamefont{Bohr}, \bibfnamefont{N.}}, \bibinfo{year}{1949},
in \bibinfo{title}{\emph{Albert Einstein: Philosopher-Scientist},
\textrm{ed. P. A. Schlipp, The Library of Living Philosophers
  (\bibinfo{publisher}{MJF Books New York})}.}

\bibitem[{\citenamefont{Bramon} \emph{et~al.}(2004)\citenamefont{Bramon,
  Garbarino, and Hiesmayr}}]{Bramon2004}
\bibinfo{author}{\bibnamefont{Bramon}, \bibfnamefont{A.}},
  \bibinfo{author}{\bibfnamefont{G.}~\bibnamefont{Garbarino}}, and
  \bibinfo{author}{\bibfnamefont{B.~C.} \bibnamefont{Hiesmayr}},
  \bibinfo{year}{2004}, \bibinfo{journal}{Phys. Rev. Lett.}
  \textbf{\bibinfo{volume}{92}}, \bibinfo{pages}{020405}.

\bibitem[{\citenamefont{Branciard}(2013)}]{Bran2013}
\bibinfo{author}{\bibnamefont{Branciard}, \bibfnamefont{C.}},
  \bibinfo{year}{2013}, \bibinfo{journal}{Proc. Natl. Acad. Sci. USA}
  \textbf{\bibinfo{volume}{110}}, \bibinfo{pages}{6742}.

\bibitem[{\citenamefont{Briegel} \emph{et~al.}(1998)\citenamefont{Briegel,
  D\"ur, Cirac, and Zoller}}]{Briegel1998}
\bibinfo{author}{\bibnamefont{Briegel}, \bibfnamefont{H.-J.}},
  \bibinfo{author}{\bibfnamefont{W.}~\bibnamefont{D\"ur}},
  \bibinfo{author}{\bibfnamefont{J.~I.} \bibnamefont{Cirac}}, and
  \bibinfo{author}{\bibfnamefont{P.}~\bibnamefont{Zoller}},
  \bibinfo{year}{1998}, \bibinfo{journal}{Phys. Rev. Lett.}
  \textbf{\bibinfo{volume}{81}}, \bibinfo{pages}{5932}.

\bibitem[{\citenamefont{de~Broglie}(1924)}]{deBr1924}
\bibinfo{author}{\bibnamefont{de~Broglie}, \bibfnamefont{L.}},
  \bibinfo{year}{1924}, \emph{\bibinfo{title}{Recherches sur la th\'{e}orie des
  quanta}}, Ph.D. Thesis, \bibinfo{school}{Paris}.

\bibitem[{\citenamefont{Brukner} \emph{et~al.}(2003)\citenamefont{Brukner, Pan,
  Simon, Weihs, and Zeilinger}}]{Brukner2003}
\bibinfo{author}{\bibnamefont{Brukner}, \bibfnamefont{{\v C}.}},
  \bibinfo{author}{\bibfnamefont{J.-W.} \bibnamefont{Pan}},
  \bibinfo{author}{\bibfnamefont{C.}~\bibnamefont{Simon}},
  \bibinfo{author}{\bibfnamefont{G.}~\bibnamefont{Weihs}}, and
  \bibinfo{author}{\bibfnamefont{A.}~\bibnamefont{Zeilinger}},
  \bibinfo{year}{2003}, \bibinfo{journal}{Phys. Rev. A}
  \textbf{\bibinfo{volume}{67}}, \bibinfo{pages}{034304}.

\bibitem[{\citenamefont{Brunner} \emph{et~al.}(2013)\citenamefont{Brunner,
  Cavalcanti, Pironio, Scarani, and S.}}]{Brun2013}
\bibinfo{author}{\bibnamefont{Brunner}, \bibfnamefont{N.}},
  \bibinfo{author}{\bibfnamefont{D.}~\bibnamefont{Cavalcanti}},
  \bibinfo{author}{\bibfnamefont{S.}~\bibnamefont{Pironio}},
  \bibinfo{author}{\bibfnamefont{V.}~\bibnamefont{Scarani}}, and
  \bibinfo{author}{\bibfnamefont{S.}~\bibnamefont{Wehner}}, \bibinfo{year}{2013},
  \bibinfo{journal}{Rev. Mod. Phys.} \textbf{\bibinfo{volume}{86}}, \bibinfo{pages}{419}.

\bibitem[{\citenamefont{Busch} \emph{et~al.}(2007)\citenamefont{Busch,
  Heinonen, and Lahti}}]{Busch2007}
\bibinfo{author}{\bibnamefont{Busch}, \bibfnamefont{P.}},
  \bibinfo{author}{\bibfnamefont{T.}~\bibnamefont{Heinonen}}, and
  \bibinfo{author}{\bibfnamefont{P.} \bibnamefont{Lahti}},
  \bibinfo{year}{2007}, \bibinfo{journal}{Phys. Rep.}
  \textbf{\bibinfo{volume}{452}}, \bibinfo{pages}{155}.

\bibitem[{\citenamefont{Busch} \emph{et~al.}(2013)\citenamefont{Busch, Lahti,
  and Werner}}]{Busc2013}
\bibinfo{author}{\bibnamefont{Busch}, \bibfnamefont{P.}},
  \bibinfo{author}{\bibfnamefont{P.} \bibnamefont{Lahti}}, and
  \bibinfo{author}{\bibfnamefont{R.~F.} \bibnamefont{Werner}},
  \bibinfo{year}{2013}, \bibinfo{journal}{Phys. Rev. Lett.}
  \textbf{\bibinfo{volume}{111}}, \bibinfo{pages}{160405}.

\bibitem[{\citenamefont{C\'{e}leri} \emph{et~al.}(2014)}]{Cele2014}
\bibinfo{author}{\bibnamefont{C\'{e}leri}, \bibfnamefont{L.~C.}},
\bibinfo{author}{\bibnamefont{R.~M.} \bibfnamefont{Gomes}},
\bibinfo{author}{\bibnamefont{R.} \bibfnamefont{Ionicioiu}},
\bibinfo{author}{\bibnamefont{T.} \bibfnamefont{Jennewein}},
\bibinfo{author}{\bibnamefont{R.~B.} \bibfnamefont{Mann}}, and
  \bibinfo{author}{\bibfnamefont{D.~R.} \bibnamefont{Terno}},
  \bibinfo{year}{2014}, \bibinfo{journal}{Found. Phys.}
  \textbf{\bibinfo{volume}{44}}, \bibinfo{pages}{576}.

\bibitem[{\citenamefont{Chen} \emph{et~al.}(2005)\citenamefont{Chen, Zhang,
  Zhao, Zhou, Lu, Peng, Yang, and Pan}}]{Chen2005}
\bibinfo{author}{\bibnamefont{Chen}, \bibfnamefont{Y.-A.}},
  \bibinfo{author}{\bibfnamefont{A.-N.} \bibnamefont{Zhang}},
  \bibinfo{author}{\bibfnamefont{Z.}~\bibnamefont{Zhao}},
  \bibinfo{author}{\bibfnamefont{X.-Q.} \bibnamefont{Zhou}},
  \bibinfo{author}{\bibfnamefont{C.-Y.} \bibnamefont{Lu}},
  \bibinfo{author}{\bibfnamefont{C.-Z.} \bibnamefont{Peng}},
  \bibinfo{author}{\bibfnamefont{T.}~\bibnamefont{Yang}}, and
  \bibinfo{author}{\bibfnamefont{J.-W.} \bibnamefont{Pan}},
  \bibinfo{year}{2005}, \bibinfo{journal}{Phys. Rev. Lett.}
  \textbf{\bibinfo{volume}{95}}, \bibinfo{pages}{200502}.

\bibitem[{\citenamefont{Clauser} \emph{et~al.}(1969)\citenamefont{Clauser,
  Horne, Shimony, and Holt}}]{Clau1969}
\bibinfo{author}{\bibnamefont{Clauser}, \bibfnamefont{J.~F.}},
  \bibinfo{author}{\bibfnamefont{M.~A.} \bibnamefont{Horne}},
  \bibinfo{author}{\bibfnamefont{A.}~\bibnamefont{Shimony}}, and
  \bibinfo{author}{\bibfnamefont{R.~A.} \bibnamefont{Holt}},
  \bibinfo{year}{1969}, \bibinfo{journal}{Phys. Rev. Lett.}
  \textbf{\bibinfo{volume}{23}}, \bibinfo{pages}{880}.

\bibitem[{\citenamefont{Coffman} \emph{et~al.}(2000)\citenamefont{Coffman,
  Kundu, and Wootters}}]{Coffman2000}
\bibinfo{author}{\bibnamefont{Coffman}, \bibfnamefont{V.}},
  \bibinfo{author}{\bibfnamefont{J.}~\bibnamefont{Kundu}}, and
  \bibinfo{author}{\bibfnamefont{W.~K.} \bibnamefont{Wootters}},
  \bibinfo{year}{2000}, \bibinfo{journal}{Phys. Rev. A}
  \textbf{\bibinfo{volume}{61}}, \bibinfo{pages}{052306}.

\bibitem[{\citenamefont{Cohen}(1999)}]{Cohen1999}
\bibinfo{author}{\bibnamefont{Cohen}, \bibfnamefont{O.}}, \bibinfo{year}{1999},
  \bibinfo{journal}{Phys. Rev. A} \textbf{\bibinfo{volume}{60}},
  \bibinfo{pages}{80}.

\bibitem[{\citenamefont{Colella} \emph{et~al.}(1975)\citenamefont{Colella,
  Overhauser, and Werner}}]{Colella1975}
\bibinfo{author}{\bibnamefont{Colella}, \bibfnamefont{R.}},
  \bibinfo{author}{\bibfnamefont{A.~W.}~\bibnamefont{Overhauser}}, and
  \bibinfo{author}{\bibfnamefont{S.~A.} \bibnamefont{Werner}},
  \bibinfo{year}{1975}, \bibinfo{journal}{Phys. Rev. Lett.}
  \textbf{\bibinfo{volume}{34}}, \bibinfo{pages}{1472}.

\bibitem[{\citenamefont{Davisson and Germer}(1927)}]{Davi1927}
\bibinfo{author}{\bibnamefont{Davisson}, \bibfnamefont{C.}} and
  \bibinfo{author}{\bibfnamefont{L.~H.} \bibnamefont{Germer}},
  \bibinfo{year}{1927}, \bibinfo{journal}{Phys. Rev.}
  \textbf{\bibinfo{volume}{30}}, \bibinfo{pages}{705}.

\bibitem[{\citenamefont{Dopfer}(1998)}]{Dopfer1998}
\bibinfo{author}{\bibnamefont{Dopfer}, \bibfnamefont{B.}},
  \bibinfo{year}{1998}, \emph{\bibinfo{title}{Zwei Experimente zur Interferenz von Zwei-Photonen Zust\"{a}nden. Ein Heisenbergmikroskop und Pendell\"{o}sung}},
  Ph.D. Thesis, \bibinfo{school}{University of Innsbruck}.

\bibitem[{\citenamefont{Duan} \emph{et~al.}(2001)\citenamefont{Duan, Lukin,
  Cirac, and Zoller}}]{Duan2001}
\bibinfo{author}{\bibnamefont{Duan}, \bibfnamefont{L.~M.}},
  \bibinfo{author}{\bibfnamefont{M.~D.} \bibnamefont{Lukin}},
  \bibinfo{author}{\bibfnamefont{J.~I.} \bibnamefont{Cirac}}, and
  \bibinfo{author}{\bibfnamefont{P.}~\bibnamefont{Zoller}},
  \bibinfo{year}{2001}, \bibinfo{journal}{Nature}
  \textbf{\bibinfo{volume}{414}}, \bibinfo{pages}{413}.

\bibitem[{\citenamefont{D{\"u}rr} \emph{et~al.}(1998)\citenamefont{D{\"u}rr,
  Nonn, and Rempe}}]{Durr1998}
\bibinfo{author}{\bibnamefont{D{\"u}rr}, \bibfnamefont{S.}},
  \bibinfo{author}{\bibfnamefont{T.}~\bibnamefont{Nonn}}, and
  \bibinfo{author}{\bibfnamefont{G.}~\bibnamefont{Rempe}},
  \bibinfo{year}{1998}, \bibinfo{journal}{Nature}
  \textbf{\bibinfo{volume}{395}}, \bibinfo{pages}{33}.

\bibitem[{\citenamefont{Eichmann} \emph{et~al.}(1993)\citenamefont{Eichmann,
  Bergquist, Bollinger, Gilligan, Itano, Wineland, and Raizen}}]{Eichmann1993}
\bibinfo{author}{\bibnamefont{Eichmann}, \bibfnamefont{U.}},
  \bibinfo{author}{\bibfnamefont{J.~C.} \bibnamefont{Bergquist}},
  \bibinfo{author}{\bibfnamefont{J.~J.} \bibnamefont{Bollinger}},
  \bibinfo{author}{\bibfnamefont{J.~M.} \bibnamefont{Gilligan}},
  \bibinfo{author}{\bibfnamefont{W.~M.} \bibnamefont{Itano}},
  \bibinfo{author}{\bibfnamefont{D.~J.} \bibnamefont{Wineland}}, and
  \bibinfo{author}{\bibfnamefont{M.~G.} \bibnamefont{Raizen}},
  \bibinfo{year}{1993}, \bibinfo{journal}{Phys. Rev. Lett.}
  \textbf{\bibinfo{volume}{70}}, \bibinfo{pages}{2359}.

\bibitem[{\citenamefont{Einstein}(1905)}]{Eins1905}
\bibinfo{author}{\bibnamefont{Einstein}, \bibfnamefont{A.}},
  \bibinfo{year}{1905}, \bibinfo{journal}{Ann. d. Phys}
  \textbf{\bibinfo{volume}{322}}, \bibinfo{pages}{23}.

\bibitem[{\citenamefont{Einstein}(1931)}]{Einstein1931}
\bibinfo{author}{\bibnamefont{Einstein}, \bibfnamefont{A.}},
  \bibinfo{year}{1931}, \bibinfo{journal}{reported in: Angewandte Chemie}
  \textbf{\bibinfo{volume}{45}}, \bibinfo{pages}{718} (1932).

\bibitem[{\citenamefont{Einstein} \emph{et~al.}(1935)\citenamefont{Einstein,
  Podolsky, and Rosen}}]{Eins1935}
\bibinfo{author}{\bibnamefont{Einstein}, \bibfnamefont{A.}},
  \bibinfo{author}{\bibfnamefont{B.}~\bibnamefont{Podolsky}}, and
  \bibinfo{author}{\bibfnamefont{N.}~\bibnamefont{Rosen}},
  \bibinfo{year}{1935}, \bibinfo{journal}{Phys. Rev.}
  \textbf{\bibinfo{volume}{47}}, \bibinfo{pages}{777}.

\bibitem[{\citenamefont{Englert} \emph{et~al.}(1995)\citenamefont{Englert, Scully,
  and Walther}}]{Englert1995}
\bibinfo{author}{\bibnamefont{Englert}, \bibfnamefont{B.-G.}},
  \bibinfo{author}{\bibfnamefont{M.~O.} \bibnamefont{Scully}}, and
  \bibinfo{author}{\bibfnamefont{H.}~\bibnamefont{Walther}},
  \bibinfo{year}{1995}, \bibinfo{journal}{Nature}
  \textbf{\bibinfo{volume}{375}}, \bibinfo{pages}{367}.

\bibitem[{\citenamefont{Englert}(1996)}]{Englert1996}
\bibinfo{author}{\bibnamefont{Englert}, \bibfnamefont{B.-G.}},
  \bibinfo{year}{1996}, \bibinfo{journal}{Phys. Rev. Lett.}
  \textbf{\bibinfo{volume}{77}}, \bibinfo{pages}{2154}.

\bibitem[{\citenamefont{Estermann and Stern}(1930)}]{Estermann1930}
\bibinfo{author}{\bibnamefont{Estermann}, \bibfnamefont{I.}} and
  \bibinfo{author}{\bibfnamefont{O.} \bibnamefont{Stern}},
  \bibinfo{year}{1930}, \bibinfo{journal}{Z. Phys.}
  \textbf{\bibinfo{volume}{61}}, \bibinfo{pages}{95}.

\bibitem[{\citenamefont{Friberg} \emph{et~al.}(1985)\citenamefont{Friberg,
  Hong, and Mandel}}]{Friberg1985}
\bibinfo{author}{\bibnamefont{Friberg}, \bibfnamefont{S.}},
  \bibinfo{author}{\bibfnamefont{C.~K.}~\bibnamefont{Hong}}, and
  \bibinfo{author}{\bibfnamefont{L.}~\bibnamefont{Mandel}},
  \bibinfo{year}{1985}, \bibinfo{journal}{Phys. Rev. Lett.}
  \textbf{\bibinfo{volume}{54}}, \bibinfo{pages}{2011}.

\bibitem[{\citenamefont{Gallicchio} \emph{et~al.}(2014)\citenamefont{Gallicchio,
  Friedman, and Kaiser}}]{Gallicchio2014}
\bibinfo{author}{\bibnamefont{Gallicchio}, \bibfnamefont{J.}},
  \bibinfo{author}{\bibfnamefont{A.~S.}~\bibnamefont{Friedman}}, and
  \bibinfo{author}{\bibfnamefont{D.~I.}~\bibnamefont{Kaiser}},
  \bibinfo{year}{2014}, \bibinfo{journal}{Phys. Rev. Lett.}
  \textbf{\bibinfo{volume}{112}}, \bibinfo{pages}{110405}.

\bibitem[{\citenamefont{Gerlich} \emph{et~al.}(2011)\citenamefont{Gerlich,
  Eibenberger, Tomandl, Nimmrichter, Hornberger, Fagan, Tuexen, Mayor, and
  Arndt}}]{Gerl2011}
\bibinfo{author}{\bibnamefont{Gerlich}, \bibfnamefont{S.}},
  \bibinfo{author}{\bibfnamefont{S.}~\bibnamefont{Eibenberger}},
  \bibinfo{author}{\bibfnamefont{M.}~\bibnamefont{Tomandl}},
  \bibinfo{author}{\bibfnamefont{S.}~\bibnamefont{Nimmrichter}},
  \bibinfo{author}{\bibfnamefont{K.}~\bibnamefont{Hornberger}},
  \bibinfo{author}{\bibfnamefont{P.~J.}~\bibnamefont{Fagan}},
  \bibinfo{author}{\bibfnamefont{J.}~\bibnamefont{Tuexen}},
  \bibinfo{author}{\bibfnamefont{M.}~\bibnamefont{Mayor}}, and
  \bibinfo{author}{\bibfnamefont{M.}~\bibnamefont{Arndt}},
  \bibinfo{year}{2011}, \bibinfo{journal}{Nature Commun.}
  \textbf{\bibinfo{volume}{2}}, \bibinfo{pages}{263}.


\bibitem[{\citenamefont{Giustina} \emph{et~al.}(2015)\citenamefont{Giustina,
  et~al}}]{Giustina2015}
\bibinfo{author}{\bibnamefont{Giustina}, \bibfnamefont{M. \emph{et~al.}}},
  \bibinfo{year}{2015}, \bibinfo{journal}{Phys. Rev. Lett.}
  \textbf{\bibinfo{volume}{115}}, \bibinfo{pages}{250401}.


\bibitem[{\citenamefont{Grangier}(1986)}]{Grangier1986a}
\bibinfo{author}{\bibnamefont{Grangier}, \bibfnamefont{P.}},
  \bibinfo{year}{1986}, \emph{\bibinfo{title}{Experimental study of
  non-classical properties of light; single-photon interferences}}, Ph.D.
  Thesis, \bibinfo{school}{Institut d'Optique et Université Paris}.

\bibitem[{\citenamefont{Greenberger}
  \emph{et~al.}(1983)\citenamefont{Greenberger, Horne, and
  Zeilinger}}]{Greenberger1983}
\bibinfo{author}{\bibnamefont{Greenberger}, \bibfnamefont{D.~M.}},
  \bibinfo{author}{\bibfnamefont{M.~A.} \bibnamefont{Horne},
  \bibfnamefont{C. G. Shull}}, and
  \bibinfo{author}{\bibfnamefont{A.}~\bibnamefont{Zeilinger}},
  \bibinfo{year}{1983}, in \bibinfo{booktitle}{\emph{International Symposium on
  foundations of Quantum Mechanics in the light of New Technology} \textrm{(Physical Society of Japan, Tokyo).}}

\bibitem[{\citenamefont{Greenberger and Yasin}(1988)}]{Greenberger1988}
\bibinfo{author}{\bibnamefont{Greenberger}, \bibfnamefont{D.~M.}} and
  \bibinfo{author}{\bibnamefont{A. Yasin}}, \bibinfo{year}{1988},
  \bibinfo{journal}{Phys. Lett. A}
  \textbf{\bibinfo{volume}{128}}, \bibinfo{pages}{391}.

\bibitem[{\citenamefont{Greenberger}
  \emph{et~al.}(2008{\natexlab{a}})\citenamefont{Greenberger, Horne, and
  Zeilinger}}]{Greenberger2008}
\bibinfo{author}{\bibnamefont{Greenberger}, \bibfnamefont{D.~M.}},
  \bibinfo{author}{\bibfnamefont{M.}~\bibnamefont{Horne}}, and
  \bibinfo{author}{\bibfnamefont{A.}~\bibnamefont{Zeilinger}},
  \bibinfo{year}{2008}{\natexlab{a}}, \bibinfo{journal}{Phys. Rev. A}
  \textbf{\bibinfo{volume}{78}}, \bibinfo{pages}{022110}.

\bibitem[{\citenamefont{Greenberger}
  \emph{et~al.}(2008{\natexlab{b}})\citenamefont{Greenberger, Horne, Zeilinger,
  and \ifmmode~\dot{Z}\else \.{Z}\fi{}ukowski}}]{Greenberger2008a}
\bibinfo{author}{\bibnamefont{Greenberger}, \bibfnamefont{D.~M.}},
  \bibinfo{author}{\bibfnamefont{M.}~\bibnamefont{Horne}},
  \bibinfo{author}{\bibfnamefont{A.}~\bibnamefont{Zeilinger}}, and
  \bibinfo{author}{\bibfnamefont{M.}~\bibnamefont{\ifmmode~\dot{Z}\else
  \.{Z}\fi{}ukowski}}, \bibinfo{year}{2008}{\natexlab{b}},
  \bibinfo{journal}{Phys. Rev. A} \textbf{\bibinfo{volume}{78}},
  \bibinfo{pages}{022111}.

\bibitem[{\citenamefont{Grover}(1997)\citenamefont{Grover}}]{Grover1997}
\bibinfo{author}{\bibnamefont{Grover}, \bibfnamefont{L. K.}},
  \bibinfo{year}{1997}, \bibinfo{journal}{Phys. Rev. Lett.}
  \textbf{\bibinfo{volume}{79}}, \bibinfo{pages}{325}.

\bibitem[{\citenamefont{G\"uhne and Toth}(2009)}]{Gueh2009}
\bibinfo{author}{\bibnamefont{G\"uhne}, \bibfnamefont{O.}} and
  \bibinfo{author}{\bibfnamefont{G.}~\bibnamefont{Toth}},
  \bibinfo{year}{2009}, \bibinfo{journal}{Phys. Rep.}
  \textbf{\bibinfo{volume}{474}}, \bibinfo{pages}{1}.

\bibitem[{\citenamefont{Halder} \emph{et~al.}(2007)\citenamefont{Halder,
  Beveratos, Gisin, Scarani, Simon, and Zbinden}}]{Halder2007}
\bibinfo{author}{\bibnamefont{Halder}, \bibfnamefont{M.}},
  \bibinfo{author}{\bibfnamefont{A.}~\bibnamefont{Beveratos}},
  \bibinfo{author}{\bibfnamefont{N.}~\bibnamefont{Gisin}},
  \bibinfo{author}{\bibfnamefont{V.}~\bibnamefont{Scarani}},
  \bibinfo{author}{\bibfnamefont{C.}~\bibnamefont{Simon}}, and
  \bibinfo{author}{\bibfnamefont{H.}~\bibnamefont{Zbinden}},
  \bibinfo{year}{2007}, \bibinfo{journal}{Nature Phys.}
  \textbf{\bibinfo{volume}{3}}, \bibinfo{pages}{692}.

\bibitem[{\citenamefont{Heisenberg}(1927)}]{Heis1927}
\bibinfo{author}{\bibnamefont{Heisenberg}, \bibfnamefont{W.}},
  \bibinfo{year}{1927}, \bibinfo{journal}{Zeitschrift f{\"u}r Physik}
  \textbf{\bibinfo{volume}{43}}, \bibinfo{pages}{172}.

\bibitem[{\citenamefont{Heisenberg}(1991)}]{Heis1991}
\bibinfo{author}{\bibnamefont{Heisenberg}, \bibfnamefont{W.}},
  \bibinfo{year}{1991}, \emph{\bibinfo{title}{Physikalische Prinzipien der
  Quantentheorie}} (\bibinfo{publisher}{B.I.-Wissenschaftsverlag,
  Mannheim / Wien / Z\"urich}).

\bibitem[{\citenamefont{Hellmuth} \emph{et~al.}(1987)\citenamefont{Hellmuth,
  Walther, Zajonc, and Schleich}}]{Hellmuth1987}
\bibinfo{author}{\bibnamefont{Hellmuth}, \bibfnamefont{T.}},
  \bibinfo{author}{\bibfnamefont{H.}~\bibnamefont{Walther}},
  \bibinfo{author}{\bibfnamefont{A.}~\bibnamefont{Zajonc}}, and
  \bibinfo{author}{\bibfnamefont{W.}~\bibnamefont{Schleich}},
  \bibinfo{year}{1987}, \bibinfo{journal}{Physical Review A}
  \textbf{\bibinfo{volume}{35}}, \bibinfo{pages}{2532}.

\bibitem[{\citenamefont{Hermann}(1935)}]{Hermann1935}
\bibinfo{author}{\bibnamefont{Hermann}, \bibfnamefont{G.}},
  \bibinfo{year}{1935}, \bibinfo{journal}{Naturwissenschaften}
  \textbf{\bibinfo{volume}{23}}, \bibinfo{pages}{718}.

\bibitem[{\citenamefont{Hensen}(2015)}]{Hensen2015}
\bibinfo{author}{\bibnamefont{Hensen}, \bibfnamefont{B., \emph{et~al.}}},
  \bibinfo{year}{2015}, \bibinfo{journal}{Nature}
  \textbf{\bibinfo{volume}{526}}, \bibinfo{pages}{682}.

\bibitem[{\citenamefont{Herzog} \emph{et~al.}(1995)\citenamefont{Herzog,
  Kwiat, Weinfurter, and Zeilinger}}]{Herzog1995}
\bibinfo{author}{\bibnamefont{Herzog}, \bibfnamefont{T.~J.}},
  \bibinfo{author}{\bibfnamefont{P.~G.} \bibnamefont{Kwiat}},
  \bibinfo{author}{\bibfnamefont{H.} \bibnamefont{Weinfurter}}, and
  \bibinfo{author}{\bibfnamefont{A.} \bibnamefont{Zeilinger}}
  \bibinfo{year}{1994}, \bibinfo{journal}{Phys. Rev. Lett.}
  \textbf{\bibinfo{volume}{75}}, \bibinfo{pages}{3034}.

\bibitem[{\citenamefont{Horodecki} \emph{et~al.}(2009)\citenamefont{Horodecki,
  Horodecki, Horodecki, and Horodecki}}]{Horo2009}
\bibinfo{author}{\bibnamefont{Horodecki}, \bibfnamefont{R.}},
  \bibinfo{author}{\bibfnamefont{P.}~\bibnamefont{Horodecki}},
  \bibinfo{author}{\bibfnamefont{M.}~\bibnamefont{Horodecki}}, and
  \bibinfo{author}{\bibfnamefont{K.}~\bibnamefont{Horodecki}},
  \bibinfo{year}{2009}, \bibinfo{journal}{Rev. Mod. Phys.}
  \textbf{\bibinfo{volume}{81}}, \bibinfo{pages}{865}.

\bibitem[{\citenamefont{Ionicioiu and Terno}(2011)}]{Ioni2011}
\bibinfo{author}{\bibnamefont{Ionicioiu}, \bibfnamefont{R.}} and
  \bibinfo{author}{\bibfnamefont{D.~R.} \bibnamefont{Terno}},
  \bibinfo{year}{2011}, \bibinfo{journal}{Phys. Rev. Lett.}
  \textbf{\bibinfo{volume}{107}}, \bibinfo{pages}{230406}.

\bibitem[{\citenamefont{Ionicioiu} \emph{et~al.}(2014)}]{Ioni2014}
\bibinfo{author}{\bibnamefont{Ionicioiu}, \bibfnamefont{R.}},
\bibinfo{author}{\bibnamefont{T.} \bibfnamefont{Jennewein}},
\bibinfo{author}{\bibnamefont{R.~B.} \bibfnamefont{Mann}}, and
  \bibinfo{author}{\bibfnamefont{D.~R.} \bibnamefont{Terno}},
  \bibinfo{year}{2014}, \bibinfo{journal}{Nature Comm.}
  \textbf{\bibinfo{volume}{5}}, \bibinfo{pages}{4997}.

\bibitem[{\citenamefont{Ionicioiu} \emph{et~al.}(2015)}]{Ioni2015}
\bibinfo{author}{\bibnamefont{Ionicioiu}, \bibfnamefont{R.}},
\bibinfo{author}{\bibnamefont{R.~B.} \bibfnamefont{Mann}}, and
  \bibinfo{author}{\bibfnamefont{D.~R.} \bibnamefont{Terno}},
  \bibinfo{year}{2015}, \bibinfo{journal}{Phys. Rev. Lett.}
  \textbf{\bibinfo{volume}{114}}, \bibinfo{pages}{060405}.

\bibitem[{\citenamefont{Jacques} \emph{et~al.}(2007)\citenamefont{Jacques, Wu,
  Grosshans, Treussart, Grangier, Aspect, and Roch}}]{Jacques2007}
\bibinfo{author}{\bibnamefont{Jacques}, \bibfnamefont{V.}},
  \bibinfo{author}{\bibfnamefont{E.}~\bibnamefont{Wu}},
  \bibinfo{author}{\bibfnamefont{F.}~\bibnamefont{Grosshans}},
  \bibinfo{author}{\bibfnamefont{F.}~\bibnamefont{Treussart}},
  \bibinfo{author}{\bibfnamefont{P.}~\bibnamefont{Grangier}},
  \bibinfo{author}{\bibfnamefont{A.}~\bibnamefont{Aspect}}, and
  \bibinfo{author}{\bibfnamefont{J.-F.} \bibnamefont{Roch}},
  \bibinfo{year}{2007}, \bibinfo{journal}{Science}
  \textbf{\bibinfo{volume}{315}}, \bibinfo{pages}{966}.

\bibitem[{\citenamefont{Jacques} \emph{et~al.}(2008)\citenamefont{Jacques, Wu,
  Grosshans, Treussart, Grangier, Aspect, and Roch}}]{Jacques2008}
\bibinfo{author}{\bibnamefont{Jacques}, \bibfnamefont{V.}},
  \bibinfo{author}{\bibfnamefont{E.}~\bibnamefont{Wu}},
  \bibinfo{author}{\bibfnamefont{F.}~\bibnamefont{Grosshans}},
  \bibinfo{author}{\bibfnamefont{F.}~\bibnamefont{Treussart}},
  \bibinfo{author}{\bibfnamefont{P.}~\bibnamefont{Grangier}},
  \bibinfo{author}{\bibfnamefont{A.}~\bibnamefont{Aspect}}, and
  \bibinfo{author}{\bibfnamefont{J.-F.} \bibnamefont{Roch}},
  \bibinfo{year}{2008}, \bibinfo{journal}{Phys. Rev. Lett.}
  \textbf{\bibinfo{volume}{100}}, \bibinfo{pages}{220402}.

\bibitem[{\citenamefont{Jaeger} \emph{et~al.}(1995)\citenamefont{Jaeger}}]{Jaeger1995}
\bibinfo{author}{\bibnamefont{Jaeger}, \bibfnamefont{G.}},
  \bibinfo{author}{\bibfnamefont{A.} \bibnamefont{Shimony}}, and
  \bibinfo{author}{\bibfnamefont{L.} \bibnamefont{Vaidman}},
  \bibinfo{year}{1993}, \bibinfo{journal}{Phys. Rev. A}
  \textbf{\bibinfo{volume}{51}}, \bibinfo{pages}{54}.

\bibitem[{\citenamefont{Jakubowicz}(1984)}]{Jakubowicz1984}
\bibinfo{author}{\bibnamefont{Jakubowicz}, \bibfnamefont{O.~G.}},
  \bibinfo{year}{1984}, \emph{\bibinfo{title}{A delayed random choice quantum
  mechanics experiment using light quanta.}}, Ph.D. Thesis,
  \bibinfo{school}{Maryland University, College Park}.

\bibitem[{\citenamefont{Jennewein} \emph{et~al.}(2000)\citenamefont{Jennewein,
  Achleitner, Weihs, Weinfurter, and Zeilinger}}]{Jennewein2000}
\bibinfo{author}{\bibnamefont{Jennewein}, \bibfnamefont{T.}},
  \bibinfo{author}{\bibfnamefont{U.}~\bibnamefont{Achleitner}},
  \bibinfo{author}{\bibfnamefont{G.}~\bibnamefont{Weihs}},
  \bibinfo{author}{\bibfnamefont{H.}~\bibnamefont{Weinfurter}}, and
  \bibinfo{author}{\bibfnamefont{A.}~\bibnamefont{Zeilinger}},
  \bibinfo{year}{2000}, \bibinfo{journal}{Rev. Sci. Instrum.}
  \textbf{\bibinfo{volume}{71}}, \bibinfo{pages}{1675}.

\bibitem[{\citenamefont{Jennewein} \emph{et~al.}(2001)\citenamefont{Jennewein,
  Weihs, Pan, and Zeilinger}}]{Jennewein2001}
\bibinfo{author}{\bibnamefont{Jennewein}, \bibfnamefont{T.}},
  \bibinfo{author}{\bibfnamefont{G.}~\bibnamefont{Weihs}},
  \bibinfo{author}{\bibfnamefont{J.-W.} \bibnamefont{Pan}}, and
  \bibinfo{author}{\bibfnamefont{A.}~\bibnamefont{Zeilinger}},
  \bibinfo{year}{2001}, \bibinfo{journal}{Phys. Rev. Lett.}
  \textbf{\bibinfo{volume}{88}},\bibinfo{pages}{017903}.

\bibitem[{\citenamefont{Jennewein}(2002)}]{Jennewein2002a}
\bibinfo{author}{\bibnamefont{Jennewein}, \bibfnamefont{T.}},
  \bibinfo{year}{2002}, \emph{\bibinfo{title}{Quantum Communication and
  Teleportation Experiments using Entangled Photon Pairs}}, Ph.D. Thesis,
  \bibinfo{school}{University of Vienna}.

\bibitem[{\citenamefont{Jeong} \emph{et~al.}(2013)\citenamefont{Jeong,
  Franco, Lim, Kim, Kim, and Kim}}]{Jeong2013}
\bibinfo{author}{\bibnamefont{Jeong}, \bibfnamefont{Y.~C}},
  \bibinfo{author}{\bibfnamefont{C.~D.}~\bibnamefont{Franco}}, and
  \bibinfo{author}{\bibfnamefont{H.~T.}~\bibnamefont{Lim}}, and
  \bibinfo{author}{\bibfnamefont{M.~S.}~\bibnamefont{Kim}}, and
  \bibinfo{author}{\bibfnamefont{Y.~H.}~\bibnamefont{Kim}},
  \bibinfo{year}{2013}, \bibinfo{journal}{Nat. Comm.}
  \textbf{\bibinfo{volume}{4}}, \bibinfo{pages}{2471}.

\bibitem[{\citenamefont{Joensson}(1961)}]{Joen1961}
\bibinfo{author}{\bibnamefont{Joensson}, \bibfnamefont{C.}},
  \bibinfo{year}{1961}, \bibinfo{journal}{Z. Phys.}
  \textbf{\bibinfo{volume}{161}}, \bibinfo{pages}{454}.

\bibitem[{\citenamefont{Joobeur} \emph{et~al.}(1994)\citenamefont{Joobeur,
  Saleh, and Teich}}]{Joobeur1994}
\bibinfo{author}{\bibnamefont{Joobeur}, \bibfnamefont{A.}},
  \bibinfo{author}{\bibfnamefont{B.~E.~A.} \bibnamefont{Saleh}}, and
  \bibinfo{author}{\bibfnamefont{M.~C.} \bibnamefont{Teich}},
  \bibinfo{year}{1994}, \bibinfo{journal}{Phys. Rev. A}
  \textbf{\bibinfo{volume}{50}}, \bibinfo{pages}{3349}.

\bibitem[{\citenamefont{Kaiser} \emph{et~al.}(2012)\citenamefont{Kaiser,
  Coudreau, Milman, Ostrowsky, and Tanzilli}}]{Kaiser2012}
\bibinfo{author}{\bibnamefont{Kaiser}, \bibfnamefont{F.}},
  \bibinfo{author}{\bibfnamefont{T.}~\bibnamefont{Coudreau}},
  \bibinfo{author}{\bibfnamefont{P.}~\bibnamefont{Milman}},
  \bibinfo{author}{\bibfnamefont{D.~B.} \bibnamefont{Ostrowsky}}, and
  \bibinfo{author}{\bibfnamefont{S.}~\bibnamefont{Tanzilli}},
  \bibinfo{year}{2012}, \bibinfo{journal}{Science}
  \textbf{\bibinfo{volume}{338}}, \bibinfo{pages}{637}.

\bibitem[{\citenamefont{Kaltenbaek}
  \emph{et~al.}(2009)\citenamefont{Kaltenbaek, Prevedel, Aspelmeyer, and
  Zeilinger}}]{Kaltenbaek2009}
\bibinfo{author}{\bibnamefont{Kaltenbaek}, \bibfnamefont{R.}},
  \bibinfo{author}{\bibfnamefont{R.}~\bibnamefont{Prevedel}},
  \bibinfo{author}{\bibfnamefont{M.}~\bibnamefont{Aspelmeyer}}, and
  \bibinfo{author}{\bibfnamefont{A.}~\bibnamefont{Zeilinger}},
  \bibinfo{year}{2009}, \bibinfo{journal}{Phys. Rev. A}
  \textbf{\bibinfo{volume}{79}}, \bibinfo{pages}{040302}.

\bibitem[{\citenamefont{Kennard}(1927)}]{Kenn1927}
\bibinfo{author}{\bibnamefont{Kennard}, \bibfnamefont{E.~H.}},
  \bibinfo{year}{1927}, \bibinfo{journal}{Z. Phys.}
  \textbf{\bibinfo{volume}{44}}, \bibinfo{pages}{326}.

\bibitem[{\citenamefont{Kim} \emph{et~al.}(2000)\citenamefont{Kim, Yu, Kulik,
  Shih, and Scully}}]{Kim2000}
\bibinfo{author}{\bibnamefont{Kim}, \bibfnamefont{Y.-H.}},
  \bibinfo{author}{\bibfnamefont{R.}~\bibnamefont{Yu}},
  \bibinfo{author}{\bibfnamefont{S.~P.} \bibnamefont{Kulik}},
  \bibinfo{author}{\bibfnamefont{Y.}~\bibnamefont{Shih}}, and
  \bibinfo{author}{\bibfnamefont{M.~O.} \bibnamefont{Scully}},
  \bibinfo{year}{2000}, \bibinfo{journal}{Phys. Rev. Lett.}
  \textbf{\bibinfo{volume}{84}}, \bibinfo{pages}{1}.

\bibitem[{\citenamefont{Kurtsiefer} \emph{et~al.}(2000)\citenamefont{Kurtsiefer et al.}}]{Kurt2000}
\bibinfo{author}{\bibnamefont{Kurtsiefer}, \bibfnamefont{C.}},
  \bibinfo{author}{\bibfnamefont{S.}~\bibnamefont{Mayer}},
  \bibinfo{author}{\bibfnamefont{P.}~\bibnamefont{Zarda}}, and
  \bibinfo{author}{\bibfnamefont{H.}~\bibnamefont{Weinfurter}},
  \bibinfo{year}{2000}, \bibinfo{journal}{Phys. Rev. Lett.}
  \textbf{\bibinfo{volume}{85}}, \bibinfo{pages}{290}.

\bibitem[{\citenamefont{Kwiat}
  \emph{et~al.}(1995)\citenamefont{Kwiat, Mattle, Weinfurter, Zeilinger, Sergienko, Shih}}]{Kwiat1995}
\bibinfo{author}{\bibnamefont{Kwait}, \bibfnamefont{P.~G.}},
  \bibinfo{author}{\bibfnamefont{K.}~\bibnamefont{Mattle}},
  \bibinfo{author}{\bibfnamefont{H.}~\bibnamefont{Weinfurter}},
  \bibinfo{author}{\bibfnamefont{A.}~\bibnamefont{Zeilinger}},
  \bibinfo{author}{\bibfnamefont{A.~V.}~\bibnamefont{Sergienko}}, and
  \bibinfo{author}{\bibfnamefont{Y.}~\bibnamefont{Shih}},
  \bibinfo{year}{1995}, \bibinfo{journal}{Phys. Rev. Lett.}
  \textbf{\bibinfo{volume}{75}}, \bibinfo{pages}{4337}.

\bibitem[{\citenamefont{Kwiat and Englert}(2004)}]{Kwiat2004}
\bibinfo{author}{\bibnamefont{Kwiat}, \bibfnamefont{P.~G.}} and
  \bibinfo{author}{\bibfnamefont{B.-G.} \bibnamefont{Englert}},
  \bibinfo{year}{2004}, in \bibinfo{title}{\emph{Science and Ultimate Reality:
  Quantum Theory, Cosmology and Complexity}}, \textrm{ed. J. D. Barrow, P. C. W. Davies, and S. C. L. Harper,
Jr (\bibinfo{publisher}{Cambridge University Press})}.

\bibitem[{\citenamefont{Lee} \emph{et~al.}(2014)\citenamefont{Lee, Lim, Hong, Jeong, Kim and Kim}}]{Lee2014}
\bibinfo{author}{\bibnamefont{Lee}, \bibfnamefont{L.~C.}}
\bibinfo{author}{\bibnamefont{Lim}, \bibfnamefont{H.~T.}}
\bibinfo{author}{\bibnamefont{Hong}, \bibfnamefont{K.~H.}}
\bibinfo{author}{\bibnamefont{Jeong}, \bibfnamefont{Y.~C.}}
\bibinfo{author}{\bibnamefont{Kim}, \bibfnamefont{M.~S.}} and
  \bibinfo{author}{\bibfnamefont{Kim} \bibnamefont{Y.~H.}},
  \bibinfo{year}{2014}, \bibinfo{journal}{Nat. Comm.}
  \textbf{\bibinfo{volume}{5}}, \bibinfo{pages}{4522}.

\bibitem[{\citenamefont{Ma} \emph{et~al.}(2009)\citenamefont{Ma, Qarry, Kofler,
  Jennewein, and Zeilinger}}]{Ma2009}
\bibinfo{author}{\bibnamefont{Ma}, \bibfnamefont{X.-S.}},
  \bibinfo{author}{\bibfnamefont{A.}~\bibnamefont{Qarry}},
  \bibinfo{author}{\bibfnamefont{J.}~\bibnamefont{Kofler}},
  \bibinfo{author}{\bibfnamefont{T.}~\bibnamefont{Jennewein}}, and
  \bibinfo{author}{\bibfnamefont{A.}~\bibnamefont{Zeilinger}},
  \bibinfo{year}{2009}, \bibinfo{journal}{Phys. Rev. A}
  \textbf{\bibinfo{volume}{79}}, \bibinfo{pages}{042101}.

\bibitem[{\citenamefont{Ma}(2010)\citenamefont{Ma}}]{Ma2010}
\bibinfo{author}{\bibnamefont{Ma}, \bibfnamefont{X.-S.}},
  \bibinfo{year}{2010}, \emph{\bibinfo{title}{Nonlocal delayed-choice experiments with entangled photons}}, Ph.D. Thesis,
  \bibinfo{school}{University of Vienna}.

\bibitem[{\citenamefont{Ma} \emph{et~al.}(2012)\citenamefont{Ma, Zotter,
  Kofler, Ursin, Jennewein, Brukner, and Zeilinger}}]{Ma2012}
\bibinfo{author}{\bibnamefont{Ma}, \bibfnamefont{X.-S.}},
  \bibinfo{author}{\bibfnamefont{S.}~\bibnamefont{Zotter}},
  \bibinfo{author}{\bibfnamefont{J.}~\bibnamefont{Kofler}},
  \bibinfo{author}{\bibfnamefont{R.}~\bibnamefont{Ursin}},
  \bibinfo{author}{\bibfnamefont{T.}~\bibnamefont{Jennewein}},
  \bibinfo{author}{\bibfnamefont{\v C}~\bibnamefont{Brukner}}, and
  \bibinfo{author}{\bibfnamefont{A.}~\bibnamefont{Zeilinger}},
  \bibinfo{year}{2012}, \bibinfo{journal}{Nature Phys.}
  \textbf{\bibinfo{volume}{8}}, \bibinfo{pages}{479}.

\bibitem[{\citenamefont{Ma} \emph{et~al.}(2013)\citenamefont{Ma, Kofler, Qarry,
  Tetik, Scheidl, Ursin, Ramelow, Herbst, Ratschbacher, Fedrizzi, Jennewein,
  and Zeilinger}}]{Ma2013a}
\bibinfo{author}{\bibnamefont{Ma}, \bibfnamefont{X.-S.}},
  \bibinfo{author}{\bibfnamefont{J.}~\bibnamefont{Kofler}},
  \bibinfo{author}{\bibfnamefont{A.}~\bibnamefont{Qarry}},
  \bibinfo{author}{\bibfnamefont{N.}~\bibnamefont{Tetik}},
  \bibinfo{author}{\bibfnamefont{T.}~\bibnamefont{Scheidl}},
  \bibinfo{author}{\bibfnamefont{R.}~\bibnamefont{Ursin}},
  \bibinfo{author}{\bibfnamefont{S.}~\bibnamefont{Ramelow}},
  \bibinfo{author}{\bibfnamefont{T.}~\bibnamefont{Herbst}},
  \bibinfo{author}{\bibfnamefont{L.}~\bibnamefont{Ratschbacher}},
  \bibinfo{author}{\bibfnamefont{A.}~\bibnamefont{Fedrizzi}},
  \bibinfo{author}{\bibfnamefont{T.}~\bibnamefont{Jennewein}}, and
  \bibinfo{author}{\bibfnamefont{A.}~\bibnamefont{Zeilinger}},
  \bibinfo{year}{2013}, \bibinfo{journal}{Proc. Natl. Acad. Sci. USA}
  \textbf{\bibinfo{volume}{110}}, \bibinfo{pages}{1221}.

\bibitem[{\citenamefont{Mach}(1892)\citenamefont{Mach}}]{Mach1892}
\bibinfo{author}{\bibnamefont{Mach}, \bibfnamefont{L.}},
  \bibinfo{year}{1892}, \bibinfo{journal}{Z. Instrumentenkd.}
  \textbf{\bibinfo{volume}{12}}, \bibinfo{pages}{89}.

\bibitem[{\citenamefont{Manning} \emph{et~al.}(2012)\citenamefont{Manning, Khakimov, Dall, and Truscott}}]{Manning2015}
\bibinfo{author}{\bibnamefont{Manning}, \bibfnamefont{A. G.}},
  \bibinfo{author}{\bibfnamefont{R. I.}~\bibnamefont{Khakimov}},
  \bibinfo{author}{\bibfnamefont{R. G.}~\bibnamefont{Dall}}, and
  \bibinfo{author}{\bibfnamefont{A. G.}~\bibnamefont{Truscott}},
  \bibinfo{year}{2015}, \bibinfo{journal}{Nature Phys.} May 2015.

\bibitem[{\citenamefont{Matsukevich}
  \emph{et~al.}(2008)\citenamefont{Matsukevich, Moehring, Olmschenk, and
  Monroe}}]{Matsukevich2008}
\bibinfo{author}{\bibnamefont{Matsukevich}, \bibfnamefont{D. N., P. Maunz}},
  \bibinfo{author}{\bibfnamefont{D.~L.} \bibnamefont{Moehring}},
  \bibinfo{author}{\bibfnamefont{S.}~\bibnamefont{Olmschenk}}, and
  \bibinfo{author}{\bibfnamefont{C.}~\bibnamefont{Monroe}},
  \bibinfo{year}{2008}, \bibinfo{journal}{Phys. Rev. Lett.}
  \textbf{\bibinfo{volume}{100}}, \bibinfo{pages}{150404}.

\bibitem[{\citenamefont{Miller}(1983)}]{Miller1983}
\bibinfo{author}{\bibnamefont{Miller}, \bibfnamefont{W.~A.}},
  \bibinfo{year}{1983}, in \bibinfo{booktitle}{\emph{International Symposium on
  foundations of Quantum Mechanics in the light of New Technology} \textrm{(Physical Society of Japan, Tokyo).}}

\bibitem[{\citenamefont{Miller and Wheeler}(1983)}]{Miller1983a}
\bibinfo{author}{\bibnamefont{Miller}, \bibfnamefont{W.~A.}} and
  \bibinfo{author}{\bibfnamefont{J.~A.} \bibnamefont{Wheeler}},
  \bibinfo{year}{1983}, in \bibinfo{booktitle}{\emph{International Symposium on
  foundations of Quantum Mechanics in the light of New Technology} \textrm{(Physical Society of Japan, Tokyo).}}

\bibitem[{\citenamefont{Mittelstaedt}(1986)}]{Mittelstaedt1986}
\bibinfo{author}{\bibnamefont{Mittelstaedt}, \bibfnamefont{P.}},
  \bibinfo{year}{1986}, in \bibinfo{booktitle}{\emph{2nd International
  Symposium on foundations of Quantum Mechanics in the light of New
  Technology} \textrm{(Physical Society of Japan, Tokyo).}}

\bibitem[{\citenamefont{Nielsen and Chuang}(2000)}]{Niel2000}
\bibinfo{author}{\bibnamefont{Nielsen}, \bibfnamefont{M.~A.}} and
  \bibinfo{author}{\bibfnamefont{I.~L.} \bibnamefont{Chuang}},
  \bibinfo{year}{2000}, \emph{\bibinfo{title}{Quantum Computation and Quantum
  Information}} (\bibinfo{publisher}{Cambridge University Press, Cambridge}).

\bibitem[{\citenamefont{Ozawa}(2004)}]{Ozaw2004}
\bibinfo{author}{\bibnamefont{Ozawa}, \bibfnamefont{M.}}, \bibinfo{year}{2004},
  \bibinfo{journal}{Phys. Lett. A} \textbf{\bibinfo{volume}{320}},
  \bibinfo{pages}{367}.

\bibitem[{\citenamefont{Pan} \emph{et~al.}(1998)\citenamefont{Pan, Bouwmeester,
  Weinfurter, and Zeilinger}}]{Pan1998}
\bibinfo{author}{\bibnamefont{Pan}, \bibfnamefont{J.-W.}},
  \bibinfo{author}{\bibfnamefont{D.}~\bibnamefont{Bouwmeester}},
  \bibinfo{author}{\bibfnamefont{H.}~\bibnamefont{Weinfurter}}, and
  \bibinfo{author}{\bibfnamefont{A.}~\bibnamefont{Zeilinger}},
  \bibinfo{year}{1998}, \bibinfo{journal}{Phys. Rev. Lett.}
  \textbf{\bibinfo{volume}{80}}, \bibinfo{pages}{3891}.

\bibitem[{\citenamefont{Pan} \emph{et~al.}(2012)}]{Pan2012}
\bibinfo{author}{\bibnamefont{Pan}, \bibfnamefont{J.-W.}},
  \bibinfo{author}{\bibfnamefont{Z.-B.}~\bibnamefont{Chen}},
  \bibinfo{author}{\bibfnamefont{C.-Y.}~\bibnamefont{Lu}},
  \bibinfo{author}{\bibfnamefont{H.}~\bibnamefont{Weinfurter}},
  \bibinfo{author}{\bibfnamefont{A.}~\bibnamefont{Zeilinger}}, and
  \bibinfo{author}{\bibfnamefont{M.}~\bibnamefont{Zukowski}},
  \bibinfo{year}{2012}, \bibinfo{journal}{Rev. Mod. Phys.}
  \textbf{\bibinfo{volume}{84}}, \bibinfo{pages}{777}.

\bibitem[{\citenamefont{Paul}(1982)}]{Paul1982}
\bibinfo{author}{\bibnamefont{Paul}, \bibfnamefont{H.}},
  \bibinfo{year}{1982}, \bibinfo{journal}{Rev. Mod. Phys.}
  \textbf{\bibinfo{volume}{54}}, \bibinfo{pages}{1061}.

\bibitem[{\citenamefont{Peres}(2000)}]{Peres2000}
\bibinfo{author}{\bibnamefont{Peres}, \bibfnamefont{A.}}, \bibinfo{year}{2000},
  \bibinfo{journal}{J. Mod. Opt.}
  \textbf{\bibinfo{volume}{47}}, \bibinfo{pages}{139}.

\bibitem[{\citenamefont{Peruzzo} \emph{et~al.}(2012)\citenamefont{Peruzzo,
  Shadbolt, Brunner, Popescu, and O'Brien}}]{Peru2012}
\bibinfo{author}{\bibnamefont{Peruzzo}, \bibfnamefont{A.}},
  \bibinfo{author}{\bibfnamefont{P.}~\bibnamefont{Shadbolt}},
  \bibinfo{author}{\bibfnamefont{N.}~\bibnamefont{Brunner}},
  \bibinfo{author}{\bibfnamefont{S.}~\bibnamefont{Popescu}}, and
  \bibinfo{author}{\bibfnamefont{J.~L.} \bibnamefont{O'Brien}},
  \bibinfo{year}{2012}, \bibinfo{journal}{Science}
  \textbf{\bibinfo{volume}{338}}, \bibinfo{pages}{634}.

\bibitem[{\citenamefont{Rarity and Tapster}(1990)}]{Rarity1990}
\bibinfo{author}{\bibnamefont{Rarity}, \bibfnamefont{J.~G.}} and
  \bibinfo{author}{\bibfnamefont{P.~R.} \bibnamefont{Tapster}},
  \bibinfo{year}{1990}, \bibinfo{journal}{Phys. Rev. Lett.}
  \textbf{\bibinfo{volume}{64}}, \bibinfo{pages}{2495}.

\bibitem[{\citenamefont{H. Rauch, W. Treimer and U. Bonse}(1974)}]{Rauch1974}
\bibinfo{author}{\bibnamefont{Rauch}, \bibfnamefont{H.}},
  \bibinfo{author}{\bibfnamefont{W.} \bibnamefont{Treimer}}, and
  \bibinfo{author}{\bibnamefont{Bonse}, \bibfnamefont{U.}}
  \bibinfo{year}{1974}, \bibinfo{journal}{Phys. Lett. A}
  \textbf{\bibinfo{volume}{47}}, \bibinfo{pages}{369}.

\bibitem[{\citenamefont{Riebe} \emph{et~al.}(2004)\citenamefont{Riebe, Haffner,
  Roos, Hansel, Benhelm, Lancaster, Korber, Becher, Schmidt-Kaler, James, and
  Blatt}}]{Riebe2004}
\bibinfo{author}{\bibnamefont{Riebe}, \bibfnamefont{M.}},
  \bibinfo{author}{\bibfnamefont{H.}~\bibnamefont{Haffner}},
  \bibinfo{author}{\bibfnamefont{C.~F.} \bibnamefont{Roos}},
  \bibinfo{author}{\bibfnamefont{W.}~\bibnamefont{Hansel}},
  \bibinfo{author}{\bibfnamefont{J.}~\bibnamefont{Benhelm}},
  \bibinfo{author}{\bibfnamefont{G.~P.~T.} \bibnamefont{Lancaster}},
  \bibinfo{author}{\bibfnamefont{T.~W.} \bibnamefont{Korber}},
  \bibinfo{author}{\bibfnamefont{C.}~\bibnamefont{Becher}},
  \bibinfo{author}{\bibfnamefont{F.}~\bibnamefont{Schmidt-Kaler}},
  \bibinfo{author}{\bibfnamefont{D.~F.~V.} \bibnamefont{James}}, and
  \bibinfo{author}{\bibfnamefont{R.}~\bibnamefont{Blatt}},
  \bibinfo{year}{2004}, \bibinfo{journal}{Nature}
  \textbf{\bibinfo{volume}{429}}, \bibinfo{pages}{734}.

\bibitem[{\citenamefont{Robertson}(1929)}]{Robe1929}
\bibinfo{author}{\bibnamefont{Robertson}, \bibfnamefont{H.~P.}},
  \bibinfo{year}{1929}, \bibinfo{journal}{Phys. Rev.}
  \textbf{\bibinfo{volume}{34}}, \bibinfo{pages}{163}.

\bibitem[{\citenamefont{Roy} \emph{et~al.}(2012)\citenamefont{Roy, Shukla, and
  Mahesh}}]{Roy2012}
\bibinfo{author}{\bibnamefont{Roy}, \bibfnamefont{S.~S.}},
  \bibinfo{author}{\bibfnamefont{A.}~\bibnamefont{Shukla}}, and
  \bibinfo{author}{\bibfnamefont{T.~S.} \bibnamefont{Mahesh}},
  \bibinfo{year}{2012}, \bibinfo{journal}{Phys. Rev. A}
  \textbf{\bibinfo{volume}{85}}, \bibinfo{pages}{022109}.

\bibitem[{\citenamefont{Sagnac}(1913)}]{Sagnac1913}
\bibinfo{author}{\bibnamefont{Sagnac}, \bibfnamefont{G.}},
  \bibinfo{year}{1913}, \bibinfo{journal}{Comptes Rendus}
  \textbf{\bibinfo{volume}{157}}, \bibinfo{pages}{708}.

\bibitem[{\citenamefont{Scarcelli} \emph{et~al.}(2007)\citenamefont{Scarcelli,
  Zhou, and Shih}}]{Scarcelli2007}
\bibinfo{author}{\bibnamefont{Scarcelli}, \bibfnamefont{G.}},
  \bibinfo{author}{\bibfnamefont{Y.}~\bibnamefont{Zhou}}, and
  \bibinfo{author}{\bibfnamefont{Y.}~\bibnamefont{Shih}}, \bibinfo{year}{2007},
  \bibinfo{journal}{EPJ D} \textbf{\bibinfo{volume}{44}},
  \bibinfo{pages}{167}.

\bibitem[{\citenamefont{Scheidl} \emph{et~al.}(2010)\citenamefont{Scheidl}}]{Scheidl}
\bibinfo{author}{\bibnamefont{Scheidl}, \bibfnamefont{T.}},
  \bibinfo{author}{\bibfnamefont{R.}~\bibnamefont{Ursin}},
  \bibinfo{author}{\bibfnamefont{J.}~\bibnamefont{Kofler}},
  \bibinfo{author}{\bibfnamefont{S.}~\bibnamefont{Ramelow}},
  \bibinfo{author}{\bibfnamefont{X.}~\bibnamefont{Ma}},
  \bibinfo{author}{\bibfnamefont{T.}~\bibnamefont{Herbst}},
  \bibinfo{author}{\bibfnamefont{L.}~\bibnamefont{Ratschbacher}},
  \bibinfo{author}{\bibfnamefont{A.}~\bibnamefont{Fedrizzi}},
  \bibinfo{author}{\bibfnamefont{N.}~\bibnamefont{Langford}},
  \bibinfo{author}{\bibfnamefont{T.}~\bibnamefont{Jennewein}}, and
  \bibinfo{author}{\bibfnamefont{A.}~\bibnamefont{Zeilinger}},
  \bibinfo{year}{2010}, \bibinfo{journal}{Proc. Natl. Acad. Sci. USA}
  \textbf{\bibinfo{volume}{107}}, \bibinfo{pages}{19708}.

\bibitem[{\citenamefont{Schleich and Walther}(1986)}]{Schleich1986}
\bibinfo{author}{\bibnamefont{Schleich}, \bibfnamefont{W.}} and
  \bibinfo{author}{\bibfnamefont{H.}~\bibnamefont{Walther}},
  \bibinfo{year}{1986}, in \bibinfo{booktitle}{\emph{2nd International
  Symposium on foundations of Quantum Mechanics in the light of New
  Technology} \textrm{(Physical Society of Japan, Tokyo).}}

\bibitem[{\citenamefont{Schr\"odinger}(1930)}]{Schr1930}
\bibinfo{author}{\bibnamefont{Schr\"odinger}, \bibfnamefont{E.}},
  \bibinfo{year}{1930}, \bibinfo{journal}{Physikalisch-mathematische Klasse}
  \textbf{\bibinfo{volume}{14}}, \bibinfo{pages}{296}.

\bibitem[{\citenamefont{Schr\"odinger}(1935)}]{Schroedinger1935}
\bibinfo{author}{\bibnamefont{Schr\"odinger}, \bibfnamefont{E.}},
  \bibinfo{year}{1935}, \bibinfo{journal}{Naturwissenschaften}
  \textbf{\bibinfo{volume}{23}}, \bibinfo{pages}{844}.

\bibitem[{\citenamefont{Sciarrino} \emph{et~al.}(2002)\citenamefont{Sciarrino,
  Lombardi, Milani, and De~Martini}}]{Sciarrino2002}
\bibinfo{author}{\bibnamefont{Sciarrino}, \bibfnamefont{F.}},
  \bibinfo{author}{\bibfnamefont{E.}~\bibnamefont{Lombardi}},
  \bibinfo{author}{\bibfnamefont{G.}~\bibnamefont{Milani}}, and
  \bibinfo{author}{\bibfnamefont{F.}~\bibnamefont{De~Martini}},
  \bibinfo{year}{2002}, \bibinfo{journal}{Phys. Rev. A}
  \textbf{\bibinfo{volume}{66}}, \bibinfo{pages}{024309}.

\bibitem[{\citenamefont{Scully and Dr\"uhl}(1982)}]{Scully1982}
\bibinfo{author}{\bibnamefont{Scully}, \bibfnamefont{M.~O.}} and
  \bibinfo{author}{\bibfnamefont{K.}~\bibnamefont{Dr\"uhl}},
  \bibinfo{year}{1982}, \bibinfo{journal}{Phys. Rev. A}
  \textbf{\bibinfo{volume}{25}}, \bibinfo{pages}{2208}.

\bibitem[{\citenamefont{Scully} \emph{et~al.}(1989)}]{Scully1989}
\bibinfo{author}{\bibnamefont{Scully}, \bibfnamefont{M.~O.}},
 \bibinfo{author}{\bibfnamefont{B.-G.} \bibnamefont{Englert}}, and
 \bibinfo{author}{\bibfnamefont{J.}~\bibnamefont{Schwinger}},
  \bibinfo{year}{1989}, \bibinfo{journal}{Phys. Rev. A}
  \textbf{\bibinfo{volume}{40}}, \bibinfo{pages}{1775}.

\bibitem[{\citenamefont{Scully} \emph{et~al.}(1991)\citenamefont{Scully,
  Englert, and Walther}}]{Scully1991}
\bibinfo{author}{\bibnamefont{Scully}, \bibfnamefont{M.~O.}},
  \bibinfo{author}{\bibfnamefont{B.-G.} \bibnamefont{Englert}}, and
  \bibinfo{author}{\bibfnamefont{H.}~\bibnamefont{Walther}},
  \bibinfo{year}{1991}, \bibinfo{journal}{Nature}
  \textbf{\bibinfo{volume}{351}}, \bibinfo{pages}{111}.

\bibitem[{\citenamefont{Shalm, L. K., et al.}(2015)}]{Shalm2015}
\bibinfo{author}{\bibnamefont{Shalm}, \bibfnamefont{L.~K., \textit{et. al.} }} 
\bibinfo{year}{2015}, \bibinfo{journal}{Phys. Rev. Lett.}
  \textbf{\bibinfo{volume}{115}}, \bibinfo{pages}{250402}.

\bibitem[{\citenamefont{Shih and Alley}(1988)}]{Shih1988}
\bibinfo{author}{\bibnamefont{Shih}, \bibfnamefont{Y.~H.}} and
  \bibinfo{author}{\bibfnamefont{C.~O.} \bibnamefont{Alley}},
  \bibinfo{year}{1988}, \bibinfo{journal}{Phys. Rev. Lett.}
  \textbf{\bibinfo{volume}{61}}, \bibinfo{pages}{2921}.

\bibitem[{\citenamefont{Simon and Irvine}(2003)}]{Simon2003}
\bibinfo{author}{\bibnamefont{Simon}, \bibfnamefont{C.}} and
  \bibinfo{author}{\bibfnamefont{W.~T.~M.} \bibnamefont{Irvine}},
  \bibinfo{year}{2003}, \bibinfo{journal}{Phys. Rev. Lett.}
  \textbf{\bibinfo{volume}{91}}, \bibinfo{pages}{110405}.

\bibitem[{\citenamefont{Stapp}(2009)}]{Stapp2009}
\bibinfo{author}{\bibnamefont{Stapp}, \bibfnamefont{H.}}, \bibinfo{year}{2009},
in \bibinfo{title}{\emph{Compendium of Quantum Physics}, \textrm{ed.
D. Greenberger, K. Hentschel, and F. Weinert
(\bibinfo{publisher}{Springer})}.}

\bibitem[{\citenamefont{Storey} \emph{et~al.}(1994)\citenamefont{Stoery,
  Tan, Collett and Walls}}]{Stoery1994}
\bibinfo{author}{\bibnamefont{Storey}, \bibfnamefont{P.}},
  \bibinfo{author}{\bibfnamefont{S.} \bibnamefont{Tan}}, \bibinfo{author}{\bibfnamefont{M.} \bibnamefont{Collett}}, and
  \bibinfo{author}{\bibfnamefont{D.}~\bibnamefont{Walls}},
  \bibinfo{year}{1994}, \bibinfo{journal}{Nature}
  \textbf{\bibinfo{volume}{367}}, \bibinfo{pages}{626}.

\bibitem[{\citenamefont{Storey} \emph{et~al.}(1995)\citenamefont{Stoery,
  Tan, Collett and Walls}}]{Stoery1995}
\bibinfo{author}{\bibnamefont{Storey}, \bibfnamefont{P.}},
  \bibinfo{author}{\bibfnamefont{S.} \bibnamefont{Tan}}, \bibinfo{author}{\bibfnamefont{M.} \bibnamefont{Collett}}, and
  \bibinfo{author}{\bibfnamefont{D.}~\bibnamefont{Walls}},
  \bibinfo{year}{1995}, \bibinfo{journal}{Nature}
  \textbf{\bibinfo{volume}{375}}, \bibinfo{pages}{368}.

\bibitem[{\citenamefont{Tang} \emph{et~al.}(2012)\citenamefont{Tang, Li, Xu,
  Xiang, Li, and Guo}}]{Tang2012}
\bibinfo{author}{\bibnamefont{Tang}, \bibfnamefont{J.-S.}},
  \bibinfo{author}{\bibfnamefont{Y.-L.} \bibnamefont{Li}},
  \bibinfo{author}{\bibfnamefont{X.-Y.} \bibnamefont{Xu}},
  \bibinfo{author}{\bibfnamefont{G.-Y.} \bibnamefont{Xiang}},
  \bibinfo{author}{\bibfnamefont{C.-F.} \bibnamefont{Li}}, and
  \bibinfo{author}{\bibfnamefont{G.-C.} \bibnamefont{Guo}},
  \bibinfo{year}{2012}, \bibinfo{journal}{Nature Photon.}
  \textbf{\bibinfo{volume}{6}}, \bibinfo{pages}{600}.

\bibitem[{\citenamefont{Taylor}(1909)}]{Tayl1909}
\bibinfo{author}{\bibnamefont{Taylor}, \bibfnamefont{G.~I.}},
  \bibinfo{year}{1909}, \bibinfo{journal}{Proc. Cam. Phil. Soc.}
  \textbf{\bibinfo{volume}{15}}, \bibinfo{pages}{114}.

\bibitem[{\citenamefont{Walborn} \emph{et~al.}(2002)\citenamefont{Walborn,
  Terra~Cunha, P\'adua, and Monken}}]{Walborn2002}
\bibinfo{author}{\bibnamefont{Walborn}, \bibfnamefont{S.~P.}},
  \bibinfo{author}{\bibfnamefont{M.~O.} \bibnamefont{Terra~Cunha}},
  \bibinfo{author}{\bibfnamefont{S.}~\bibnamefont{P\'adua}}, and
  \bibinfo{author}{\bibfnamefont{C.~H.} \bibnamefont{Monken}},
  \bibinfo{year}{2002}, \bibinfo{journal}{Phys. Rev. A}
  \textbf{\bibinfo{volume}{65}}, \bibinfo{pages}{033818}.

\bibitem[{\citenamefont{Weisz} \emph{et~al.}(2014)\citenamefont{Weisz, Choi, Sivan, Heiblum, Gefen, Mahalu, and Umansky}}]{Weisz2014}
\bibinfo{author}{\bibnamefont{Weisz}, \bibfnamefont{E.}},
  \bibinfo{author}{\bibfnamefont{H.~K.}~\bibnamefont{Choi}},
  \bibinfo{author}{\bibfnamefont{I.} \bibnamefont{Sivan}},
  \bibinfo{author}{\bibfnamefont{M.} \bibnamefont{Heiblum}},
  \bibinfo{author}{\bibfnamefont{Y.} \bibnamefont{Gefen}},
  \bibinfo{author}{\bibfnamefont{D.} \bibnamefont{Mahalu}}, and
  \bibinfo{author}{\bibfnamefont{V.} \bibnamefont{Umansky}},
  \bibinfo{year}{2014}, \bibinfo{journal}{Science}
  \textbf{\bibinfo{volume}{344}}, \bibinfo{pages}{1363}.

\bibitem[{\citenamefont{Weizs{\"a}cker}(1941)}]{Weiz1941}
\bibinfo{author}{\bibnamefont{Weizs{\"a}cker}, \bibfnamefont{C.~F.~v.}},
  \bibinfo{year}{1941}, \bibinfo{journal}{Z. Phys.} \textbf{\bibinfo{volume}{118}},
  \bibinfo{pages}{489}.

\bibitem[{\citenamefont{Weizs{\"a}cker}(1931)}]{Weiz1931}
\bibinfo{author}{\bibnamefont{Weizs{\"a}cker}, \bibfnamefont{K.~F.~v.}},
  \bibinfo{year}{1931}, \bibinfo{journal}{Z. Phys.} \textbf{\bibinfo{volume}{70}},
  \bibinfo{pages}{114}.

\bibitem[{\citenamefont{Weyl}(1928)}]{Weyl1928}
\bibinfo{author}{\bibnamefont{Weyl}, \bibfnamefont{H.}}, \bibinfo{year}{1928},
  \emph{\bibinfo{title}{Gruppentheorie und Quantenmechanik}}
  (\bibinfo{publisher}{Leipzig, Hirzel}).

\bibitem[{\citenamefont{Wheeler}(1978)}]{Wheeler1978}
\bibinfo{author}{\bibnamefont{Wheeler}, \bibfnamefont{J.~A.}},
  \bibinfo{year}{1978}, in \bibinfo{title}{\emph{Mathematical Foundations
  of Quantum Theory}}, \textrm{ed. R. Marlow
  (\bibinfo{publisher}{Academic Press, New York})}.

\bibitem[{\citenamefont{Wheeler}(1984)}]{Wheeler1984}
\bibinfo{author}{\bibnamefont{Wheeler}, \bibfnamefont{J.~A.}},
  \bibinfo{year}{1984}, in \bibinfo{title}{\emph{Quantum Theory and
  Measurement}}, \textrm{ed. J. A. Wheeler and W. H. Zurek
  (\bibinfo{publisher}{Princeton University Press})}.

\bibitem[{\citenamefont{Wiseman and Harrison}(1995)\citenamefont{Wiseman and Harrison}}]{Wiseman1995}
\bibinfo{author}{\bibnamefont{Wiseman}, \bibfnamefont{H.~M.}} and
  \bibinfo{author}{\bibfnamefont{F.~E.}~\bibnamefont{Harrison}},
  \bibinfo{year}{1995}, \bibinfo{journal}{Nature}
  \textbf{\bibinfo{volume}{377}}, \bibinfo{pages}{584}.

\bibitem[{\citenamefont{Wiseman} \emph{et~al.}(1997)\citenamefont{Wiseman, Harrison, Collett, Tan, Walls, and Killip}}]{Wiseman1997}
\bibinfo{author}{\bibnamefont{Wiseman}, \bibfnamefont{H.~M.}},
  \bibinfo{author}{\bibfnamefont{F.~E.}~\bibnamefont{Harrison}},
  \bibinfo{author}{\bibfnamefont{M.~J.}~\bibnamefont{Collett}},
  \bibinfo{author}{\bibfnamefont{S.~M.}~\bibnamefont{Tan}},
  \bibinfo{author}{\bibfnamefont{D.~F.}~\bibnamefont{Walls}}, and
  \bibinfo{author}{\bibfnamefont{R.~B.}~\bibnamefont{Killip}},
  \bibinfo{year}{1997}, \bibinfo{journal}{Phys. Rev. A}
  \textbf{\bibinfo{volume}{56}}, \bibinfo{pages}{55}.

\bibitem[{\citenamefont{Wootters and Zurek}(1979)}]{Wootters1979}
\bibinfo{author}{\bibnamefont{Wootters}, \bibfnamefont{W.~K.}} and
  \bibinfo{author}{\bibfnamefont{W.~H.} \bibnamefont{Zurek}},
  \bibinfo{year}{1979}, \bibinfo{journal}{Phys. Rev. D}
  \textbf{\bibinfo{volume}{19}}, \bibinfo{pages}{473}.

\bibitem[{\citenamefont{Wootters and Fields}(1989)}]{Woot1989}
\bibinfo{author}{\bibnamefont{Wootters}, \bibfnamefont{W. K.}} and
  \bibinfo{author}{\bibfnamefont{B. D.}~\bibnamefont{Fields}},
  \bibinfo{year}{1989}, \bibinfo{journal}{Ann. Phys.}
  \textbf{\bibinfo{volume}{191}}, \bibinfo{pages}{363}.

\bibitem[{\citenamefont{Young}(1804)\citenamefont{Young}}]{Young1804}
\bibinfo{author}{\bibnamefont{Young}, \bibfnamefont{T.}},
  \bibinfo{year}{1804}, \bibinfo{journal}{Phil. Trans. R. Soc. Lond.}
  \textbf{\bibinfo{volume}{94}}, \bibinfo{pages}{1}.

\bibitem[{\citenamefont{Yuan} \emph{et~al.}(2008)\citenamefont{Yuan, Chen,
  Zhao, Chen, Schmiedmayer, and Pan}}]{Yuan2008}
\bibinfo{author}{\bibnamefont{Yuan}, \bibfnamefont{Z.-S.}},
  \bibinfo{author}{\bibfnamefont{Y.-A.} \bibnamefont{Chen}},
  \bibinfo{author}{\bibfnamefont{B.}~\bibnamefont{Zhao}},
  \bibinfo{author}{\bibfnamefont{S.}~\bibnamefont{Chen}},
  \bibinfo{author}{\bibfnamefont{J.}~\bibnamefont{Schmiedmayer}}, and
  \bibinfo{author}{\bibfnamefont{J.-W.} \bibnamefont{Pan}},
  \bibinfo{year}{2008}, \bibinfo{journal}{Nature}
  \textbf{\bibinfo{volume}{454}},
  \bibinfo{pages}{1098}.

\bibitem[{\citenamefont{Zehnder}(1891)\citenamefont{Zehnder}}]{Zehnder1891}
\bibinfo{author}{\bibnamefont{Zehnder}, \bibfnamefont{L.}},
  \bibinfo{year}{1891}, \bibinfo{journal}{Z. Instrumentenkd.}
  \textbf{\bibinfo{volume}{11}}, \bibinfo{pages}{275}.

\bibitem[{\citenamefont{Zeilinger}(1999)}]{Zeil1999}
\bibinfo{author}{\bibnamefont{Zeilinger}, \bibfnamefont{A.}},
  \bibinfo{year}{1999}, \bibinfo{journal}{Rev. Mod. Phys.}
  \textbf{\bibinfo{volume}{71}}, \bibinfo{pages}{S288}.

\bibitem[{\citenamefont{Zeilinger}(2005)}]{Zeil2005}
\bibinfo{author}{\bibnamefont{Zeilinger}, \bibfnamefont{A.}},
\bibinfo{author}{\bibnamefont{G.}, \bibfnamefont{Weihs}},
\bibinfo{author}{\bibnamefont{T.}, \bibfnamefont{Jennewein}}, and
\bibinfo{author}{\bibnamefont{M.}, \bibfnamefont{Aspelmeyer}},
  \bibinfo{year}{2005}, \bibinfo{journal}{Nature}
  \textbf{\bibinfo{volume}{433}}, \bibinfo{pages}{230}.

  \bibitem[{\citenamefont{Zukowski} \emph{et~al.}(1993)\citenamefont{Zukowski,
  Zeilinger, Horne, and Ekert}}]{Zukowski1993}
\bibinfo{author}{\bibnamefont{Zukowski}, \bibfnamefont{M.}},
  \bibinfo{author}{\bibfnamefont{A.}~\bibnamefont{Zeilinger}},
  \bibinfo{author}{\bibfnamefont{M.~A.} \bibnamefont{Horne}}, and
  \bibinfo{author}{\bibfnamefont{A.~K.} \bibnamefont{Ekert}},
  \bibinfo{year}{1993}, \bibinfo{journal}{Phys. Rev. Lett.}
  \textbf{\bibinfo{volume}{71}}, \bibinfo{pages}{4287}.

\end{thebibliography}

\end{document}